\documentclass[10pt,graphicx,subfigure,axodraw,cleverref]{article}
\usepackage{bm}
\usepackage{multirow}
\usepackage{amsfonts}
\usepackage{amssymb}
\usepackage{epsfig}
\usepackage{float}
\restylefloat{table}

\usepackage{cleveref}

\usepackage[usenames,dvipsnames]{color}
\newcommand{\beqn}{\begin{eqnarray}}
\newcommand{\eeqn}{\end{eqnarray}}
\newcommand{\be}{\begin{equation}}
\newcommand{\ee}{\end{equation}}

\def\s1{$s_{\alpha}$}
\def\s2{$s_{\gamma}$}
\def\s3{$s_{\delta}$}
\def\c1{$c_{\alpha}$}
\def\c2{$c_{\gamma}$}
\def\c3{$c_{\delta}$}

\def\45{\overline{45}}
\def\5{\overline{5}}
\def\70{\overline{70}}
\def\50{\overline{50}}

\newcommand{\ov}{\overline }

\def\c{\acute{c}}

\begin{document}
\baselineskip 18pt
\begin{titlepage}

\begin{flushright}
\end{flushright}

\begin{center}
{\bf {\ {\textsf{\Large{
 {\boldmath}
    }}}}}

\vskip 0.5 true cm \vspace{0cm}
\renewcommand{\thefootnote}
{\fnsymbol{footnote}}

{\small{\bf \Large
An Analysis of {\boldmath${B-L=-2}$} Operators from  Matter-Higgs Interactions in a Class of Supersymmetric
{\boldmath${SO(10)}$} Models}\\~\\
Pran Nath$^a$\footnote{Email: nath@neu.edu} and Raza M. Syed$^{a,b}$\footnote{
Email:  rsyed@aus.edu}\vskip 0.5 true cm}
\end{center}
\begin{center}
{\small\noindent
$^a$\textit{Department of Physics, Northeastern University,
Boston, MA 02115-5000, USA} \\
$^b$\textit{Department of Physics, American University of Sharjah,
P.O. Box 26666,
Sharjah, UAE\footnote{Permanent address}}}\\
\end{center}

\small{
\vskip 1.0 true cm \centerline{\bf Abstract}

\noindent Recently interest in GUT baryogenesis has been resurrected due to the observation that
$\mathsf{B}$-violating dimension seven operators that arise in grand unified theories that also violate
$\mathsf{B-L}$  produce baryon asymmetry {that} cannot be wiped out
by sphaleron processes.
While a general analysis of such higher dimensional operators from a bottom up approach
exists in the literature, a full analysis of them derived from  grand
unification does not exist. In this work we present a complete analysis of ${\mathsf{B-L}}=-2$ operators
within a realistic $\mathsf{SO(10)}$ grand unification where the doublet-triplet splitting {is} automatic
{via a missing partner mechanism}.  {Specifically we compute all allowed dimension five}, dimension seven
{and dimension nine operators arising from matter-Higgs interactions}.
The relative strength of all the allowed ${\mathsf{B-L}}=-2$ operators is given.
Such interactions are useful in  the study of  neutrino masses, baryogenesis,  proton decay
and {$n-\bar n$  oscillations} within a common  realistic grand unification framework.
}
\medskip
\noindent

\end{titlepage}

{\small
\section{Introduction \label{sec1}}

Analyses of higher dimensional operators within an effective field theory framework to explore physics beyond the
Standard Model has a long history
\cite{Weinberg:1979sa,Wilczek:1979hc,Weinberg:1980bf,Weldon:1980gi,Mohapatra:1980qe,Chang:1980qw,Kuo:1980ew,Nieves:1981tv,Rao:1982gt,Caswell:1982qs,Rao:1983sd,Leung:1984ni,Buchmuller:1985jz,Babu:2001ex,deGouvea:2007xp,Bonnet:2009ej,Grzadkowski:2010es,delAguila:2012nu,Angel:2012ug,Krauss:2013gy,Degrande:2013kka,Chalons:2013mya,Lehman:2014jma}. Included in such effective theories are $\mathsf{B-L}$ violating operators.
Such operators {already appear in the study of see saw neutrino masses and in the study of $n-\bar n$
oscillation and}
have gained further interest recently in the context of GUT scale baryogenesis
~\cite{Enomoto:2011py,Babu:2012iv,Babu:2012vb}.
  Thus while
baryogenesis arising from baryon and lepton number violating but $\mathsf{B-L}$ preserving interactions  from GUT
models is wiped out by sphaleron interactions which violate $\mathsf{B+L}$ and preserve $\mathsf{B-L}$, this is not the case for $\mathsf{B-L}$ {violating}
interactions. The simplest GUT model $\mathsf{SU(5)}$ {with renormalizable interactions
and $\mathsf{R}$ parity conservation}
has only $\mathsf{B-L}$ preserving interactions and is not a desirable
 model for GUT scale baryogenesis. However, $\mathsf{SO(10)}$ models ~\cite{georgi,Fritzsch:1974nn}
 can generate $\mathsf{B-L}$ violating interactions.
 {While a significant amount of work has been done recently in the study of baryogenesis within
 $\mathsf{SO(10)}$ using $\mathsf{B-L}$ violating interactions ~\cite{Babu:2012iv,Babu:2012vb},
to our knowledge there is as yet no complete analysis of $\mathsf{B-L}$ violating interactions that arise
in $\mathsf{SO(10)}$.}
In this work we give a full  and rigorous analysis of such interactions and compute dimension 5, 7 and 9
${\mathsf{B-L}}=-2$ interactions within a class of $\mathsf{SO(10)}$ models where the $\mathsf{B-L}$ violating operators arise from matter-Higgs
interactions. The model we consider has a natural doublet-triplet splitting within the framework of a missing partner
mechanism \cite{Masiero:1982fe,Grinstein:1982um,Babu:2006nf,Babu:2011tw}
{(for a recent application of $\mathsf{SO(10)}$ missing partner model see \cite{Du:2013nza})}.
While our analysis is done within  a specific model,  it is likely to be applicable
to a broader class of models where the light spectrum is that of MSSM.\\

 The outline of the rest of the paper is as follows:
In Sec.(\ref{sec2}), we give the details of the $\mathsf{SO(10)}$ model.
Here, we also  discuss the  spontaneous breaking of
the $\mathsf{SO(10)}$ GUT symmetry down to the symmetry of the Standard Model gauge group, i.e., the symmetry $\mathsf{SU(3)_C\times SU(2)_L\times U(1)_Y}$
using  $\mathsf{126+\overline{126}+210}$ multiplets.
{The GUT sector of this model is the same as discussed in \cite{1} }.
The symmetry breaking in these models was investigated in \cite{2} where  a cubic
equation for spontaneous symmetry breaking was obtained.
The models of \cite{1} had in addition 10-plets of Higgs
and were later extended to include $\mathsf{120}$-plet of Higgs \cite{3}. Further applications of such models were
made in a number of works \cite{4,5,6}.  While the GUT sector of the model considered here is identical
to the previous works the model overall is  different for the following reasons: In the usual GUT models to which the
works listed above belong, the choice of the Higgs content is arbitrary. For example, in the previous models one can
add any number of additional Higgs fields, such as one or more $\mathsf{10}$-plets and $\mathsf{120}$-plets as there is no principle that
restricts it. In the missing partner model the Higgs sector is strictly constrained and in $\mathsf{SO(10)}$ only few examples
are known ~\cite{Babu:2006nf,Babu:2011tw}. Specifically the GUT sector must be enchored either in $\mathsf{126}+
\overline{\mathsf{126}}$ or $\mathsf{560}+\overline{\mathsf{560}}$.  Even more stringent is the constraint on the light Higgs sector, i.e., the
sector which provides a component to the light Higgs doublet. Thus once an anchor in the GUT sector is assumed
the light sector cannot be chosen in an arbitrary fashion as is possible in the usual GUT models. This is needed
to satisfy two constraints: first to ensure that the light sector has an excess number of Higgs doublets by one over the
 heavy Higgs sector while there is an exact match of the Higgs triplets/anti-triplets between the light and the heavy sectors.
 This ensures that all Higgs doublets will become heavy except one and all Higgs triplets will become heavy because
 of mixing between the light and the heavy sectors. Second, that all the exotic fields in the light Higgs sector will become
 heavy as a result of mixing with the heavy fields. These constrains are strong enough to eliminate a large number of
 models except the ones listed in ~\cite{Babu:2006nf,Babu:2011tw}. For example, for the case when the GUT sector
 is assumed  to be $560+\overline{560}$, which breaks the GUT symmetry down to $\mathsf{SU(3)_C}\times \mathsf{SU(2)_L}\times \mathsf{U(1)_Y}$,
 the light sector must consist only of the fields $2\times \mathsf{10}+\mathsf{320}$. Further, no masses are allowed for the light sector
 and the mass generation is allowed only via mixings with the heavy sector. This automatically requires several
 couplings to vanish. The model we consider here
requires us to choose in an unambiguous manner a light sector which is $2\times \mathsf{10}+\mathsf{120}$.
The doublet-triplet splitting has been an Achilles heel of grand unification and the missing partner mechanism is
one of the ways the problem can be redressed. Thus $\mathsf{SO(10)}$ model which contains many desirable features
coupled with the missing partner mechanism provides a natural framework for grand unification.\\

 In Sec.(\ref{sec4}), we give details of the doublet-triplet splitting and determine the linear combination of the  fields in the $2\times \mathsf{10+120}$
 plet of Higgs
 that produce a pair of light Higgs doublets.
 In Sec.(\ref{sec:5}), we give analysis of the ${\mathsf{B-L}}=-2$ operators arising from
 matter-Higgs interactions. A discussion of results  is given in Sec.(\ref{sec6}).
 {Conclusions are given in Sec.(6).}
 Further details of the analysis are given in several appendices. In Appendix A, we define the notation and give the decomposition of the
 $\mathsf{SO(10)}$ multiplets in terms of the $\mathsf{SU(5)}$ multiplets. {Appendix B contains the reduction of
 $\mathsf{24, 45, 50, 75}$ plets of
 $\mathsf{SU(5)}$ fields  in terms  of component states with
 $\mathsf{SU(3)_C\times SU(2)_L \times U(1)_Y}$  quantum numbers.  These fields enter in
  the spontaneous breaking of GUT and electroweak symmetry.}
 In Appendix {C}, we give additional details of the GUT symmetry breaking.
In Appendix {D}, we give a  further discussion of the following sets of $\mathsf{SO(10)}$ Higgs couplings
{in $SU(5)\times U(1)$ decomposition}:
 $\mathsf{10\cdot126\cdot210}$,   $\mathsf{10\cdot\overline{126}\cdot210}$, $\mathsf{120\cdot126\cdot210}$ and $\mathsf{120\cdot\overline{126}\cdot210}$.  These couplings enter in the doublet-triplet splitting.
 The analysis of these couplings
 is based on the oscillator mechanism~\cite{Mohapatra:1979nn,Wilczek:1981iz}
 using techniques developed in~\cite{Nath:2001uw,Nath:2001yj,Nath:2003rc,Syed:2005gd,Babu:2005gx,Nath:2005bx,Babu:2006rp}.
 {An analysis of such couplings in Pati-Salam subgroups was given in earlier works of
 \cite{4,Aulakh:2002zr}   using techniques of ~\cite{Aulakh:2002zr}.}
 Finally in Appendix {F} we exhibit some of the coefficients of $\mathsf{B-L=-2}$ operators.
  In this work we do not discuss $\mathsf{B-L}$ violating interactions that arise from four-point
 Higgs interactions. There are a large number of such interactions and they include  couplings of the type
 $(\mathsf{126}\times \mathsf{126})_r\cdot(X\times Y)_r$ and $(\mathsf{126}\times \mathsf{\overline{126}})_r\cdot(X\times Y)_r$
where $X,Y=\mathsf{10},~\mathsf{45},~\mathsf{54},~\mathsf{120},~\mathsf{126}, ~\mathsf{\overline{126}}, ~\mathsf{210}$. Many of these operators are discussed in~\cite{Babu:2012iv}.
  A full analysis of the $\mathsf{B-L}$ violating interactions from this set is outside the scope of this  work and requires a separate analysis.

\section{The {\boldmath$\mathsf{SO(10)}$} model \label{sec2}}
The $\mathsf{SO(10)}$ model we discuss has the following particle content ~\cite{Babu:2011tw}
\begin{equation}
\mathsf{126}(\Delta_{\mu\nu\rho\sigma\lambda}),
 ~~~\overline{\mathsf{126}}(\overline{\Delta}_{\mu\nu\rho\sigma\lambda}), ~~~\mathsf{210}(\Phi_{\mu\nu\rho\sigma}),~~~
 \mathsf{10_1}({}^{1}\Gamma_\mu), ~~\mathsf{10_2}({}^{2}\Gamma_\mu), ~~
\mathsf{120}(\Sigma_{\mu\nu\lambda}).
\label{multiplets}
\end{equation}
Here the fields   $\mathsf{126+\overline{126}+210}$ constitute the heavy sector while the fields $2\times \mathsf{10+120}$ constitute the
light sector.
Additionally, the model contains three generations of  matter fields which reside in three copies of  $\mathsf{16}${-plet} spinor representation of $\mathsf{SO(10)}$.
The light fields consisting of $2\times \mathsf{10 + 120}$, together with the heavy sector,
generate the desired doublet-triplet  splitting  and make all the triplets and doublets heavy except for one pair of Higgs doublets.
This will be discussed in detail in the next section. Here we discuss the breaking of the $\mathsf{SO(10)}$  GUT symmetry to the Standard Model
gauge {group}.
{{The spontaneous breaking of the GUT symmetry for this model was first discussed in~\cite{2}.}
Here we discuss it to set up the framework for the discussion  in the following sections.}
Thus the breaking comes about as follows:  The
$\mathsf{126+\ov{126}}$ multiplets reduces the rank of the group and the $\mathsf{210}$ plet
breaks the rest of the gauge symmetry down to the Standard Model gauge group. The superpotential that breaks the
GUT symmetry is
\begin{equation}
  W_{{GUT}}= {m_{\Phi}}\Phi_{\mu\nu\sigma\xi}\Phi_{\mu\nu\sigma\xi}+{m_{\Delta}}\Delta_{\mu\nu\sigma\xi\zeta}\overline{\Delta}_{\mu\nu\sigma\xi\zeta}
  +{\lambda}
  \Phi_{\mu\nu\sigma\xi}\Phi_{\sigma\xi\rho\tau}\Phi_{\rho\tau\mu\nu}+ {\eta}\Phi_{\mu\nu\sigma\xi}\Delta_{\mu\nu\rho\tau\zeta}\overline{\Delta}_{\sigma\xi\rho\tau\zeta},
   \label{superpotential gut so(10)tensor}
\end{equation}
where $m_{\Phi}$ is the mass of the $\mathsf{210}$-plet field and $m_{\Delta}$ is the mass  of the $\mathsf{126 +\ov{126}}$ multiplets.
The fields that develop VEVs  {are $\mathsf{SU(3)_C\times SU(2)_L\times U(1)_Y}$ singlets: ${\mathbf S}_{1_{126}}$, ${\bf S}_{1_{\ov{126}}}$, $\mathbf S_{1_{_{{210}}}}                  $,              $\mathbf S_{24_{_{{210}}}}                  $, $\mathbf S_{75_{_{{210}}}}  $. {Here for example  $\mathbf S_{24_{_{{210}}}}$ means that the  Standard Model singlet is in the $\mathsf{24}$ plet of $\mathsf{SU(5)}$ contained in the $210$ plet of $\mathsf{SO(10)}$
{(See Appendix B for further details)}.
 The superpotential, when expanded in terms of {these} Standard Model singlets, gives (see Appendix {C} for details)
\begin{eqnarray}
  W_{{GUT}}&=&{m_{\Phi}}\left(\frac{3}{4}\mathbf S_{75_{_{{210}}}}                  ^2+\frac{5}{12}\mathbf S_{24_{_{{210}}}}                  ^2+\frac{3}{80}\mathbf S_{1_{_{{210}}}}                  ^2+\cdots\right)+{m_{\Delta}}\left(\frac{15}{2}\mathbf S_{1_{_{{126}}}}                  \mathbf S_{1_{_{\overline{126}}}}     +\cdots\right)\nonumber\\
  &&+\lambda\left({\frac{1}{18}}\mathbf S_{75_{_{{210}}}}                  ^3-{\frac{1}{18}}\mathbf S_{75_{_{{210}}}}                  ^2\mathbf S_{24_{_{{210}}}}                  +\frac{25}{864}\mathbf S_{75_{_{{210}}}}                  \mathbf S_{24_{_{{210}}}}                  ^2+\frac{1}{40}\mathbf S_{75_{_{{210}}}}                  ^2\mathbf S_{1_{_{{210}}}}                  \right.\nonumber\\
  &&~~~~~~\left.-\frac{35}{3888}\mathbf S_{24_{_{{210}}}}                  ^3-\frac{1}{192}\mathbf S_{24_{_{{210}}}}                  ^2\mathbf S_{1_{_{{210}}}}                  -\frac{3}{3200}\mathbf S_{1_{_{{210}}}}                  ^3+\cdots\right)\nonumber\\
  &&+\eta\left(-\frac{3}{16}\mathbf S_{1_{_{{210}}}}                  \mathbf S_{1_{_{{126}}}}                  \mathbf S_{1_{_{\overline{126}}}}     +\cdots\right).\label{superpotential gut su(5)tensor}
\end{eqnarray}
Vanishing of the D-terms implies $\mathbf S_{1_{_{{126}}}}                  =\mathbf S_{1_{_{\overline{126}}}}     $, while the F-terms yield
\begin{eqnarray}
\frac{3}{40}{m_{\Phi}}\mathbf S_{1_{_{{210}}}}                  +\lambda\left(\frac{1}{40}\mathbf S_{75_{_{{210}}}}                  ^2-\frac{1}{192}\mathbf S_{24_{_{{210}}}}                  ^2-\frac{9}{3200}\mathbf S_{1_{_{{210}}}}                  ^2\right)-\frac{3}{16}\eta\mathbf S_{1_{_{{126}}}}                  ^2&=&0,\nonumber\\
\frac{5}{6}{m_{\Phi}}\mathbf S_{24_{_{{210}}}}                  +\lambda\left(-\frac{1}{{18}}\mathbf S_{75_{_{{210}}}}                  ^2+\frac{25}{432}\mathbf S_{75_{_{{210}}}}                  \mathbf S_{24_{_{{210}}}}                  -\frac{35}{1296}\mathbf S_{24_{_{{210}}}}                  ^2-\frac{1}{96}\mathbf S_{24_{_{{210}}}}                  \mathbf S_{1_{_{{210}}}}                  \right)&=&0,\nonumber\\
\frac{3}{2}{m_{\Phi}}\mathbf S_{75_{_{{210}}}}                  +\lambda\left(\frac{1}{{6}}\mathbf S_{75_{_{{210}}}}                  ^2-\frac{1}{{9}}\mathbf S_{75_{_{{210}}}}                  \mathbf S_{24_{_{{210}}}}                  +\frac{25}{864}\mathbf S_{24_{_{{210}}}}                  ^2+\frac{1}{20}\mathbf S_{75_{_{{210}}}}                  \mathbf S_{1_{_{{210}}}}                  \right)&=&0,\nonumber\\
15{m_{\Delta}}\mathbf S_{1_{_{{126}}}}                  -\eta\frac{3}{8}\mathbf S_{1_{_{{210}}}}                  \mathbf S_{1_{_{{126}}}}                  &=&0.\label{F Terms}
\end{eqnarray}
For the sake of brevity, we define
\begin{eqnarray}
{\cal{M}}_{\Delta}\equiv\frac{{m_{\Delta}}}{\eta},~~~~~~~{\cal{M}}_{\Phi}\equiv\frac{{m_{\Phi}}}{\lambda}.\label{Redefine parameters}
\end{eqnarray}
Note that $\mathbf S_{1_{_{{210}}}}                  $ can be immediately solved for and is given by
\begin{eqnarray}
\mathbf S_{1_{_{{210}}}}                  =40{\cal{M}}_{\Delta}.\label{Determination S_1_210}
\end{eqnarray}
Except for a trivial solution, $\mathbf S_{24_{_{{210}}}}                  $ satisfies a cubic equation
\begin{eqnarray}
{9}\mathbf S_{24_{_{{210}}}}                  ^3+{24}\mathbf S_{24_{_{{210}}}}                  ^2\left({28}{\cal{M}}_{\Delta}{-45}{\cal{M}}_{\Phi}\right)+ {64}\mathbf S_{24_{_{{210}}}}                  \left({320}{\cal{M}}_{\Delta}^2{-279}{\cal{M}}_{\Delta}{\cal{M}}_{\Phi}\right.\nonumber\\
\left.+{972}{\cal{M}}_{\Phi}^2\right)+{13824}\left({\cal{M}}_{\Delta}-2{\cal{M}}_{\Phi}\right)\left(4{\cal{M}}_{\Delta}+3{\cal{M}}_{\Phi}\right)^2=0.\label{Determination S_24_210}
\label{cubic-equation}
\end{eqnarray}
Once $\mathbf S_{24_{_{{210}}}}                  $ is determined, $\mathbf S_{75_{_{{210}}}}                  $ and $\mathbf S_{1_{_{{126}}}}                  =\mathbf S_{1_{_{\overline{126}}}}     $ are given by
\begin{eqnarray}
\mathbf S_{75_{_{{210}}}}                  &=&\frac{{5}\left[\mathbf S_{24_{_{{210}}}}                  ^2+24\mathbf S_{24_{_{{210}}}}                  \left({\cal{M}}_{\Delta}-2{\cal{M}}_{\Phi}\right)\right]}
{{6}\left[{1}\mathbf S_{24_{_{{210}}}}                  +{8}\left(4{\cal{M}}_{\Delta}+3{\cal{M}}_{\Phi}\right)\right]},\label{Determination S_75_210}\\
\mathbf S_{1_{_{{126}}}}                  \cdot\mathbf S_{1_{_{\overline{126}}}}     &=&\left(\frac{\lambda}{\eta}\right)
\frac{1}{{{216}\left({1}\mathbf S_{24_{_{{210}}}}                  +{32}{\cal{M}}_{\Delta}+{24}{\cal{M}}_{\Phi}\right)}}
\left[{5}\mathbf S_{24_{_{{210}}}}                  ^3
-{32}\mathbf S_{24_{_{{210}}}}                  ^2\left(8{\cal{M}}_{\Delta}+{39}{\cal{M}}_{\Phi}\right)\right.\nonumber\\
&&\left.~-{1728}\mathbf S_{24_{_{{210}}}}                  \left({7}{\cal{M}}_{\Delta}^2-{7}{\cal{M}}_{\Delta}{\cal{M}}_{\Phi}{-6}{\cal{M}}_{\Phi}^2\right)\right.\nonumber\\
&&\left.~-{13824}{\cal{M}}_{\Delta}\left(3{\cal{M}}_{\Delta}-2{\cal{M}}_{\Phi}\right)\left(4{\cal{M}}_{\Delta}+3{\cal{M}}_{\Phi}\right)\right].\label{Determination S_1_126}
\end{eqnarray}
Thus all the VEVs, i.e., $\mathbf S_{1_{_{{210}}}}                  , ~\mathbf S_{75_{_{{210}}}}                  , ~\mathbf S_{1_{_{{126}}}}                  , ~\mathbf S_{1_{_{\overline{126}}}}     $ can be determined in terms of just one
VEV, i.e., $\mathbf S_{24_{_{{210}}}}                  $  using the minimization conditions
 Eq.(\ref{F Terms}).
 {We note that  the VEVs depend on all the four parameters $m_{\Delta},~m_{\Phi},~\lambda$ and $\eta$.}
 Table \ref{Estimation of SM singlets} gives the numerical estimates of these Standard Model singlets. {{Note that in Table \ref{Estimation of SM singlets} corresponding to each set of ${\cal{M}}_{\Delta}$ and ${\cal{M}}_{\Phi}$, there exists three solutions for $\mathbf S_{24_{_{{210}}}}                  $ as given by Eq.(\ref{Determination S_24_210})}.}
{We note that the cubic equation for spontaneous symmetry breaking exhibited in Eq.(\ref{cubic-equation})
 was first obtained in the work of \cite{2}. }\\

\section{Light and heavy Higgs fields after spontaneous breaking of the GUT symmetry
  \label{sec4}}

As mentioned in Sec.(\ref{sec2}), the doublet-triplet splitting arises  as a consequence of mixing between the light
sector consisting of $2\times \mathsf{10+ 120}$ plets of Higgs fields and the heavy sector consisting of $\mathsf{126+\ov{126} + 210}$ of
Higgs fields. The interactions mixing the light and the heavy fields are given in Appendix {D}.
As discussed in~\cite{Babu:2011tw}
  the light sector contains four pairs of
Higgs doublets and four pairs of Higgs triplets/anti-triplets while the heavy sector consists of three pairs of Higgs doublets
and four pairs of Higgs {triplets/antitriplets}
leaving only one  pair of light Higgs doublet. This light Higgs doublet pair is, in general,
a linear combination of the
{Higgs doublets in the $\mathsf{10}$-plets, in the $\mathsf{120}$-plet and in the $\mathsf{\overline{126}}$-plet of Higgs fields and there is no component of it
in the heavy Higgs sector which breaks the GUT symmetry. Actually it turns out that the light Higgs field in the
present model is  a  linear combination only of the Higgs doublets that arise from the $\mathsf{10_1+ 10_2}$ plets
and from the $\mathsf{120}$ plet of $\mathsf{SO(10)}$ Higgs fields. } In this section, we discuss the details of the analysis to determine
the exact combination of these doublets in the residual light doublet pair.
The superpotential that enters in the doublet-triplet splitting is given by
\begin{eqnarray}
W_{{DT}}=\mathrm{A}~{}^{1}{\Gamma}_{\mu}\Delta_{\mu\nu\sigma\xi\zeta}\Phi_{\mu\sigma\xi\zeta}+
\mathrm{B_r}~{}^{r}{\Gamma}_{\mu}\overline{\Delta}_{\mu\nu\sigma\xi\zeta}\Phi_{\mu\sigma\xi\zeta}+\mathrm{C}~\Sigma_{\mu\nu\sigma}\Delta_{\nu\sigma\xi\zeta\rho}
\Phi_{\mu\xi\zeta\rho}+\overline{\mathrm{C}}~\Sigma_{\mu\nu\sigma}\overline{\Delta}_{\nu\sigma\xi\zeta\rho}
\Phi_{\mu\xi\zeta\rho},
\label{wdt.1}
\end{eqnarray}
where $\mathrm{r}=1,2$.
Next, we
 exhibit the Higgs doublet pairs ({{\textnormal{D}}}) consisting of up- and  down-type Higgs and Higgs triplet/anti-triplet pairs ({{\textnormal{T}}}) that participate in the missing partner mechanism.
 {Note that we could have added the  following set of terms to $W_{DT}$
 allowed by the gauge invariance of the theory

\begin{eqnarray}
\Delta W_{DT}= m_{rs}~ \mathsf{10}^{(r)} \cdot \mathsf{10}^{(s)}+
\lambda_\Sigma~ \mathsf{120}\cdot \mathsf{120}
+   \lambda^{(r)}_{{10}\cdot {120}\cdot {210}}~ \mathsf{10}^{(r)}\cdot \mathsf{120}\cdot \mathsf{210}
+ \lambda_{120^2\cdot210}~ \mathsf{120}\cdot \mathsf{120}\cdot \mathsf{210}
\end{eqnarray}
 The missing partner mechanism requires
 \begin{eqnarray}
m_{rs} =0, ~  \lambda_\Sigma=0,~ \lambda^{(r)}_{10\cdot120\cdot210}=0,
 ~\lambda_{120^2\cdot210}=0.
\end{eqnarray}
 which represents a significant reduction of parameters.
 }

{{\small
\begin{table}[htb]
\begin{center}
\begin{tabular}{|||c|c|||c|c|||}
\hline
&&&\\
 \small $\mathsf{SU(2)_L}$ Doublet & \small$\mathsf{SU(3)_C}$ & \small$\mathsf{SU(2)_L}$ Doublet  &\small$\mathsf{SU(3)_C}$ \\
 \small(Up-Type) &\small Triplet & \small(Down-Type)  &  \small Anti-Triplet\\
   \hline
    \hline
   &&&\\
  ${}^{({5}_{10_1})}\!{\mathsf D}^{a}$& ${}^{({5}_{10_1})}\!{\mathsf T}^{\alpha}$ & ${}^{({\overline{5}}_{10_1})}\!{\mathsf D}_{a}$& ${}^{({\overline{5}}_{10_1})}\!{\mathsf T}_{\alpha}$\\
    \hline
    &&&\\
   ${}^{({5}_{10_2})}\!{\mathsf D}^{a}$ & ${}^{({5}_{10_2})}\!{\mathsf T}^{\alpha}$& ${}^{({\overline{5}}_{10_2})}\!{\mathsf D}_{a}$ & ${}^{({\overline{5}}_{10_2})}\!{\mathsf T}_{\alpha}$\\
    \hline
    &&&\\
    ${}^{({5}_{120})}\!{\mathsf D}^{a}$& ${}^{({5}_{120})}\!{\mathsf T}^{\alpha}$ & ${}^{({\overline{5}}_{120})}\!{\mathsf D}_{a}$& ${}^{({\overline{5}}_{120})}\!{\mathsf T}_{\alpha}$ \\
    \hline
    &&&\\
     ${}^{({5}_{\overline{126}})}\!{\mathsf D}^{a}$& ${}^{({5}_{\overline{126}})}\!{\mathsf T}^{\alpha}$ &${}^{({\overline{5}}_{{126}})}\!{\mathsf D}_{a}$& ${}^{({\overline{5}}_{{126}})}\!{\mathsf T}_{\alpha}$\\
    \hline
    &&&\\
    ${}^{({5}_{210})}\!{\mathsf D}^{a}$ &${}^{({5}_{210})}\!{\mathsf T}^{\alpha}$ &${}^{({\overline{5}}_{210})}\!{\mathsf D}_{a}$ &${}^{({\overline{5}}_{210})}\!{\mathsf T}_{\alpha}$ \\
    \hline
    &&&\\
   ${}^{({45}_{120})}\!{\mathsf D}^{a}$ & ${}^{({45}_{120})}\!{\mathsf T}^{\alpha}$ &${}^{({\overline{45}}_{120})}\!{\mathsf D}_{a}$ & ${}^{({\overline{45}}_{120})}\!{\mathsf T}_{\alpha}$\\
    \hline
    &&&\\
   ${}^{({45}_{126})}\!{\mathsf D}^{a}$ &${}^{({45}_{126})}\!{\mathsf T}^{\alpha}$&${}^{({\overline{45}}_{\overline{126}})}\!{\mathsf D}_{a}$ &${}^{({\overline{45}}_{\overline{126}})}\!{\mathsf T}^{\alpha}$\\
    \hline
     &&&\\
    $-$&${}^{({50}_{\overline{126}})}\!{\mathsf T}^{\alpha}$  &$-$&  ${}^{({\overline{50}}_{{126}})}\!{\mathsf T}_{\alpha}$ \\
    \hline
\end{tabular}\\~\\

\caption{\small
{{Symbolic representation}} of up--type and down--type Higgs doublets,
and Higgs triplet and anti-triplet pairs in the $\mathsf{SO(10)}$ missing partner model discussed in this work.}\label{t2}
\end{center}
\end{table}
}}
\vskip 0.5cm

 \begin{enumerate}
 \item {Pairs of {{\textnormal{D}}} and {{\textnormal{T}}} in the Heavy Sector}
 \begin{eqnarray*}
 \mathsf{126+\overline{126}}&\supset& 2{{\textnormal{D}}}\left\{\left({}^{({5}_{\overline{126}})}\!{\mathsf D}^{a}, {}^{(\overline{5}_{{126}})}\!{\mathsf D}_{a}\right); \left({}^{({45}_{{126}})}\!{\mathsf D}^{a}, {}^{(\overline{45}_{\overline{126}})}\!{\mathsf D}_{a}\right)\right\}+3{{\textnormal{T}}}\left\{\left({}^{({5}_{\overline{126}})}\!{\mathsf T}^{\alpha}, {}^{(\overline{5}_{{126}})}\!{\mathsf T}_{\alpha}\right)\right.;\\
&&\left.\left({}^{({45}_{{126}})}\!{\mathsf T}^{\alpha}, {}^{(\overline{45}_{\overline{126}})}\!{\mathsf T}_{\alpha}\right);\left({}^{({50}_{\overline{126}})}\!{\mathsf T}^{\alpha}, {}^{(\overline{50}_{{126}})}\!{\mathsf T}_{\alpha}\right)\right\} \\
 \mathsf{210} &\supset& 1{{\textnormal{D}}}\left({}^{({5}_{{210}})}\!{\mathsf D}^{a}, {}^{(\overline{5}_{{210}})}\!{\mathsf D}_{a}\right)+1{{\textnormal{T}}}\left({}^{({5}_{{210}})}\!{\mathsf T}^{\alpha}, {}^{(\overline{5}_{{210}})}\!{\mathsf T}_{\alpha}\right)
 \end{eqnarray*}
  \item {Pairs of {{\textnormal{D}}} and {{\textnormal{T}}} in the Light Sector}
  \begin{eqnarray*}
 2\times\mathsf{10}&\supset& 2{{\textnormal{D}}}\left\{\left({}^{({5}_{{10_1}})}\!{\mathsf D}^{a}, {}^{(\overline{5}_{{10_1}})}\!{\mathsf D}_{a}\right); \left({}^{({5}_{{10_2}})}\!{\mathsf D}^{a}, {}^{(\overline{5}_{{10_2}})}\!{\mathsf D}_{a}\right)\right\}+2{{\textnormal{T}}}\left\{\left({}^{({5}_{{10_1}})}\!{\mathsf T}^{\alpha}, {}^{(\overline{5}_{{10_1}})}\!{\mathsf T}_{\alpha}\right);\right.\\
 &&\left.\left({}^{({5}_{{10_2}})}\!{\mathsf T}^{\alpha}, {}^{(\overline{5}_{{10_2}})}\!{\mathsf T}_{\alpha}\right)\right\}\\
 \mathsf{120} &\supset& 2{{\textnormal{D}}}\left\{\left({}^{({5}_{{120}})}\!{\mathsf D}^{a}, {}^{(\overline{5}_{{120}})}\!{\mathsf D}_{a}\right); \left({}^{({45}_{{120}})}\!{\mathsf D}^{a}, {}^{(\overline{45}_{{120}})}\!{\mathsf D}_{a}\right)\right\}+2{{\textnormal{T}}}\left\{\left({}^{({5}_{{120}})}\!{\mathsf T}^{\alpha}, {}^{(\overline{5}_{{120}})}\!{\mathsf T}_{\alpha}\right);\right.\\
 &&\left.\left({}^{({45}_{{120}})}\!{\mathsf T}^{\alpha}, {}^{(\overline{45}_{{120}})}\!{\mathsf T}_{\alpha}\right)\right\}
  \end{eqnarray*}
   \item {Residual Set of Light modes}: 1{{\textnormal{D}}}
\end{enumerate}

 \vskip 0.5cm
\noindent  {Here, for example,  ${}^{({45}_{120})}\!{\mathsf D}^{a}$  means that the doublet  is in the $\mathsf{45}$ plet of $\mathsf{SU(5)}$ contained in the $120$ plet of $\mathsf{SO(10)}$.} Here $\alpha=1,2,3$  and $a=4,5$ represent $\mathsf{SU(3)}$ color and $\mathsf{SU(2)}$ weak indices, respectively.
{The result of the above analysis is summarized  in Table \ref{t2}}.
 The mass terms for the $\mathsf{SU(2)}$ doublets in the superpotential can be written as

 {
\begin{eqnarray}
\left({}^{({5}_{10_1})}\!{\mathsf D}^{a},{}^{({5}_{10_2})}\!{\mathsf D}^{a},{}^{({5}_{120})}\!{\mathsf D}^{a},{}^{({5}_{\overline{126}})}\!{\mathsf D}^{a},{}^{({5}_{{210}})}\!{\mathsf D}^{a},{}^{({45}_{120})}\!{\mathsf D}^{a},{}^{({45}_{{126}})}\!{\mathsf D}^{a} \right){M_d}\left(\matrix{{}^{(\overline{5}_{10_1})}\!{\mathsf D}_{a}\cr {}^{(\overline{5}_{10_2})}\!{\mathsf D}_{a}\cr{}^{(\overline{5}_{120})}\!{\mathsf D}_{a}\cr{}^{(\overline{5}_{{126}})}\!{\mathsf D}_{a} \cr{}^{(\overline{5}_{{210}})}\!{\mathsf D}_{a} \cr{}^{(\overline{45}_{120})}\!{\mathsf D}_{a} \cr{}^{(\overline{45}_{\overline{126}})}\!{\mathsf D}_{a}} \right),
\end{eqnarray}\label{Mass term 1}
}
where the doublet mass matrix ${M_d}$ {receives contributions from \cref{superpotential gut so(10)tensor} and
  Eq.(\ref{wdt.1}) and is given by}

{\small
 {\beqn
{M_d}=
\left(\matrix{
0& 0& 0&\displaystyle\mathsf{d_2} &&&\displaystyle\mathsf{d_1} &&0 &\displaystyle\left(\frac{b_1}{a}\right)\mathsf{d_3}\cr\cr
0& 0&0 & 0&&& 0&& 0&\displaystyle\left(\frac{b_2}{a}\right)\mathsf{d_3} \cr\cr
0& 0& 0&\displaystyle\mathsf{d_5} &&&\displaystyle\mathsf{d_4} && 0&\displaystyle\left(\frac{\overline{c}}{c}\right)\mathsf{d_6} \cr\cr
 \displaystyle\left(\frac{b_1}{a}\right)\mathsf{d_2}&  \displaystyle\left(\frac{b_2}{a}\right)\mathsf{d_2}&\displaystyle\left(\frac{\overline{c}}{c}\right)\mathsf{d_5}& {\mathsf{d_9}}&&& {\mathsf{d_{11}}}&&\displaystyle\left(\frac{\overline{c}}{c}\right)\mathsf{d_7} &0\cr\cr
 \displaystyle\left(\frac{b_1}{a}\right)\left(\frac{\mathbf S_{1_{_{\overline{126}}}}}{\mathbf S_{1_{_{126}}}}\right)\mathsf{d_1}& \displaystyle \left(\frac{b_2}{a}\right)\left(\frac{\mathbf S_{1_{_{\overline{126}}}}     }{\mathbf S_{1_{_{126}}}}\right)\mathsf{d_1}&\displaystyle\left(\frac{\overline{c}}{c}\right)\left(\frac{\mathbf S_{1_{_{\overline{126}}}}     }{\mathbf S_{1_{_{126}}}}\right)\mathsf{d_4}& {\displaystyle\left(\frac{\mathbf S_{1_{_{\overline{126}}}}}{\mathbf S_{1_{_{126}}}}\right)\mathsf{d_{11}}}&&& {\displaystyle\mathsf{d_{10}}}&&0 &0\cr\cr
 0& 0& 0&\displaystyle\mathsf{d_7} &&&0 && 0&\displaystyle\left(\frac{\overline{c}}{c}\right)\displaystyle\mathsf{d_8} \cr\cr
 \displaystyle\mathsf{d_3}& 0&\displaystyle \mathsf{d_6}& 0 &&&0 &&\displaystyle\mathsf{d_8}&{\displaystyle\mathsf{d_{12}}}\cr\cr
 }\right),~~~\label{doublet mass matrix}
\eeqn }}
and
\begin{eqnarray}\label{elements of doublet mass matrix}
a&\equiv&\frac{\imath}{5!}\mathrm{A};~~b_{1,2}\equiv\frac{\imath}{5!}\mathrm{B_{1,2}};~~c\equiv\frac{\imath}{5!}\mathrm{C};~~\overline{c}\equiv\frac{\imath}{5!}\overline{\mathrm{C}},\nonumber\\
\mathsf{d_1}&\equiv&\frac{a}{2\sqrt{5}}\mathbf S_{1_{_{{126}}}}                  ,\nonumber\\
\mathsf{d_2}&\equiv&-a\left[\frac{\sqrt{3}}{10}\mathbf S_{1_{_{{210}}}}                  +\frac{\sqrt{3}}{20}\mathbf S_{24_{_{{210}}}}                  \right],\nonumber\\
\mathsf{d_3}&\equiv&a\left[-\frac{1}{4\sqrt{6}}\mathbf S_{24_{_{{210}}}}                  +\frac{1}{4\sqrt{15}}\mathbf S_{75_{_{{210}}}}                  \right],\nonumber\\
{\mathsf{d_4}}&\equiv&{-\frac{c}{\sqrt{30}}}\mathbf S_{1_{_{{126}}}}                  ,\nonumber\\
\mathsf{d_5}&\equiv&c\left[-\frac{1}{10\sqrt{2}}\mathbf S_{1_{_{{210}}}}                  +\frac{3}{40\sqrt{2}}\mathbf S_{24_{_{{210}}}}                  \right],\nonumber\\
{\mathsf{d_6}}&\equiv&-c\left[{\frac{1}{48}}\mathbf S_{24_{_{{210}}}}                  +\frac{1}{12\sqrt{10}}\mathbf S_{75_{_{{210}}}}                  \right],\nonumber\\
\mathsf{d_7}&\equiv&c\left[\frac{1}{48\sqrt{3}}\mathbf S_{24_{_{{210}}}}                  +\frac{1}{12\sqrt{30}}\mathbf S_{75_{_{{210}}}}                  \right],\nonumber\\
\mathsf{d_8}&\equiv&-c\left[\frac{1}{20\sqrt{6}}\mathbf S_{1_{_{{210}}}}                  +\frac{1}{240\sqrt{6}}\mathbf S_{24_{_{{210}}}}                  +\frac{1}{12\sqrt{15}}\mathbf S_{75_{_{{210}}}} \right],\nonumber\\
{\mathsf{d_9}}&{\equiv}&{{2}m_{\Delta}-\eta\left[{\frac{2}{5\sqrt{15}}}\mathbf S_{1_{_{{210}}}}+{\frac{3}{{20}}\sqrt{\frac{3}{5}}}\mathbf S_{24_{_{{210}}}}                  \right]},\nonumber\\
{\mathsf{d_{10}}}&{\equiv}&{2m_{\Phi}-\lambda\left[{\frac{3}{10\sqrt{2}}}\mathbf S_{1_{_{{210}}}}+\frac{1}{2}\sqrt{\frac{3}{5}}\mathbf S_{24_{_{{210}}}}                  \right]},
\nonumber\\
{\mathsf{d_{11}}}&{\equiv}&{\frac{1}{5}\eta\mathbf S_{1_{_{{126}}}}},\nonumber\\
{\mathsf{d_{12}}}&{\equiv}&{m_{\Delta}+\eta\left[-{\frac{1}{6\sqrt{15}}}\mathbf S_{24_{_{{210}}}} +{\frac{1}{15\sqrt{6}}}\mathbf S_{75_{_{{210}}}} \right]}.
\end{eqnarray}

Substituting the values of  $\mathbf{S_{1_{210}}},~ \mathbf S_{75_{210}},~ \mathbf S_{1_{126}},~ \mathbf S_{1_{\ov{126}}}$  in terms of just one
VEV, i.e., $\mathbf S_{24_{210}}$, all the matrix elements of ${M_{d}}$ can be determined in terms of a single
VEV, i.e., $\mathbf S_{24_{210}}$.
{An illustrative example of the numerical sizes of $\mathbf{S_{1_{210}}},~
 \mathbf S_{24_{210}},~ \mathbf S_{75_{210}},~ \mathbf S_{1_{126}}$ is} given in Table \ref{Estimation of SM singlets}.
The
matrix ${M_d}$ is non-symmetric and is diagonalized by two $7\times7$ unitary matrices ${U_d}$ and ${V_d}$:
\begin{equation}
{U_d}^{\dagger}{M_d}{V_d}=\textnormal{diag}\left(0,\mathsf{m}_{d_2},\mathsf{m}_{d_3},\cdots,\mathsf{m}_{d_7}\right).\label{eigenvalues}
\end{equation}
The columns of the matrices ${U_d}$ and ${V_d}$ are the eigenvectors of matrices ${M_d}^{\dagger}{M_d}$  and ${M_dM_d}^{\dagger}$ respectively,
\begin{equation}
{U_d^{\dagger}}[M_d^{\dagger}M_d]{U_d}=\textnormal{diag}\left(0,\mathsf{m}_{d_2}^2,\mathsf{m}_{d_3}^2,\cdots,\mathsf{m}_{d_7}^2\right)={V_d^{\dagger}}
[M_dM_d^{\dagger}]{V_d}.\label{eigenvaluessquared}
\end{equation}
The mass {eigenstates of the}  Higgs doublet fields are expressed in terms of the primitive Higgs doublet fields through

\begin{equation}
 \left(\matrix{{}^{1}\!{\mathsf D}_{a}^{\prime}\cr{}^{2}\!{\mathsf D}_{a}^{\prime}\cr{}^{3}\!{\mathsf D}_{a}^{\prime}\cr{}^{4}\!{\mathsf D}_{a}^{\prime}\cr{}^{5}\!{\mathsf D}_{a}^{\prime}\cr{}^{6}\!{\mathsf D}_{a}^{\prime}\cr{}^{7}\!{\mathsf D}_{a}^{\prime}}\right)= {V_d}^{\dagger}\left(\matrix{{}^{(\overline{5}_{10_1})}\!{\mathsf D}_{a}\cr{}^{(\overline{5}_{10_2})}\!{\mathsf D}_{a}\cr{}^{(\overline{5}_{120})}\!{\mathsf D}_{a}\cr{}^{(\overline{5}_{{126}})}\!{\mathsf D}_{a}\cr{}^{(\overline{5}_{{210}})}\!{\mathsf D}_{a}\cr{}^{(\overline{45}_{120})}\!{\mathsf D}_{a}\cr{}^{(\overline{45}_{\overline{126}})}\!{\mathsf D}_{a}}\right);~~\left(\matrix{{}^{1}\!{\mathsf D}^{\prime a}\cr {}^{2}\!{\mathsf D}^{\prime a}\cr{}^{3}\!{\mathsf D}^{\prime a}\cr{}^{4}\!{\mathsf D}^{\prime a} \cr{}^{5}\!{\mathsf D}^{\prime a} \cr{}^{6}\!{\mathsf D}^{\prime a} \cr{}^{7}\!{\mathsf D}^{\prime a}} \right)={U_d}^{\dagger}\left(\matrix{{}^{({5}_{10_1})}\!{\mathsf D}^{a}\cr {}^{({5}_{10_2})}\!{\mathsf D}^{a}\cr{}^{({5}_{120})}\!{\mathsf D}^{a}\cr{}^{({5}_{\overline{126}})}\!{\mathsf D}^{a} \cr{}^{({5}_{{210}})}\!{\mathsf D}^{a} \cr{}^{({45}_{120})}\!{\mathsf D}^{a} \cr{}^{({45}_{126})}\!{\mathsf D}^{a}} \right).\label{doublet mass eigenstates}
\end{equation}

{We identify the
 light Higgs doublet pair to be $\left({}^{1}\!{\mathsf D}_{a}^{\prime},{}^{1}\!{\mathsf D}^{\prime a}\right)\equiv({\mathbf{H_d}}_a,{\mathbf{H_u}}^a)$ while all the remaining mass eigenstates of the
Higgs doublets are superheavy.}
Thus, the inverse transformation of Eq.(\ref{doublet mass eigenstates}) gives
\begin{eqnarray}
{}^{(\overline{5}_{10_1})}\!{\mathsf D}_{a}={V_{d{_{11}}}}{\mathbf{H_d}}_a+\cdots,~~~&{}^{(\overline{5}_{10_2})}\!{\mathsf D}_{a}={V_{d{_{21}}}}{\mathbf{H_d}}_a+\cdots,&~~~{}^{(\overline{5}_{120})}\!{\mathsf D}_{a}={V_{d{_{31}}}}{\mathbf{H_d}}_a+\cdots,\nonumber\\
{}^{(\overline{5}_{{126}})}\!{\mathsf D}_{a}={V_{d{_{41}}}}{\mathbf{H_d}}_a+\cdots,~~~&
{}^{(\overline{5}_{{210}})}\!{\mathsf D}_{a}={V_{d{_{51}}}}{\mathbf{H_d}}_a+\cdots,&~~~{}^{(\overline{45}_{120})}\!{\mathsf D}_{a}={V_{d{_{61}}}}{\mathbf{H_d}}_a+\cdots,\nonumber\\
&{}^{(\overline{45}_{\overline{126}})}\!{\mathsf D}_{a}={V_{d{_{71}}}}{\mathbf{H_d}}_a+\cdots,&
\end{eqnarray}
and
\begin{eqnarray}
{}^{({5}_{10_1})}\!{\mathsf D}^{a}={U_{d{_{11}}}}{\mathbf{H_u}}^a+\cdots,~~~&{}^{({5}_{10_2})}\!{\mathsf D}^{a}={U_{d{_{21}}}}{\mathbf{H_u}}^a+\cdots,&~~~{}^{({5}_{120})}\!{\mathsf D}^{a}={U_{d{_{31}}}}{\mathbf{H_u}}^a+\cdots,\nonumber\\
{}^{({5}_{\overline{126}})}\!{\mathsf D}^{a}={U_{d{_{41}}}}{\mathbf{H_u}}^a+\cdots,~~~&
{}^{({5}_{{210}})}\!{\mathsf D}^{a}={U_{d{_{51}}}}{\mathbf{H_u}}^a+\cdots,&~~~{}^{({45}_{120})}\!{\mathsf D}^{a}={U_{d{_{61}}}}{\mathbf{H_u}}^a+\cdots,\nonumber\\
&{}^{({45}_{{126}})}\!{\mathsf D}^{a}={U_{d{_{71}}}}{\mathbf{H_u}}^a+\cdots.&
\end{eqnarray}
{where
$+\cdots$ stand for heavy Higgs  doublet fields. The heavy Higgs doublets are to be integrated out and do not appear
 in the effective  low energy theory}. {The matrix elements ${U_{d_{k1}}}$  and ${V_{d_{k1}}}$ are functions of $\mathbf S_{24_{_{{210}}}}$ except for $k=4,~5,~7$ for which ${U_{d_{k1}}}=0={V_{d_{k1}}}$.}
The numerical values of ${U_{d_{k1}}}$  and ${V_{d_{k1}}},~ {k}=1,~2,~3,~6$ are given in {Table \ref{Estimation of elements of U matrix} and Table \ref{Estimation of elements of V matrix}}. {Again note that in Tables {\ref{Estimation of elements of U matrix} and \ref{Estimation of elements of V matrix}} corresponding to each set of ${\cal{M}}_{\Delta}$ and ${\cal{M}}_{\Phi}$, there exists three solutions for  ${U_{d_{k1}}}$  and ${V_{d_{k1}}}$. This is simply because $\mathbf S_{24_{_{{210}}}}                  $ satisfies
{a cubic equation  Eq.(\ref{Determination S_24_210})}}.

In summary, all the eigenvalues of the doublet Higgs mass  matrix given by Eq.(\ref{doublet mass matrix}) are superheavy except
one pair which is massless and corresponds to the electroweak Higgs doublets of MSSM. The zero eigenmode
is determined by transformation matrix elements ${U_{d_{k1}}}$  and ${V_{d_{k1}}}$ where ${k}=1,\cdots, 7$.

{{\small
\begin{table}[t]
\begin{center}
{
\begin{tabular}{|c|c|c|c|c|c|}
\multicolumn{6}{c}{}\cr \hline \hline
{\scriptsize${\cal{M}}_{\Delta}~({\rm GeV})$}&\scriptsize{${\cal{M}}_{\Phi}~({\rm GeV})$}&\scriptsize $\mathbf S_{1_{_{{210}}}}                  ~({\rm GeV})$&
 \scriptsize$\mathbf S_{24_{_{{210}}}}                  ~({\rm GeV})$&\scriptsize$\mathbf S_{75_{_{{210}}}}                  ~({\rm GeV})$&\scriptsize$\mathbf S_{1_{_{{126}}}}                  ~({\rm GeV})$
 \\
 \hline\hline
\multirow{3}{*} {\scriptsize{$10^{15}$}}&\multirow{3}{*} {\scriptsize{$10^{15}$}} & {\scriptsize{$4\times10^{16}$}}&
{\scriptsize{$1.10\times 10^{16}$}}&{\scriptsize{$-1.78\times 10^{15}$}}&{\scriptsize{$\imath1.28\times 10^{16}$}}\\
\cline{3-6}
& & {\scriptsize{$4\times10^{16}$}}&
{\scriptsize{$\left(1.72-\imath8.08\right)\times 10^{16}$}}&
{\scriptsize{$\left(-2.94-\imath4.20\right)\times 10^{16}$}}&
{\scriptsize{$\left(1.38+\imath2.75\right)\times 10^{16}$}} \\
\cline{3-6}
& & {\scriptsize{$4\times10^{16}$}}&
{\scriptsize{$\left(1.72+\imath8.08\right)\times 10^{16}$}}&
{\scriptsize{$\left(-2.94+\imath4.20\right)\times 10^{16}$}}&
{\scriptsize{$\left(1.38-\imath2.75\right)\times 10^{16}$}} \\
\hline\hline
\multirow{3}{*} {\scriptsize{$6.67\times 10^{15}$}}&\multirow{3}{*} {\scriptsize{$6.67\times 10^{15}$}} & {\scriptsize{$2.67\times10^{17}$}}&
{\scriptsize{$7.36\times 10^{16}$}}&{\scriptsize{$-1.19\times 10^{16}$}}&{\scriptsize{$\imath8.57\times 10^{16}$}}\\
\cline{3-6}
& & {\scriptsize{$2.67\times 10^{17}$}}&
{\scriptsize{$\left(1.14-\imath5.39\right)\times 10^{17}$}}&
{\scriptsize{$\left(-1.96-\imath2.80\right)\times 10^{17}$}}&
{\scriptsize{$\left(9.23+\imath18.4\right)\times 10^{16}$}} \\
\cline{3-6}
& & {\scriptsize{$2.67\times 10^{17}$}}&
{\scriptsize{$\left(1.14+\imath5.39\right)\times 10^{17}$}}&
{\scriptsize{$\left(-1.96+\imath2.80\right)\times 10^{17}$}}&
{\scriptsize{$\left(9.23-\imath18.4\right)\times 10^{16}$}} \\
\hline\hline
\multirow{3}{*} {\scriptsize{$1.25\times 10^{16}$}}&\multirow{3}{*} {\scriptsize{$1.25\times 10^{16}$}} & {\scriptsize{$5\times10^{17}$}}&
{\scriptsize{$1.38\times 10^{17}$}}&{\scriptsize{$-2.22\times 10^{16}$}}&{\scriptsize{$\imath1.61\times 10^{17}$}}\\
\cline{3-6}
& & {\scriptsize{$5\times10^{17}$}}&
{\scriptsize{$\left(2.14-\imath10.10\right)\times 10^{17}$}}&
{\scriptsize{$\left(-3.67-\imath5.24\right)\times 10^{17}$}}&
{\scriptsize{$\left(1.72+\imath3.44\right)\times 10^{17}$}} \\
\cline{3-6}
& & {\scriptsize{$5\times10^{17}$}}&
{\scriptsize{$\left(2.14+\imath10.10\right)\times 10^{17}$}}&
{\scriptsize{$\left(-3.67+\imath5.24\right)\times 10^{17}$}}&
{\scriptsize{$\left(1.72-\imath3.44\right)\times 10^{17}$}} \\
\hline\hline
\multirow{3}{*} {\scriptsize{$6.67\times 10^{15}$}}&\multirow{3}{*} {\scriptsize{$2\times 10^{16}$}} & {\scriptsize{$2.67\times10^{17}$}}&
{\scriptsize{$1.66\times 10^{17}$}}&{\scriptsize{$-1.02\times 10^{17}$}}&{\scriptsize{$\imath6.51\times 10^{16}$}}\\
\cline{3-6}
& & {\scriptsize{$2.67\times10^{17}$}}&
{\scriptsize{$\left(8.68-\imath12.49\right)\times 10^{17}$}}&
{\scriptsize{$\left(-1.84-\imath7.71\right)\times 10^{17}$}}&
{\scriptsize{$\left(1.65+\imath3.17\right)\times 10^{17}$}} \\
\cline{3-6}
& & {\scriptsize{$2.67\times10^{17}$}}&
{\scriptsize{$\left(8.68+\imath12.49\right)\times 10^{17}$}}&
{\scriptsize{$\left(-1.84+\imath7.71\right)\times 10^{17}$}}&
{\scriptsize{$\left(1.65-\imath3.17\right)\times 10^{17}$}} \\
\hline\hline

\multirow{3}{*} {\scriptsize{$2\times 10^{16}$}}&\multirow{3}{*} {\scriptsize{$6.67\times 10^{15}$}} & {\scriptsize{$8\times10^{17}$}}&
{\scriptsize{$-1.15\times 10^{17}$}}&{\scriptsize{$-6.24\times 10^{15}$}}&{\scriptsize{$2.41\times 10^{17}$}}\\
\cline{3-6}
& & {\scriptsize{$8\times10^{17}$}}&
{\scriptsize{$\left(-2.89-\imath8.96\right)\times 10^{17}$}}&
{\scriptsize{$\left(-5.69-\imath3.87\right)\times 10^{17}$}}&
{\scriptsize{$\left(2.87+\imath5.36\right)\times 10^{17}$}} \\
\cline{3-6}
& & {\scriptsize{$8\times10^{17}$}}&
{\scriptsize{$\left(-2.89+\imath8.96\right)\times 10^{17}$}}&
{\scriptsize{$\left(-5.69+\imath3.87\right)\times 10^{17}$}}&
{\scriptsize{$\left(2.87-\imath5.36\right)\times 10^{17}$}} \\
\hline\hline
\end{tabular}
}
\caption{\small{{A Numerical estimate of the VEVs of the Standard Model singlets in $\mathsf{210}$, $\mathsf{126}$ and  $\mathsf{\overline{126}}$-plets {arising} in the spontaneous breaking of the $\mathsf{SO(10)}$ GUT gauge symmetry under the assumption {$\lambda=\eta\sim 1$ and $\mathbf S_{1_{_{{126}}}}=\mathbf S_{1_{_{\overline{126}}}}$}.}}}\label{Estimation of SM singlets}
\end{center}
\end{table}
}}

{{\tiny
\begin{table}[t]
\begin{center}
{
\begin{tabular}{|c|c|c|c|c|c|}
\multicolumn{6}{c}{}\cr \hline \hline
{\scriptsize${\cal{M}}_{\Delta}~({\rm GeV})$}&\scriptsize{${\cal{M}}_{\Phi}~({\rm GeV})$}& \scriptsize$U_{d{_{11}}}$&
 \scriptsize${U_{d{_{21}}}}$&\scriptsize${U_{d{_{31}}}}$&\scriptsize${U_{d{_{61}}}}$
 \\
 \hline\hline
\multirow{3}{*} {\scriptsize{$10^{15}$}}&\multirow{3}{*} {\scriptsize{$10^{15}$}} & {\scriptsize{$0.388 + \imath0.273$}}&
{\scriptsize{$-0.394- \imath0.277 $}}&{\scriptsize{$-0.00732-\imath 0.00516$}}& {\scriptsize{$-0.602 -\imath 0.424$}}\\
\cline{3-6}
& & {\scriptsize{$0.133-\imath 0.0527$}}&
{\scriptsize{$-0.203 + \imath0.174 $}}&
{\scriptsize{$-0.0847+\imath 0.149$}}&{\scriptsize{$-0.821 -\imath 0.453$}}
\\
\cline{3-6}
& & {\scriptsize{$0.133+\imath 0.0527$}}&
{\scriptsize{$-0.203 - \imath0.174 $}}&
{\scriptsize{$-0.0847-\imath 0.149$}}&{\scriptsize{$-0.821 +\imath 0.453$}}\\
\hline\hline
\multirow{3}{*} {\scriptsize{$6.67\times 10^{15}$}}&\multirow{3}{*} {\scriptsize{$6.67\times 10^{15}$}} & {\scriptsize{$-0.341-\imath 0.330$}}&
{\scriptsize{$0.347 + \imath0.335$}}&{\scriptsize{$0.00644+\imath 0.00622$}}&
{\scriptsize{$0.530+ \imath0.512$}}\\
\cline{3-6}
& &  {\scriptsize{$0.0126-\imath 0.143$}}&
{\scriptsize{$0.0654+\imath 0.259$}}&{\scriptsize{$0.0954 + \imath0.142$}}&
{\scriptsize{$-0.772+\imath0.532$}}\\
\cline{3-6}
& & {\scriptsize{$0.0126+\imath 0.143$}}&
{\scriptsize{$0.0654-\imath 0.259$}}&{\scriptsize{$0.0954 - \imath0.142$}}&
{\scriptsize{$-0.772-\imath0.532$}}\\
\hline\hline
\multirow{3}{*} {\scriptsize{$1.25\times 10^{16}$}}&\multirow{3}{*} {\scriptsize{$1.25\times 10^{16}$}} & {\scriptsize{$-0.474 + \imath0.0142$}}&
{\scriptsize{$0.482 - \imath0.0144$}}& {\scriptsize{$0.00895 - \imath0.000268$}} &{\scriptsize{$0.736- \imath0.0220$}}\\
\cline{3-6}
& & {\scriptsize{$0.132+\imath0.0571$}}&
{\scriptsize{$-0.267 - \imath0.0200$}} &
{\scriptsize{$-0.165+\imath 0.0455$}}&{\scriptsize{$-0.260 -\imath 0.900$}}\\
\cline{3-6}
& & {\scriptsize{$0.132-\imath0.0571$}}&
{\scriptsize{$-0.267 +\imath0.0200$}} &
{\scriptsize{$-0.165-\imath 0.0455$}}&{\scriptsize{$-0.260 +\imath 0.900$}}\\
\hline\hline
\multirow{3}{*} {\scriptsize{$6.67\times 10^{15}$}}&\multirow{3}{*} {\scriptsize{$2\times 10^{16}$}} & {\scriptsize{$-0.125+\imath 0.0773$}}&
{\scriptsize{$0.130-\imath 0.0804$}}&{\scriptsize{$-0.00623 - \imath0.00386$}} &{\scriptsize{$0.831 - \imath0.515$}}\\
\cline{3-6}
& & {\scriptsize{$-0.0806 - \imath0.197$}}&
{\scriptsize{$0.131+\imath 0.436$}}&
{\scriptsize{$0.0612 +\imath 0.293$}}&{\scriptsize{$-0.576 +\imath 0.571$}}\\
\cline{3-6}
& & {\scriptsize{$-0.0806 + \imath0.197$}}&
{\scriptsize{$0.131-\imath 0.436$}}&
{\scriptsize{$0.0612 -\imath 0.293$}}&{\scriptsize{$-0.576 -\imath 0.571$}}\\
\hline\hline
\multirow{3}{*} {\scriptsize{$2\times 10^{16}$}}&\multirow{3}{*} {\scriptsize{$6.67\times 10^{15}$}} & {\scriptsize{$-0.631$}}&
{\scriptsize{$0.634$}}&{\scriptsize{$0.00396$}}&{\scriptsize{$-0.448$}}\\
\cline{3-6}
& & {\scriptsize{$0.0792 + \imath0.0411$}}&
{\scriptsize{$-0.0542 +\imath 0.0561$}}&
{\scriptsize{$0.0306+\imath 0.119$}}&{\scriptsize{$-0.947 - 0.273$}}\\
\cline{3-6}
& & {\scriptsize{$0.0792 - \imath0.0411$}}&
{\scriptsize{$-0.0542 -\imath 0.0561$}}&
{\scriptsize{$0.0306-\imath 0.119$}}&{\scriptsize{$-0.947 +0.273$}}\\
\cline{3-6}
\hline\hline
\end{tabular}
}
\caption{{\small{A Numerical estimate of the elements of the zero mode eigenvectors using the analysis of Table \ref{Estimation of SM singlets} {and under the additional assumption $a=b_{1,2}=c=\overline{c}\sim 1$.}}}}\label{Estimation of elements of U matrix}
\end{center}

\end{table}
}}
{{\tiny
\begin{table}[t]
\begin{center}
{
\begin{tabular}{|c|c|c|c|c|c|}
\multicolumn{6}{c}{}\cr \hline \hline
{\scriptsize${\cal{M}}_{\Delta}~({\rm GeV})$}&\scriptsize{${\cal{M}}_{\Phi}~({\rm GeV})$}& \scriptsize$V_{d{_{11}}}$&
 \scriptsize${V_{d{_{21}}}}$&\scriptsize${V_{d{_{31}}}}$&\scriptsize${V_{d{_{61}}}}$
 \\
 \hline\hline
\multirow{3}{*} {\scriptsize{$10^{15}$}}&\multirow{3}{*} {\scriptsize{$10^{15}$}} & {\scriptsize{$-0.00831$}}&
{\scriptsize{$0.547$}}&{\scriptsize{$-0.0102$}}& {\scriptsize{$-0.837$}}\\
\cline{3-6}
& & {\scriptsize{$-0.140$}}&
{\scriptsize{$0.252 - \imath0.0900$}}&
{\scriptsize{$-0.171$}}&{\scriptsize{$-0.0119+\imath 0.0.938$}}
\\
\cline{3-6}
& & {\scriptsize{$-0.140$}}&
{\scriptsize{$0.252 +\imath0.0900$}}&
{\scriptsize{$-0.171$}}&{\scriptsize{$-0.0119-\imath 0.0.938$}}\\
\hline\hline
\multirow{3}{*} {\scriptsize{$6.67\times 10^{15}$}}&\multirow{3}{*} {\scriptsize{$6.67\times 10^{15}$}} & {\scriptsize{$0.00831$}}&
{\scriptsize{$-0.547$}}&{\scriptsize{$0.0102$}}&{\scriptsize{$0.837$}}\\
\cline{3-6}
& & {\scriptsize{$0.140$}}&
{\scriptsize{$-0.252 +\imath 0.0900$}}&
{\scriptsize{$0.171$}}&{\scriptsize{$0.0119 -\imath 0.938$}}\\
\cline{3-6}
& & {\scriptsize{$0.140$}}&
{\scriptsize{$-0.252 - \imath0.0900$}}&{\scriptsize{$0.171$}}&
{\scriptsize{$0.0119 + \imath0.938$}} \\
\hline\hline
\multirow{3}{*} {\scriptsize{$1.25\times 10^{16}$}}&\multirow{3}{*} {\scriptsize{$1.25\times 10^{16}$}} & {\scriptsize{$0.00831$}}&
{\scriptsize{$-0.547$}}&{\scriptsize{$0.0102$}}&{\scriptsize{$0.837$}}\\
\cline{3-6}
& & {\scriptsize{$-0.140$}}&
{\scriptsize{$0.252-\imath 0.0900$}}&
{\scriptsize{$-0.171$}}&{\scriptsize{$-0.0119+\imath 0.938$}}\\
\cline{3-6}
& & {\scriptsize{$-0.140$}}&
{\scriptsize{$0.252 +\imath 0.0900$}}&
{\scriptsize{$-0.171$}}&{\scriptsize{$-0.0119-\imath 0.938$}}\\
\hline\hline
\multirow{3}{*} {\scriptsize{$6.67\times 10^{15}$}}&\multirow{3}{*} {\scriptsize{$2\times 10^{16}$}} & {\scriptsize{$0.00605$}}&
{\scriptsize{$-0.154$}}&{\scriptsize{$0.00741$}}&{\scriptsize{$0.988$}}\\
\cline{3-6}
& & {\scriptsize{$0.243$}}&
{\scriptsize{$-0.451- \imath0.0385$}}&
{\scriptsize{$0.298$}}&{\scriptsize{$0.438-\imath 0.675$}}\\
\cline{3-6}
& & {\scriptsize{$0.243$}}&
{\scriptsize{$-0.451+ \imath0.0385$}}&
{\scriptsize{$0.298$}}&{\scriptsize{$0.438+\imath 0.675$}}\\
\hline\hline
\multirow{3}{*} {\scriptsize{$2\times 10^{16}$}}&\multirow{3}{*} {\scriptsize{$6.67\times 10^{15}$}} & {\scriptsize{$-0.00417$}}&
{\scriptsize{$0.817$}}&{\scriptsize{$-0.00511$}}&{\scriptsize{$0.577$}}\\
\cline{3-6}
& & {\scriptsize{$0.100$}}&
{\scriptsize{$-0.0408 +\imath 0.0664$}}&
{\scriptsize{$0.123$}}&{\scriptsize{$-0.500-\imath 0.848$}}\\
\cline{3-6}
& & {\scriptsize{$0.100$}}&
{\scriptsize{$-0.0408 -\imath 0.0664$}}&
{\scriptsize{$0.123$}}&{\scriptsize{$-0.500+\imath 0.848$}}\\
\cline{3-6}
\hline\hline
\end{tabular}
}
\caption{{\small{A Numerical estimate of the elements of the zero mode eigenvectors using the analysis of Table \ref{Estimation of SM singlets} {and under the additional assumption $a=b_{1,2}=c=\overline{c}\sim 1$.}}}}\label{Estimation of elements of V matrix}
\end{center}
\end{table}
}}
\vskip 0.5cm

\noindent The mass terms for the $\mathsf{SU(3)}$ color triplets in the superpotential can be written as
{{
\begin{eqnarray}
\left({}^{({5}_{10_1})}\!{\mathsf T}^{\alpha},{}^{({5}_{10_2})}\!{\mathsf T}^{\alpha},{}^{({5}_{120})}\!{\mathsf T}^{\alpha},{}^{({5}_{\overline{126}})}\!{\mathsf T}^{\alpha},{}^{({5}_{{210}})}\!{\mathsf T}^{\alpha},{}^{({45}_{120})}\!{\mathsf T}^{\alpha},{}^{({45}_{{126}})}\!{\mathsf T}^{\alpha},{}^{({50}_{\overline{126}})}\!{\mathsf T}^{\alpha}\right){M_t}\left(\matrix{{}^{(\overline{5}_{10_1})}\!{\mathsf T}_{\alpha}\cr {}^{(\overline{5}_{10_2})}\!{\mathsf T}_{\alpha}\cr{}^{(\overline{5}_{120})}\!{\mathsf T}_{\alpha}\cr{}^{(\overline{5}_{{126}})}\!{\mathsf T}_{\alpha} \cr{}^{(\overline{5}_{{210}})}\!{\mathsf T}_{\alpha} \cr{}^{(\overline{45}_{120})}\!{\mathsf T}_{\alpha} \cr{}^{(\overline{45}_{\overline{126}})}\!{\mathsf T}_{\alpha}  \cr{}^{(\overline{50}_{{126}})}\!{\mathsf T}_{\alpha}}\right),\label{Mass term 2}
\end{eqnarray}
}}
{where the triplet mass matrix ${M_t}$ {receives contributions from \cref{superpotential gut so(10)tensor} and
  eq.(\ref{wdt.1}) and is given by}}

{\small
 { \beqn
 {M_t}=\left(\matrix{
0& 0& 0&\displaystyle\mathsf{t_2} &&&&\displaystyle\mathsf{t_1} &&0 &\displaystyle\left(\frac{b_1}{a}\right)\mathsf{t_3}&&\mathsf{t_9}\cr\cr
0& 0&0 & 0&&&& 0&& 0&\displaystyle\left(\frac{b_2}{a}\right)\mathsf{t_3}&&0 \cr\cr
0& 0& 0&\displaystyle\mathsf{t_5} &&&&\displaystyle\mathsf{t_4} && 0&\displaystyle\left(\frac{\overline{c}}{c}\right)\mathsf{t_6}&&0 \cr\cr
 \displaystyle\left(\frac{b_1}{a}\right)\mathsf{t_2}&  \displaystyle\left(\frac{b_2}{a}\right)\mathsf{t_2}&\displaystyle\left(\frac{\overline{c}}{c}\right)\mathsf{t_5}& \displaystyle{\mathsf{t_{11}}}&&&& \displaystyle{\mathsf{t_{13}}}&&\displaystyle\left(\frac{\overline{c}}{c}\right)\mathsf{t_7} &0&&\displaystyle{\mathsf{t_{16}}}\cr\cr
 \displaystyle\left(\frac{b_1}{a}\right)\left(\frac{\mathbf S_{1_{_{\overline{126}}}}     }{\mathbf S_{1_{_{126}}}}\right)\mathsf{t_1}& \displaystyle \left(\frac{b_2}{a}\right)\left(\frac{\mathbf S_{1_{_{\overline{126}}}}}{\mathbf S_{1_{_{126}}}}\right)\mathsf{t_1}&\displaystyle\left(\frac{\overline{c}}{c}\right)\left(\frac{\mathbf S_{1_{_{\overline{126}}}}     }{\mathbf S_{1_{_{126}}}}\right)\mathsf{t_4}& \displaystyle{\left(\frac{\mathbf S_{1_{_{\overline{126}}}}}{\mathbf S_{1_{_{126}}}}\right){\mathsf{t_{13}}}}&&&& \displaystyle{{\mathsf{t_{12}}}}&&0 &0&&0\cr\cr
 0& 0& 0&\displaystyle\mathsf{t_7} &&&&0 && 0&\displaystyle\left(\frac{\overline{c}}{c}\right)\mathsf{t_8}&&\mathsf{t_{10}} \cr\cr
 \displaystyle\mathsf{t_3}& 0&\displaystyle \mathsf{t_6}& 0 &&&&0 &&\displaystyle\mathsf{t_8}&{\displaystyle\mathsf{t_{14}}}&&0\cr\cr
 \displaystyle\left(\frac{b_1}{a}\right)\mathsf{t_9}&\displaystyle\left(\frac{b_2}{a}\right)\mathsf{t_9}&0& \displaystyle{\mathsf{t_{16}}} &&&&0 &&\displaystyle\left(\frac{\overline{c}}{c}\right)\displaystyle\mathsf{t_{10}}&0&&\displaystyle{\mathsf{t_{15}}}\cr\cr
 }\right)\,.~~~\label{triplet mass matrix}
\eeqn}}
{Here}
\begin{eqnarray}\label{triplet mass matrix elements}
\mathsf{t_1}&\equiv&\frac{a}{2\sqrt{5}}\mathbf S_{1_{_{{126}}}}                  ,\nonumber\\
\mathsf{t_2}&\equiv&a\left[-\frac{\sqrt{3}}{10}\mathbf S_{1_{_{{210}}}}                  +\frac{1}{10\sqrt{3}}\mathbf S_{24_{_{{210}}}}                  \right],\nonumber\\
{\mathsf{t_3}}&\equiv&-a\left[\frac{1}{6\sqrt{2}}\mathbf S_{24_{_{{210}}}}                  {+\frac{1}{12\sqrt{5}}\mathbf S_{75_{_{{210}}}}}                  \right],\nonumber\\
{\mathsf{t_4}}&\equiv&{-\frac{c}{\sqrt{30}}\mathbf S_{1_{_{{126}}}}}                  ,\nonumber\\
\mathsf{t_5}&\equiv&-c\left[\frac{1}{10\sqrt{2}}\mathbf S_{1_{_{{210}}}}                  +\frac{1}{20\sqrt{2}}\mathbf S_{24_{_{{210}}}}                  \right],\nonumber\\
{\mathsf{t_6}}&\equiv&-c\left[{\frac{1}{24\sqrt{3}}\mathbf S_{24_{_{210}}}+\frac{1}{12\sqrt{30}}\mathbf S_{75_{_{210}}}}\right]                  ,\nonumber\\
\mathsf{t_7}&\equiv&-c\left[\frac{1}{72}\mathbf S_{24_{_{{210}}}}                  +\frac{1}{36\sqrt{10}}\mathbf S_{75_{_{{210}}}}                  \right],\nonumber\\
\mathsf{t_8}&\equiv&-c\left[\frac{1}{20\sqrt{6}}\mathbf S_{1_{_{{210}}}}                  +\frac{1}{40\sqrt{6}}\mathbf S_{24_{_{{210}}}}                  \right],\nonumber\\
{\mathsf{t_9}}&\equiv&{\frac{a}{12\sqrt{5}}\mathbf S_{75_{_{{210}}}}},\nonumber\\
\mathsf{t_{10}}&\equiv&c\left[-\frac{1}{60\sqrt{6}}\mathbf S_{24_{_{{210}}}}{+\frac{1}{36\sqrt{15}}\mathbf S_{75_{_{210}}}}\right],\nonumber\\
{\mathsf{t_{11}}}&{\equiv}&{{2}m_{\Delta}+\eta\left[-{\frac{2}{5\sqrt{15}}}\mathbf S_{1_{_{{210}}}}+{\frac{1}{10}\sqrt{\frac{3}{5}}}\mathbf S_{24_{_{{210}}}}                  \right]},\nonumber\\
{\mathsf{t_{12}}}&{\equiv}&{2m_{\Phi}+\lambda\left[{-\frac{3}{10\sqrt{2}}}\mathbf S_{1_{_{{210}}}}+\frac{1}{\sqrt{15}}\mathbf S_{24_{_{{210}}}}                  \right]},
\nonumber\\
{\mathsf{t_{13}}}&{\equiv}&{\frac{1}{5}\eta\mathbf S_{1_{_{{126}}}}},\nonumber\\
{\mathsf{t_{14}}}&{\equiv}&{m_{\Delta}},\nonumber\\
{\mathsf{t_{15}}}&{\equiv}&{\frac{1}{6}m_{\Delta}+\eta\left[\frac{1}{60\sqrt{15}}\mathbf S_{1_{_{210}}}{+\frac{1}{180\sqrt{15}}\mathbf S_{24_{_{210}}}}+{\frac{1}{60\sqrt{6}}}\mathbf S_{75_{_{210}}}\right]},\nonumber\\
{\mathsf{t_{16}}}&{\equiv}&{-{\frac{2}{45}}\eta\mathbf S_{75_{_{{210}}}}}.
\end{eqnarray}
{Substitution of  $\mathbf{S_{1_{210}}}, ~\mathbf S_{75_{210}}, ~\mathbf S_{1_{126}}, ~\mathbf S_{1_{\ov{126}}}$  in terms of
 $\mathbf S_{24_{210}}$,  gives all the matrix elements of the Higgs triplet mass matrix in terms of one  VEV.
The Higgs triplet mass
matrix ${M_t}$  is diagonalized by an  $8\times8$ biunitary transformation}
\begin{equation}
{U_t}^{\dagger}{M_t}{V_t}=\textnormal{diag}\left(\mathsf{m}_{t_1},\mathsf{m}_{t_2},\cdots,\mathsf{m}_{t_8}\right).\label{eigenvalues2}
\end{equation}
There is no zero mode in the triplet mass matrix {and} all the eigenvalues of this  matrix are superheavy.
The triplet Higgs mass spectrum is, of course, central to the study of baryon and lepton number violating dimension five
operators leading to proton decay (for a review see~\cite{Nath:2006ut,Babu:2013jba}), which can act as a discriminant for a variety
of GUT and string models (see, e.g., \cite{Arnowitt:1993pd}).\\

\section{{\boldmath$\mathsf{B-L}=-2$}  operators from cubic matter-Higgs interactions  \label{sec:5}}

{In this section we compute the $\mathsf{B-L}=-2$ interactions arising in the model discussed in section 2.
The $\mathsf{B-L}$ violating interactions arise as a consequence of the singlets of $\mathsf{126}$ and $\overline{\mathsf{126}}$
gaining VEVs. In turn this VEV formation gives mass to the singlets of the $\mathsf{16}$-plets of matter.
Thus the heavy fields in the model after spontaneous breaking of the GUT  symmetry consist of
all of the Higgs fields except for a pair of  light Higgs doublets and in addition the singlet fields
arising from the $\mathsf{16}$-plets of matter.
From the couplings of Higgs with matter
we
are interested in pulling out only the parts that give $\mathsf{B-L}=-2$. To obtain a low energy effective Lagrangian
which contains $\mathsf{B-L}=-2$ violations, we integrate  on the  heavy fields which can generate
such interactions. { In this analysis we will focus on $B-L=-2$ interactions arising from the elimination
of  $\mathsf{5}+\overline{\mathsf{5}},~ \mathsf{45}+\overline{\mathsf{45}}$ fields (excluding the light modes)
and the matter singlets.
The elimination of  $\mathsf{10}+\overline{\mathsf{10}}$ will not be considered as this requires a further
overlapping analysis of  Goldstones in the $SO(10)$ symmetry breaking and their  absorption in the $\mathsf{10}+\overline{\mathsf{10}}$
gauge vector bosons to make them heavy in the symmetry breaking of $SO(10)$ and a full analysis of this is outside the scope of this work.
Returning to the integration over $\mathsf{5}+\overline{\mathsf{5}},~\mathsf{45}+\overline{\mathsf{45}}$ extra care is needed
in handling the integration.}}
This  is due to a mixing between the doublets and the triplets arising from $\mathsf{5}+\overline{\mathsf{5}}$ and
$\mathsf{45}+\overline{\mathsf{45}}$. Further, the doublet mass matrix has a zero mode which must be extracted
before integration on the heavy Higgs doublets can be performed. Similarly, integration on the
$\mathsf{45}+\overline{\mathsf{45}}$ requires that we first extract out the doublet and the triplets modes
before integration on them. We follow the following path in integration of the heavy fields:
 {First we integrate on the matter singlets and then integrate on the remaining heavy Higgs
fields. Another integration path is found to give the same result.}\\

We begin by displaying the cubic matter-Higgs couplings which consists of
  $\mathsf{16\cdot16\cdot10}$, $\mathsf{16\cdot16\cdot120}$ and $\mathsf{16\cdot16\cdot\overline{126}}$  couplings. In  $\mathsf{SU(5)}$ decomposition
  (for notation see Appendix A)   they are given by ~\cite{Nath:2001uw}

 \begin{eqnarray}
 \label{5.1a}
  W^{(16\cdot16\cdot10)}&=&\imath2\sqrt 2f^{(10_r+)}_{\acute{x}\acute{y}}\left(\mathsf{M}^{ij}_{\acute{x}}~\mathsf{M}_{\acute{y}i}~\mathsf{H}^{(10_r)}_{j}
  -\mathsf{M}_{\acute{x}}~\mathsf{M}_{\acute{y}i}~\mathsf{H}^{(10_r)i} + \frac{1}{8}\epsilon_{ijklm}~\mathsf{M}^{ij}_{\acute{x}}~\mathsf{M}^{kl}_{\acute{y}}~ \mathsf{H}^{(10_r)m}\right),\\
W^{(16\cdot16\cdot120)}&=&\imath\frac{2}{\sqrt 3}f_{\acute{x}\acute{y}}^{(120-)}\left(2 \mathsf{M}_{\acute{x}} ~\mathsf{M}_{\acute{y}i}~\mathsf{H}^{(120)i} +
\mathsf{M}^{ij}_{\acute{x}}~\mathsf{M}_{\acute{y}}~\mathsf{H}_{ij}^{(120)}
+\mathsf{M}_{\acute{x}i}~\mathsf{M}_{\acute{y}j}~\mathsf{H}^{(120)ij}\right.\nonumber\\
&&\left.-\mathsf{M}^{ij}_{\acute{x}}~\mathsf{M}_{\acute{y}i}~\mathsf{H}_j^{(120)}+  \mathsf{M}_{\acute{x}i}~\mathsf{M}_{\acute{y}}^{jk}~\mathsf{H}^{(120)i}_{jk}
-\frac{1}{4}\epsilon_{ijklm}~\mathsf{M}_{\acute{x}}^{ij}~\mathsf{M}_{\acute{y}}^{mn}~\mathsf{H}^{(120)kl}_n\right),\\
W^{(16\cdot16\cdot\overline{126})}&=& \imath\sqrt{\frac{2}{{15}}}f_{\acute{x}\acute{y}}^{(\overline{126}+)} \left(-\sqrt 2
\mathsf{M}_{\acute{x}}~ \mathsf{M}_{\acute{y}}~\mathsf{H}^{(\overline{126})}-\sqrt 3  \mathsf{M}_{\acute{x}}~ \mathsf{M}_{\acute{y}i}~\mathsf{H}^{(\overline{126})i}
+\mathsf{M}_{\acute{x}}~\mathsf{M}^{ij}_{\acute{y}}~\mathsf{H}_{ij}^{(\overline{126})}\right.\nonumber\\
&&\left.-\frac{1}{8\sqrt 3}\epsilon_{ijklm}~\mathsf{M}_{\acute{x}}^{ij}~\mathsf{M}_{\acute{y}}^{kl}~\mathsf{H}^{(\overline{126})m}
-\mathsf{M}_{\acute{x}i}~\mathsf{M}_{\acute{y}j}~\mathsf{H}^{(\overline{126})ij}_{(S)}+\mathsf{M}^{ij}_{\acute{x}}~\mathsf{M}_{\acute{y}k}~\mathsf{H}^{(\overline{126})k}_{ij}\right.\nonumber\\
&&\left.-\frac{1}{12\sqrt 2}\epsilon_{ijklm}~\mathsf{M}_{\acute{x}}^{lm}~\mathsf{M}_{\acute{y}}^{rs}~\mathsf{H}^{(\overline{126})ijk}_{rs}\right),
\label{5.1c}
\end{eqnarray}
where
the front factors $f_{\acute{x}\acute{y}}^{(\cdot\pm)}$ in Eq.(\ref{5.1a})-Eq.(\ref{5.1c})   exhibit
  the  symmetry and anti-symmetry in the generation indices:  $f_{\acute{x}\acute{y}}^{(\cdot\pm)}=\frac{1}{2}\left(f_{\acute{x}\acute{y}}^{(\cdot)}\pm f_{\acute{y}\acute{x}}^{(\cdot)}\right)$.\\

Next we assume that because of spontaneous symmetry breaking the singlet field in the $\mathsf{\overline{126}}$-plet of Higgs field develops a VEV,
 i.e., $<\mathsf{H}^{(\overline{126})}>~\equiv~\mathbf S_{1_{_{\overline{126}}}}     ~\neq~ 0$, which gives mass to the singlets in the $\mathsf{16}$-plet of matter fields.
 Collecting the terms which contain the singlet fields of matter from Eq.(\ref{5.1a})-Eq.(\ref{5.1c})
 we have
 \begin{eqnarray}
W&=&  \mathsf{M}_{\acute{x}}\left\{-\imath2\sqrt 2f^{(10_r+)}_{\acute{x}\acute{y}} \mathsf{M}_{\acute{y}i}~\mathsf{H}^{(10_r)i}
  + \imath\frac{4}{\sqrt 3}f_{\acute{x}\acute{y}}^{(120-)}\mathsf{M}_{\acute{y}i}~\mathsf{H}^{(120)i}\right.\nonumber\\
  &&\left.-\imath\sqrt{\frac{2}{{5}}}f_{\acute{x}\acute{y}}^{(\overline{126}+)}\mathsf{M}_{\acute{y}i}~\mathsf{H}^{(\overline{126})i}
\right\}+\frac{1}{2}\mathsf{M}_{\acute{x}}\left\{-\imath{\frac{4}{\sqrt{15}}}f_{\acute{x}\acute{y}}^{(\overline{126}+)}    \mathbf S_{1_{_{\overline{126}}}}     \right\}\mathsf{M}_{\acute{y}}
\end{eqnarray}
In the above equation the mass term for $\mathsf{1_{16}}$ violates $\mathsf{B-L}$.
Next eliminating  $\mathsf{M}_{\acute{x}}$ through $\displaystyle\frac{\partial{W}}{\partial{\mathsf{M}_{\acute{x}}}}=0$, we get the following 4-point matter-Higgs interactions:
\begin{equation}
W=  \sum_{i=1}^6 W_i,
\label{4.11}
\end{equation}
where

\begin{eqnarray}
W_1=\frac{\displaystyle 1}{\displaystyle\mathbf S_{1_{_{\overline{126}}}}     }
\imath\sqrt{15}~\mathsf{M}_{\acute{x}i}~\mathsf{H}^{(10_r)i}\left[f^{(10_r+)}f^{(\overline{126}+)^{-1}}f^{(10_s+)}\right]_{_{\acute{x}\acute{y}}}     \mathsf{M}_{\acute{y}j}~\mathsf{H}^{(10_s)j},
\label{4.13}
\end{eqnarray}

\begin{eqnarray}
W_2=-\frac{\displaystyle 1}{\displaystyle\mathbf S_{1_{_{\overline{126}}}}     }
\imath2\sqrt{\frac{5}{3}}~\mathsf{M}_{\acute{x}i}~\mathsf{H}^{(120)i}\left[f^{(120-)}f^{(\overline{126}+)^{-1}}f^{(120-)}\right]_{_{\acute{x}\acute{y}}}     \mathsf{M}_{\acute{y}j}~\mathsf{H}^{(120)j},
\label{4.14}
\end{eqnarray}

\begin{eqnarray}
W_3=\frac{\displaystyle 1}{\displaystyle\mathbf S_{1_{_{\overline{126}}}}     }
\frac{\imath}{4}\sqrt{\frac{3}{5}}~\mathsf{M}_{\acute{x}i}~\mathsf{H}^{(\overline{126})i}~f^{(\overline{126}+)}_{_{\acute{x}\acute{y}}}     ~\mathsf{M}_{\acute{y}j}~\mathsf{H}^{(\overline{126})j},
\label{4.15}
\end{eqnarray}

\begin{eqnarray}
W_4=-\frac{\displaystyle1}{\displaystyle\mathbf S_{1_{_{\overline{126}}}}     }
\imath2\sqrt{10}~\mathsf{M}_{\acute{x}i}~\mathsf{H}^{(10_r)i}\left[f^{(10_r+)}f^{(\overline{126}+)^{-1}}f^{(120-)}\right]_{_{\acute{x}\acute{y}}}     \mathsf{M}_{\acute{y}j}~\mathsf{H}^{(120)j},
\label{4.16}
\end{eqnarray}

\begin{eqnarray}
W_5=\frac{\displaystyle 1}{\displaystyle\mathbf S_{1_{_{\overline{126}}}}     }
\imath\sqrt{2}~\mathsf{M}_{\acute{x}i}~\mathsf{H}^{(120)i}~f^{(120-)}_{_{\acute{x}\acute{y}}}~\mathsf{M}_{\acute{y}j}~\mathsf{H}^{(\overline{126})j},
\label{4.17}
\end{eqnarray}

\begin{eqnarray}
W_6=\frac{\displaystyle1}{\displaystyle\mathbf S_{1_{_{\overline{126}}}}     }
\imath\sqrt{3}~\mathsf{M}_{\acute{x}i}~\mathsf{H}^{(10_r)i}~f^{(10_r+)}_{_{\acute{x}\acute{y}}}~\mathsf{M}_{\acute{y}j}~
\mathsf{H}^{(\overline{126})j}.
\label{4.18}
\end{eqnarray}

The $\mathsf{5}$-plets of Higgs can produce
heavy Higgs doublets and triplets and their decays violate $\mathsf{B-L}$.
 Further their couplings
carry new sources of CP violation not  subject to CKM constraints and can be large.
Thus these decays can be used to
produce GUT scale baryogenesis using standard techniques (see, for e.g., \cite{Enomoto:2011py,Babu:2012iv,Feng:2013zda}).
As pointed out in \cite{Enomoto:2011py,Babu:2012iv} a baryon number excess  produced this way in the early universe will not be
washed away by sphaleron interactions at the electroweak scale. We now compute the relevant $d=5$, $d=7$ and $d=9$ operators arising from the above matter-Higgs interactions.

\subsection{Operators arising from \boldmath{$\mathsf{M}_{\acute{x}i}~\mathsf{H}^{(10_r)i}~\mathsf{M}_{\acute{y}j}~\mathsf{H}^{(10_s)j}$}}
Here the Higgs fields could be doublets or triplets. Thus we have three possibilities, i.e., that both the
fields are doublets, both are triplets or one is a doublet and the other a triplet. Thus we write}

\begin{eqnarray}
W_1&=&\frac{\displaystyle 1}{\displaystyle\mathbf S_{1_{_{\overline{126}}}}     }
\imath\sqrt{15}~\mathsf{M}_{\acute{x}i}~\mathsf{H}^{(10_r)i}\left[f^{(10_r+)}f^{(\overline{126}+)^{-1}}f^{(10_s+)}\right]_{_{\acute{x}\acute{y}}}     \mathsf{M}_{\acute{y}j}~\mathsf{H}^{(10_s)j}\nonumber\\
&=&W_1^{DD}+W_1^{TT}+W_1^{DT},
\end{eqnarray}
where
\begin{eqnarray}
W_1^{DD}&\equiv &{\cal E}_{\acute{x}\acute{y}}^{(rs)}~
\mathsf{M}_{\acute{x}a}~\mathsf{M}_{\acute{y}b}~~\mathsf{H}^{(10_r)a}~\mathsf{H}^{(10_s)b},\\
W_1^{TT}&\equiv &{\cal E}_{\acute{x}\acute{y}}^{(rs)}~
\mathsf{M}_{\acute{x}\alpha}~\mathsf{M}_{\acute{y}\beta}~~\mathsf{H}^{(10_r)\alpha}~\mathsf{H}^{(10_s)\beta},\\
W_1^{DT}&\equiv &2{\cal E}_{\acute{x}\acute{y}}^{(rs)}~
\mathsf{M}_{\acute{x}a}~\mathsf{M}_{\acute{y}\alpha}~~\mathsf{H}^{(10_r)a}~\mathsf{H}^{(10_s)\alpha},
\end{eqnarray}
and {where}
\begin{equation}
{\cal E}_{\acute{x}\acute{y}}^{(rs)}\equiv\frac{\displaystyle 1}{\displaystyle\mathbf S_{1_{_{\overline{126}}}}     }\left(\imath\sqrt{15}\right)\left[f^{(10_r+)}
f^{(\overline{126}+)^{-1}}f^{(10_s+)}\right]_{_{\acute{x}\acute{y}}}.
\end{equation}
{Next we obtain effective operators at low energy from each of the terms $W_1^{DD}, W_1^{TT},
W_1^{DT}$.}

\subsubsection{Evaluating \boldmath{$W_1^{DD}$}}
\begin{eqnarray}
\label{54}
W_{1}^{DD\prime}&=&\sum_{r,s=1}^{2}{\cal E}_{\acute{x}\acute{y}}^{(rs)}~\mathsf{M}_{\acute{x}a}~\mathsf{M}_{\acute{y}b}
\left[{{U}_{d_{r1}}}~{\mathbf{H_u}}^a+\sum_{M=2}^{7}{{U}_{d_{rM}}}~{\cal{H}}_{\mathbf{u}M}^{a}\right]
\left[{{U}_{d_{s1}}}~{\mathbf{H_u}}^b+\sum_{N=2}^{7}{{U}_{d_{sN}}}~{\cal{H}}_{\mathbf{u}N}^{b}\right]\nonumber\\
&&+\frac{\imath}{2\sqrt{2}}\epsilon_{ijkla}~\mathsf{M}_{\acute{x}}^{ij}~\mathsf{M}_{\acute{y}}^{kl}
\sum_{N=2}^{7}\sum_{r=1}^{2}f^{(10_r+)}_{_{\acute{x}\acute{y}}}{{U}_{d_{rN}}}~{\cal{H}}_{\mathbf{u}N}^{a}\nonumber\\
&&+\imath 2\sqrt{2}~\mathsf{M}_{\acute{x}}^{ia}~\mathsf{M}_{\acute{y}i}\sum_{N=2}^{7}\sum_{r=1}^{2}f^{(10_r+)}_{_{\acute{x}\acute{y}}}{{V}_{d_{rN}}}
~{\cal{H}}_{\mathbf{d}Na}\nonumber\\
&&+\frac{1}{2}\sum_{N=2}^{7}{\mathsf{m}}_{d_N}~{\cal{H}}_{\mathbf{u}N}^{a}~{\cal{H}}_{\mathbf{d}Na}.
\end{eqnarray}

{In  Eq.(\ref{54}), for example, ${{\cal{H}}_{\mathbf{d}2}}_a\equiv{}^{2}\!{\mathsf D}_{a}^{\prime},~~{{\cal{H}}_{\mathbf{d}3}}_{a}\equiv {}^{3}\!{\mathsf D}_{a}^{\prime},......,{\cal{H}}_{\mathbf{u}4}^{a}\equiv {}^{4}\!{\mathsf D}^{\prime a},~~  {\cal{H}}_{\mathbf{u}5}^{a}\equiv  {}^{5}\!{\mathsf D}^{\prime a},É.$} and that the second and third lines  come from the $\mathsf{16\cdot16\cdot10}$ coupling.
Eliminating ${\cal{H}}_{\mathbf{u}N}^{a}$ and ${\cal{H}}_{\mathbf{d}Na}$ through $\displaystyle\frac{\partial W_{1}^{DD\prime}}{\partial{{\cal{H}}_{\mathbf{u}N}^{a}}}=0$ and $\displaystyle\frac{\partial W_{1}^{DD\prime}}{\partial{{\cal{H}}_{\mathbf{d}Na}}}=0$, we get {$\mathsf{B-L=-2}$ operators with four fields consisting of two matter fields and two
light Higgs fields, operators with five fields one of which is a light Higgs field and the other matter fields,
 and operators with six matter fields. In addition we also get $\mathsf{B=0}$, $\mathsf{L=0}$ operators with four matter fields.
Thus we have }

\begin{eqnarray}
\label{55}
W_{1}^{DD\prime}=&&\sum_{r,s=1}^{2}{\cal E}_{\acute{x}\acute{y}}^{(rs)}{{U}_{d_{r1}}}{{U}_{d_{s1}}}\mathbf{L}_{\acute{x}a}~\mathbf{L}_{\acute{y}b}~{\mathbf{H_u}}^a~{\mathbf{H_u}}^b\nonumber\\
&&+16\left[\mathbf{L}_{\acute{u}a}~\mathbf{L}_{\acute{v}b}~\mathbf{Q}^{a\alpha}_{\acute{w}}
~\mathbf{D}^{\mathsf c}_{\acute{x}\alpha}~\mathbf{Q}^{b\beta}_{\acute{y}}~\mathbf{D}^{\mathsf c}_{\acute{z}\beta}+\epsilon^{ac}\epsilon^{bd}~\mathbf{L}_{\acute{u}a}~\mathbf{L}_{\acute{v}b}~\mathbf{E}^{\mathsf c}_{\acute{w}}
~\mathbf{L}_{\acute{x}c}~\mathbf{E}^{\mathsf c}_{\acute{y}}~\mathbf{L}_{\acute{z}d}\right.\nonumber\\
&&\left.+\epsilon^{bc}~\mathbf{L}_{\acute{u}a}~\mathbf{L}_{\acute{v}b}~\mathbf{Q}^{a\alpha}_{\acute{w}}
~\mathbf{D}^{\mathsf c}_{\acute{x}\alpha}~\mathbf{E}^{\mathsf c}_{\acute{y}}~\mathbf{L}_{\acute{z}c}
+\epsilon^{ac}~\mathbf{L}_{\acute{u}a}~\mathbf{L}_{\acute{v}b}~\mathbf{E}^{\mathsf c}_{\acute{w}}~\mathbf{L}_{\acute{x}c}~\mathbf{Q}^{b\alpha}_{\acute{y}}
~\mathbf{D}^{\mathsf c}_{\acute{z}\alpha}\right]\nonumber\\
&&\times\sum_{p,q=1}^{2}\left(\sum_{M=2}^{7}\frac{{U}_{d_{pM}}
\left\{\sum_{s=1}^{2}f^{(10_s+)}_{_{\acute{w}\acute{x}}}{V}_{d_{sM}}\right\}}{{\mathsf{m}_{d_M}}}\right){\cal E}_{\acute{u}\acute{v}}^{(pq)} \left(\sum_{N=2}^{7}\frac{{{U}_{d_{qN}}}
\left\{\sum_{r=1}^{2}f^{(10_r+)}_{_{\acute{y}\acute{z}}}{V}_{d_{rN}}\right\}}{{\mathsf{m}_{d_N}}}\right)\nonumber\\
&&+\imath 8\sqrt{2}\left[\mathbf{L}_{\acute{w}a}~\mathbf{L}_{\acute{x}b}~\mathbf{Q}^{b\alpha}_{\acute{y}}
~\mathbf{D}^{\mathsf c}_{\acute{z}\alpha}~{\mathbf{H_u}}^a + \epsilon^{bc}~\mathbf{L}_{\acute{w}a}~\mathbf{L}_{\acute{x}b}~\mathbf{E}^{\mathsf c}_{\acute{y}}
~\mathbf{L}_{\acute{z}c}~{\mathbf{H_u}}^a\right]\nonumber\\
&&\times\sum_{N=2}^{7}\frac{\left\{\sum_{p,q=1}^{2}{U}_{d_{p1}}{\cal E}_{\acute{w}\acute{x}}^{(pq)}{{U}_{d_{qN}}}\right\}\left\{\sum_{s=1}^{2}f^{(10_s+)}_{_{\acute{y}\acute{z}}}{{V}_{d_{sN}}}\right\}}{{\mathsf{m}_{d_N}}}\nonumber\\
&&-16\left[\mathbf{E}^{\mathsf c}_{\acute{w}}~\mathbf{L}_{\acute{x}a}~\mathbf{U}^{\mathsf c}_{\acute{y}\alpha}~\mathbf{Q}^{a\alpha}_{\acute{z}}
+ \epsilon_{ab}~\mathbf{Q}^{a\alpha}_{\acute{w}}~\mathbf{D}^{\mathsf c}_{\acute{x}\alpha}~\mathbf{U}^{\mathsf c}_{\acute{y}\beta}~\mathbf{Q}^{b\beta}_{\acute{z}}
\right]\nonumber\\
&&\times \sum_{N=2}^{7}\frac{\left\{\sum_{r=1}^{2}f^{(10_r+)}_{_{\acute{w}\acute{x}}}{U}_{d_{rN}}\right\}
\left\{\sum_{s=1}^{2}f^{(10_s+)}_{_{\acute{y}\acute{z}}}{V}_{d_{sN}}\right\}}
{{\mathsf{m}_{d_N}}}.
\end{eqnarray}
\subsubsection{Evaluating \boldmath{$W_1^{TT}$}}
{For the case when both the Higgs fields are triplet, they are both superheavy and their elimination
leads to the following set of terms }

\begin{eqnarray}
\label{56}
W_{1}^{TT\prime}=&&
16\left[\epsilon^{\alpha\gamma\delta}\epsilon^{\beta\mu\nu}~\mathbf{D}^{\mathsf c}_{\acute{u}\alpha}~\mathbf{D}^{\mathsf c}_{\acute{v}\beta}~\mathbf{U}^{\mathsf c}_{\acute{w}\gamma}~\mathbf{D}^{\mathsf c}_{\acute{x}\delta}~\mathbf{U}^{\mathsf c}_{\acute{y}\mu}~\mathbf{D}^{\mathsf c}_{\acute{z}\nu}+\mathbf{D}^{\mathsf c}_{\acute{u}\alpha}~\mathbf{D}^{\mathsf c}_{\acute{v}\beta}~\mathbf{Q}^{a\alpha}_{\acute{w}}~\mathbf{L}_{\acute{x}a}~\mathbf{Q}^{b\beta}_{\acute{y}}~\mathbf{L}_{\acute{z}b}\right.\nonumber\\
&&\left.+\epsilon^{\alpha\gamma\delta}~\mathbf{D}^{\mathsf c}_{\acute{u}\alpha}~\mathbf{D}^{\mathsf c}_{\acute{v}\beta}~\mathbf{U}^{\mathsf c}_{\acute{w}\gamma}~\mathbf{D}^{\mathsf c}_{\acute{x}\delta}~\mathbf{Q}^{a\beta}_{\acute{y}}~\mathbf{L}_{\acute{z}a}+\epsilon^{\beta\gamma\delta}~\mathbf{D}^{\mathsf c}_{\acute{u}\alpha}~\mathbf{D}^{\mathsf c}_{\acute{v}\beta}~\mathbf{Q}^{a\alpha}_{\acute{w}}~\mathbf{L}_{\acute{x}a}~\mathbf{U}^{\mathsf c}_{\acute{y}\gamma}~\mathbf{D}^{\mathsf c}_{\acute{z}\delta}\right]\nonumber\\
&&\times\sum_{p,q=1}^{2}\left(\sum_{M=1}^{8}\frac{{U}_{t_{pM}}
\left\{\sum_{s=1}^{2}f^{(10_s+)}_{_{\acute{w}\acute{x}}}{V}_{t_{sM}}\right\}}{{\mathsf{m}_{t_M}}}\right){\cal E}_{\acute{u}\acute{v}}^{(pq)} \left(\sum_{N=1}^{8}\frac{{{U}_{t_{qN}}}
\left\{\sum_{r=1}^{2}f^{(10_r+)}_{_{\acute{y}\acute{z}}}{V}_{t_{rN}}\right\}}{{\mathsf{m}_{t_N}}}\right)\nonumber\\
&&+8\left[2\epsilon^{\alpha\beta\gamma}~\mathbf{U}^{\mathsf c}_{\acute{w}\alpha}~\mathbf{D}^{\mathsf c}_{\acute{x}\beta}~\mathbf{E}^{\mathsf c}_{\acute{y}}~\mathbf{U}^{\mathsf c}_{\acute{z}\gamma}+2\epsilon_{ab}~\mathbf{U}^{\mathsf c}_{\acute{w}\alpha}~\mathbf{D}^{\mathsf c}_{\acute{x}\beta}~\mathbf{Q}^{a\beta}_{\acute{y}}~\mathbf{Q}^{b\alpha}_{\acute{z}}\right.\nonumber\\
&&\left.+2~\mathbf{Q}^{a\alpha}_{\acute{w}}~\mathbf{L}_{\acute{x}a}~\mathbf{E}^{\mathsf c}_{\acute{y}}~\mathbf{U}^{\mathsf c}_{\acute{z}\alpha}
-\epsilon_{bc}\epsilon_{\alpha\beta\gamma}
~\mathbf{Q}^{a\alpha}_{\acute{w}}~\mathbf{L}_{\acute{x}a}~\mathbf{Q}^{b\beta}_{\acute{y}}~\mathbf{Q}^{c\gamma}_{\acute{z}}\right]\nonumber\\
&&\times\sum_{N=1}^{8}\frac{\left\{\sum_{r=1}^{2}f^{(10_r+)}_{_{\acute{w}\acute{x}}}{U}_{t_{rN}}\right\}
\left\{\sum_{s=1}^{2}f^{(10_s+)}_{_{\acute{y}\acute{z}}}{V}_{t_{sN}}\right\}}{{\mathsf{m}_{t_N}}}.
\end{eqnarray}
{Here the top two lines give us the $\mathsf{B-L=-2}$ operators with six matter fields while the bottom two lines
give us  $\mathsf{B-L=0}$ operators with four matter fields which include $\mathsf{B}$ violating and $\mathsf{L}$ violating operators.}

\subsubsection{Evaluating \boldmath{$W_1^{DT}$}}
{Here we have one Higgs field which is a doublet while the other field is a triplet.
Since the doublet fields have both  light and heavy modes while the triplets are all heavy, we get a combination of $\mathsf{B-L=-2}$
operators with five fields and with six fields. Additionally we get $\mathsf{B-L=0}$ operators with four fields as follows}

\begin{eqnarray}
\label{57}
W_1^{DT\prime}=&&-\imath 8\sqrt{2}\left[\epsilon^{\alpha\beta\gamma}~\mathbf{D}^{\mathsf c}_{\acute{w}\alpha}~\mathbf{L}_{\acute{x}a}~\mathbf{U}^{\mathsf c}_{\acute{y}\beta}~\mathbf{D}^{\mathsf c}_{\acute{z}\gamma}~{\mathbf{H_u}}^a+\mathbf{D}^{\mathsf c}_{\acute{w}\alpha}~\mathbf{L}_{\acute{x}a}~\mathbf{Q}^{b\alpha}_{\acute{y}}~\mathbf{L}_{\acute{z}b}~{\mathbf{H_u}}^a\right]\nonumber\\
&&\times\sum_{N=1}^{8}\frac{\left\{\sum_{p,q=1}^{2}{U}_{d_{p1}}{\cal E}_{\acute{w}\acute{x}}^{(pq)} {U}_{t_{qN}}\right\}
\left\{\sum_{s=1}^{2}f^{(10_s+)}_{_{\acute{y}\acute{z}}}{V}_{t_{sN}}\right\}}{{\mathsf{m}_{t_N}}}\nonumber\\
&&-64\left[\epsilon^{ab}\mathbf{L}_{\acute{u}a}~\mathbf{D}^{\mathsf c}_{\acute{v}\alpha}~\mathbf{E}^{\mathsf c}_{\acute{w}}
~\mathbf{L}_{\acute{x}b}~\mathbf{Q}^{c\alpha}_{\acute{y}}~\mathbf{L}_{\acute{z}c}
+\epsilon^{\alpha\beta\gamma}~\mathbf{L}_{\acute{u}a}~\mathbf{D}^{\mathsf c}_{\acute{v}\alpha}~\mathbf{Q}^{a\rho}_{\acute{w}}
~\mathbf{D}^{\mathsf c}_{\acute{x}\rho}~\mathbf{U}^{\mathsf c}_{\acute{y}\beta}~\mathbf{D}^{\mathsf c}_{\acute{z}\gamma}\right.\nonumber\\
&&\left.+\epsilon^{\alpha\beta\gamma}\epsilon^{ab}~\mathbf{L}_{\acute{u}a}~\mathbf{D}^{\mathsf c}_{\acute{v}\alpha}~\mathbf{E}^{\mathsf c}_{\acute{w}}
~\mathbf{L}_{\acute{x}b}~\mathbf{U}^{\mathsf c}_{\acute{y}\beta}~\mathbf{D}^{\mathsf c}_{\acute{z}\gamma}+\mathbf{L}_{\acute{u}a}~\mathbf{D}^{\mathsf c}_{\acute{v}\alpha}~\mathbf{Q}^{a\beta}_{\acute{w}}
~\mathbf{D}^{\mathsf c}_{\acute{x}\beta}~\mathbf{Q}^{b\alpha}_{\acute{y}}~\mathbf{L}_{\acute{z}b}\right]\nonumber\\
&&\times\sum_{p,q=1}^{2}\left(\sum_{M=2}^{7}\frac{{U}_{d_{pM}}
\left\{\sum_{s=1}^{2}f^{(10_s+)}_{_{\acute{w}\acute{x}}}{V}_{d_{sM}}\right\}}{{\mathsf{m}_{d_M}}}\right){\cal E}_{\acute{u}\acute{v}}^{(pq)} \left(\sum_{N=1}^{8}\frac{{{U}_{t_{qN}}}
\left\{\sum_{r=1}^{2}f^{(10_r+)}_{_{\acute{y}\acute{z}}}{V}_{t_{rN}}\right\}}{{\mathsf{m}_{t_N}}}\right)\nonumber\\
&&+8\left[2\epsilon^{\alpha\beta\gamma}~\mathbf{U}^{\mathsf c}_{\acute{w}\alpha}~\mathbf{D}^{\mathsf c}_{\acute{x}\beta}~\mathbf{E}^{\mathsf c}_{\acute{y}}~\mathbf{U}^{\mathsf c}_{\acute{z}\gamma}+2\epsilon_{ab}~\mathbf{U}^{\mathsf c}_{\acute{w}\alpha}~\mathbf{D}^{\mathsf c}_{\acute{x}\beta}~\mathbf{Q}^{a\beta}_{\acute{y}}~\mathbf{Q}^{b\alpha}_{\acute{z}}\right.\nonumber\\
&&\left.+2~\mathbf{Q}^{a\alpha}_{\acute{w}}~\mathbf{L}_{\acute{x}a}~\mathbf{E}^{\mathsf c}_{\acute{y}}~\mathbf{U}^{\mathsf c}_{\acute{z}\alpha}
-\epsilon_{bc}\epsilon_{\alpha\beta\gamma}
~\mathbf{Q}^{a\alpha}_{\acute{w}}~\mathbf{L}_{\acute{x}a}~\mathbf{Q}^{b\beta}_{\acute{y}}~\mathbf{Q}^{c\gamma}_{\acute{z}}\right]\nonumber\\
&&\times\sum_{N=1}^{8}\frac{\left\{\sum_{r=1}^{2}f^{(10_r+)}_{_{\acute{w}\acute{x}}}{U}_{t_{rN}}\right\}
\left\{\sum_{s=1}^{2}f^{(10_s+)}_{_{\acute{y}\acute{z}}}{V}_{t_{sN}}\right\}}{{\mathsf{m}_{t_N}}}\nonumber\\
&&-16\left[\mathbf{E}^{\mathsf c}_{\acute{w}}~\mathbf{L}_{\acute{x}a}~\mathbf{U}^{\mathsf c}_{\acute{y}\alpha}~\mathbf{Q}^{a\alpha}_{\acute{z}}
+ \epsilon_{ab}~\mathbf{Q}^{a\alpha}_{\acute{w}}~\mathbf{D}^{\mathsf c}_{\acute{x}\alpha}~\mathbf{U}^{\mathsf c}_{\acute{y}\beta}~\mathbf{Q}^{b\beta}_{\acute{z}}
\right]\nonumber\\
&&\times \sum_{N=2}^{7}\frac{\left\{\sum_{r=1}^{2}f^{(10_r+)}_{_{\acute{w}\acute{x}}}{U}_{d_{rN}}\right\}
\left\{\sum_{s=1}^{2}f^{(10_s+)}_{_{\acute{y}\acute{z}}}{V}_{d_{sN}}\right\}}
{{\mathsf{m}_{d_N}}}.
\end{eqnarray}

\subsection{Operators arising from \boldmath{$\mathsf{M}_{\acute{x}i}~\mathsf{H}^{(120)i}~\mathsf{M}_{\acute{y}j}~\mathsf{H}^{(120)j}$}}

{The analysis of this case is very similar to that of $W_2$ and the two Higgs fields can be either both doublets,
both triplets or one doublet and one triplet. Results are as below}

\begin{eqnarray}
W_2&=&-\frac{\displaystyle 1}{\displaystyle\mathbf S_{1_{_{\overline{126}}}}     }
\imath2\sqrt{\frac{5}{3}}~\mathsf{M}_{\acute{x}i}~\mathsf{H}^{(120)i}\left[f^{(120-)}f^{(\overline{126}+)^{-1}}f^{(120-)}\right]_{_{\acute{x}\acute{y}}}     \mathsf{M}_{\acute{y}j}~\mathsf{H}^{(120)j}\nonumber\\
&=&W_2^{DD}+W_2^{TT}+W_2^{DT},
\end{eqnarray}
where
\begin{eqnarray}
W_2^{DD}&\equiv &{\cal F}_{\acute{x}\acute{y}}~
\mathsf{M}_{\acute{x}a}~\mathsf{M}_{\acute{y}b}~~\mathsf{H}^{(120)a}~\mathsf{H}^{(120)b},\\
W_2^{TT}&\equiv &{\cal F}_{\acute{x}\acute{y}}~
\mathsf{M}_{\acute{x}\alpha}~\mathsf{M}_{\acute{y}\beta}~~\mathsf{H}^{(120)\alpha}~\mathsf{H}^{(120)\beta},\\
W_2^{DT}&\equiv &2{\cal F}_{\acute{x}\acute{y}}~
\mathsf{M}_{\acute{x}a}~\mathsf{M}_{\acute{y}\alpha}~~\mathsf{H}^{(120)a}~\mathsf{H}^{(120)\alpha},
\end{eqnarray}
and
\begin{equation}
{\cal F}_{\acute{x}\acute{y}}\equiv-\frac{\displaystyle 1}{\displaystyle\mathbf S_{1_{_{\overline{126}}}}}\left(\imath2\sqrt{\frac{5}{3}}\right)\left[f^{(120-)}
f^{(\overline{126}+)^{-1}}f^{(120-)}\right]_{_{\acute{x}\acute{y}}}.
\end{equation}
\subsubsection{Evaluating \boldmath{$W_2^{DD}$}}
\begin{eqnarray}
\label{63}
W_2^{DD\prime}=&&{U}_{d_{31}}^2{\cal F}_{\acute{x}\acute{y}}~\mathbf{L}_{\acute{x}a}~\mathbf{L}_{\acute{y}b}~{\mathbf{H_u}}^a~{\mathbf{H_u}}^b\nonumber\\
&&+\frac{8}{3}{\cal F}_{\acute{u}\acute{v}}f^{(120-)}_{_{\acute{w}\acute{x}}}f^{(120-)}_{_{\acute{y}\acute{z}}}
\left[\mathbf{L}_{\acute{u}a}~\mathbf{L}_{\acute{v}b}~\mathbf{Q}^{a\alpha}_{\acute{w}}
~\mathbf{D}^{\mathsf c}_{\acute{x}\alpha}~\mathbf{Q}^{b\beta}_{\acute{y}}~\mathbf{D}^{\mathsf c}_{\acute{z}\beta}+\epsilon^{ac}\epsilon^{bd}~\mathbf{L}_{\acute{u}a}~\mathbf{L}_{\acute{v}b}~\mathbf{E}^{\mathsf c}_{\acute{w}}
~\mathbf{L}_{\acute{x}c}~\mathbf{E}^{\mathsf c}_{\acute{y}}~\mathbf{L}_{\acute{z}d}\right.\nonumber\\
&&\left.+2\epsilon^{bc}~\mathbf{L}_{\acute{u}a}~\mathbf{L}_{\acute{v}b}~\mathbf{Q}^{a\alpha}_{\acute{w}}
~\mathbf{D}^{\mathsf c}_{\acute{x}\alpha}~\mathbf{E}^{\mathsf c}_{\acute{y}}~\mathbf{L}_{\acute{z}c}\right]\left(\sum_{N=2}^{7}\frac{{{V}_{d_{3N}}}
{{U}_{d_{3N}}}}{{\mathsf{m}_{d_N}}}\right)^2\nonumber\\
&&-\frac{\imath 8}{\sqrt{3}}{U}_{d_{31}}{\cal F}_{\acute{w}\acute{x}}f^{(120-)}_{_{\acute{y}\acute{z}}}\left[\mathbf{L}_{\acute{w}a}~\mathbf{L}_{\acute{x}b}~\mathbf{Q}^{b\alpha}_{\acute{y}}
~\mathbf{D}^{\mathsf c}_{\acute{z}\alpha}~{\mathbf{H_u}}^a + \epsilon^{bc}~\mathbf{L}_{\acute{w}a}~\mathbf{L}_{\acute{x}b}~\mathbf{E}^{\mathsf c}_{\acute{y}}
~\mathbf{L}_{\acute{z}c}~{\mathbf{H_u}}^a\right]\sum_{N=2}^{7}\frac{{{V}_{d_{3N}}}
{{U}_{d_{3N}}}}{{\mathsf{m}_{d_N}}}.
\end{eqnarray}
\subsubsection{Evaluating \boldmath{$W_2^{TT}$}}
\begin{eqnarray}
\label{64}
W_{2}^{TT\prime}=&&\frac{8}{3}{\cal F}_{\acute{u}\acute{v}}f^{(120-)}_{_{\acute{w}\acute{x}}}f^{(120-)}_{_{\acute{y}\acute{z}}}\left[\mathbf{D}^{\mathsf c}_{\acute{u}\alpha}~\mathbf{D}^{\mathsf c}_{\acute{v}\beta}~\mathbf{Q}^{a\alpha}_{\acute{w}}
~\mathbf{L}_{\acute{x}a}~\mathbf{Q}^{b\beta}_{\acute{y}}~\mathbf{L}_{\acute{z}b}+\epsilon^{\alpha\beta\gamma}\epsilon^{\rho\sigma\lambda}~\mathbf{D}^{\mathsf c}_{\acute{u}\alpha}~\mathbf{D}^{\mathsf c}_{\acute{v}\rho}~\mathbf{U}^{\mathsf c}_{\acute{w}\beta}~\mathbf{D}^{\mathsf c}_{\acute{x}\gamma}~\mathbf{U}^{\mathsf c}_{\acute{y}\sigma}~\mathbf{D}^{\mathsf c}_{\acute{z}\lambda}\right.\nonumber\\
&&\left.+2\epsilon^{\alpha\beta\gamma}~\mathbf{D}^{\mathsf c}_{\acute{u}\rho}~\mathbf{D}^{\mathsf c}_{\acute{v}\alpha}~\mathbf{Q}^{a\rho}_{\acute{w}}
~\mathbf{L}_{\acute{x}a}~\mathbf{U}^{\mathsf c}_{\acute{y}\beta}~\mathbf{D}^{\mathsf c}_{\acute{z}\gamma}\right]\left(\sum_{N=1}^{8}\frac{{{V}_{t_{3N}}}
{{U}_{t_{3N}}}}{{\mathsf{m}_{t_N}}}\right)^2.
\end{eqnarray}
\subsubsection{Evaluating \boldmath{$W_2^{DT}$}}
\begin{eqnarray}
\label{65}
W_2^{DT\prime}=&&\frac{\imath8}{\sqrt{3}}{U}_{d_{31}}{\cal F}_{\acute{w}\acute{x}}f^{(120-)}_{_{\acute{y}\acute{z}}}\left[\epsilon^{\alpha\beta\gamma}\mathbf{L}_{\acute{w}a}~\mathbf{D}^{\mathsf c}_{\acute{x}\alpha}~\mathbf{U}^{\mathsf c}_{\acute{y}\beta}
~\mathbf{D}^{\mathsf c}_{\acute{z}\gamma}~{\mathbf{H_u}}^a+\mathbf{L}_{\acute{w}a}
~\mathbf{D}^{\mathsf c}_{\acute{x}\alpha}~\mathbf{Q}^{b\alpha}_{\acute{w}}~\mathbf{L}_{\acute{z}b}~{\mathbf{H_u}}^a\right]\sum_{N=1}^{8}\frac{{{V}_{t_{3N}}}
{{U}_{t_{3N}}}}{{\mathsf{m}_{t_N}}}\nonumber\\
&&-\frac{32}{\sqrt{3}}{\cal F}_{\acute{u}\acute{v}}f^{(120-)}_{_{\acute{w}\acute{x}}}f^{(120-)}_{_{\acute{y}\acute{z}}}\left[\epsilon^{ab}\mathbf{L}_{\acute{u}a}~\mathbf{D}^{\mathsf c}_{\acute{v}\alpha}~\mathbf{E}^{\mathsf c}_{\acute{w}}
~\mathbf{L}_{\acute{x}b}~\mathbf{Q}^{c\alpha}_{\acute{y}}~\mathbf{L}_{\acute{z}c}+\epsilon^{\alpha\beta\gamma}~\mathbf{L}_{\acute{u}a}~\mathbf{D}^{\mathsf c}_{\acute{v}\alpha}~\mathbf{Q}^{a\rho}_{\acute{w}}
~\mathbf{D}^{\mathsf c}_{\acute{x}\rho}~\mathbf{U}^{\mathsf c}_{\acute{y}\beta}~\mathbf{D}^{\mathsf c}_{\acute{z}\gamma}\right.\nonumber\\
&&\left.+\epsilon^{\alpha\beta\gamma}\epsilon^{ab}~\mathbf{L}_{\acute{u}a}~\mathbf{D}^{\mathsf c}_{\acute{v}\alpha}~\mathbf{E}^{\mathsf c}_{\acute{w}}
~\mathbf{L}_{\acute{x}b}~\mathbf{U}^{\mathsf c}_{\acute{y}\beta}~\mathbf{D}^{\mathsf c}_{\acute{z}\gamma}+\mathbf{L}_{\acute{u}a}~\mathbf{D}^{\mathsf c}_{\acute{v}\alpha}~\mathbf{Q}^{a\beta}_{\acute{w}}
~\mathbf{D}^{\mathsf c}_{\acute{x}\beta}~\mathbf{Q}^{b\alpha}_{\acute{y}}~\mathbf{L}_{\acute{z}b}\right]\nonumber\\
&&\times\left(\sum_{M=2}^{7}\frac{{{V}_{d_{3M}}}
{{U}_{d_{3M}}}}{{\mathsf{m}_{d_M}}}\right)\left(\sum_{N=1}^{8}\frac{{{V}_{t_{3N}}}
{{U}_{t_{3N}}}}{{\mathsf{m}_{t_N}}}\right).
\end{eqnarray}
\subsection{Operators arising from \boldmath{$\mathsf{M}_{\acute{x}i}~\mathsf{H}^{(\overline{126})i}~\mathsf{M}_{\acute{y}j}~\mathsf{H}^{(\overline{126})j}$}}

\begin{eqnarray}
W_3=\frac{\displaystyle 1}{\displaystyle\mathbf S_{1_{_{\overline{126}}}}     }
\frac{\imath}{4}\sqrt{\frac{3}{5}}~\mathsf{M}_{\acute{x}i}~\mathsf{H}^{(\overline{126})i}~f^{(\overline{126}+)}_{_{\acute{x}\acute{y}}}     ~\mathsf{M}_{\acute{y}j}~\mathsf{H}^{(\overline{126})j}.
\end{eqnarray}
No contribution. Firstly, because ${U}_{d_{41}} =0$ and secondly because there is no $\mathsf{\overline{5}}$ of $\mathsf{SU(5)}$ in $\mathsf{\overline{126}}$ and hence a mass term involving $\mathsf{5}$ and  $\mathsf{\overline{5}}$ cannot be written.

\subsection{Operators arising from \boldmath{$\mathsf{M}_{\acute{x}i}~\mathsf{H}^{(10_r)i}~\mathsf{M}_{\acute{y}j}~\mathsf{H}^{(120)j}$}}
{The analysis of this case is similar to that of $W_2$ and $W_3$. Thus without
further explanation we give the analysis below}
\begin{eqnarray}
W_4&=&-\frac{\displaystyle1}{\displaystyle\mathbf S_{1_{_{\overline{126}}}}     }
\imath2\sqrt{10}~\mathsf{M}_{\acute{x}i}~\mathsf{H}^{(10_r)i}\left[f^{(10_r+)}f^{(\overline{126}+)^{-1}}f^{(120-)}\right]_{_{\acute{x}\acute{y}}}     \mathsf{M}_{\acute{y}j}~\mathsf{H}^{(120)j}\nonumber\\
&=&W_4^{DD}+W_4^{TT}+W_4^{DT}+W_4^{TD},
\end{eqnarray}
where
\begin{eqnarray}
W_4^{DD}&\equiv&{\cal G}_{\acute{x}\acute{y}}^{(r)}~
\mathsf{M}_{\acute{x}a}~\mathsf{M}_{\acute{y}b}~~\mathsf{H}^{(10_r)a}~\mathsf{H}^{(120)b},\\
W_4^{TT}&\equiv &{\cal G}_{\acute{x}\acute{y}}^{(r)}~
\mathsf{M}_{\acute{x}\alpha}~\mathsf{M}_{\acute{y}\beta}~~\mathsf{H}^{(10_r)\alpha}~\mathsf{H}^{(120)\beta},\\
W_4^{DT}&\equiv &{\cal G}_{\acute{x}\acute{y}}^{(r)}~
\mathsf{M}_{\acute{x}a}~\mathsf{M}_{\acute{y}\alpha}~~\mathsf{H}^{(10_r)a}~\mathsf{H}^{(120)\alpha},\\
W_4^{TD}&\equiv &{\cal G}_{\acute{x}\acute{y}}^{(r)}~
\mathsf{M}_{\acute{x}\alpha}~\mathsf{M}_{\acute{y}a}~~\mathsf{H}^{(10_r)\alpha}~\mathsf{H}^{(120)a},
\end{eqnarray}
and
\begin{equation}
{\cal G}_{\acute{x}\acute{y}}^{(r)}\equiv-\frac{\displaystyle 1}{\displaystyle\mathbf S_{1_{_{\overline{126}}}}}\left(\imath2\sqrt{10}\right)\left[f^{(10_r+)}
f^{(\overline{126}+)^{-1}}f^{(120-)}\right]_{_{\acute{x}\acute{y}}}.
\end{equation}
\subsubsection{Evaluating \boldmath{$W_4^{DD}$}}
\begin{eqnarray}
\label{74}
W_4^{DD\prime}=&&{U}_{d_{31}}\sum_{r=1}^{2}{U}_{d_{r1}}{\cal G}_{\acute{x}\acute{y}}^{(r)}~\mathbf{L}_{\acute{x}a}~\mathbf{L}_{\acute{y}b}
~{\mathbf{H_u}}^a~{\mathbf{H_u}}^b\nonumber\\
&&+8{\cal I}_{\acute{w}\acute{x}, \acute{y}\acute{z}}\left[\epsilon^{ac}\epsilon^{bd}~\mathbf{L}_{\acute{u}a}~\mathbf{L}_{\acute{v}b}~\mathbf{E}^{\mathsf c}_{\acute{w}}~\mathbf{L}_{\acute{x}c}~\mathbf{E}^{\mathsf c}_{\acute{y}}~\mathbf{L}_{\acute{z}d}+\mathbf{L}_{\acute{u}a}~\mathbf{L}_{\acute{v}b}~\mathbf{Q}^{a\alpha}_{\acute{w}}~\mathbf{D}^{\mathsf c}_{\acute{x}\alpha}~\mathbf{Q}^{b\beta}_{\acute{w}}~\mathbf{D}^{\mathsf c}_{\acute{z}\beta}\right.\nonumber\\
&&\left.+\epsilon^{ac}~\mathbf{L}_{\acute{u}a}~\mathbf{L}_{\acute{v}b}~\mathbf{E}^{\mathsf c}_{\acute{w}}~\mathbf{L}_{\acute{x}c}~\mathbf{Q}^{b\alpha}_{\acute{y}}~\mathbf{D}^{\mathsf c}_{\acute{z}\alpha}+\epsilon^{bc}~\mathbf{L}_{\acute{u}a}~\mathbf{L}_{\acute{v}b}~\mathbf{Q}^{a\alpha}_{\acute{w}}~\mathbf{D}^{\mathsf c}_{\acute{x}\alpha}~\mathbf{E}^{\mathsf c}_{\acute{y}}~\mathbf{L}_{\acute{z}c}\right]\nonumber\\
&&+\imath 4{U}_{d_{31}}\left[\epsilon^{ab}~\mathbf{L}_{\acute{w}a}~\mathbf{L}_{\acute{x}c}~\mathbf{E}^{\mathsf c}_{\acute{y}}~\mathbf{L}_{\acute{z}b}~{\mathbf{H_u}}^c+\mathbf{L}_{\acute{w}a}~\mathbf{L}_{\acute{x}b}~\mathbf{Q}^{a\alpha}_{\acute{y}}~\mathbf{D}^{\mathsf c}_{\acute{z}\alpha}~{\mathbf{H_u}}^b\right]\nonumber\\
&&\times\left[\sqrt{2}\sum_{N=2}^{7}\frac{\left(\sum_{r=1}^{2}{\cal G}_{\acute{w}\acute{x}}^{(r)}{{U}_{d_{rN}}}\right)\left(\sum_{s=1}^{2}f^{(10_s+)}_{_{\acute{y}\acute{z}}}{{V}_{d_{sN}}}\right)}{\mathsf{m}_{d_N}}
-\frac{1}{\sqrt{3}}f^{(120-)}_{_{\acute{y}\acute{z}}}\sum_{N=2}^{7}\frac{\left(\sum_{r=1}^{2}{\cal G}_{\acute{w}\acute{x}}^{(r)}{{U}_{d_{rN}}}\right){{V}_{d_{3N}}}}{\mathsf{m}_{d_N}}\right]\nonumber\\
&&+\imath 4\sum_{r=1}^{2}{U}_{d_{r1}}{\cal G}_{\acute{x}\acute{y}}^{(r)}\left[\epsilon^{bc}~\mathbf{L}_{\acute{w}a}~\mathbf{L}_{\acute{x}b}~\mathbf{E}^{\mathsf c}_{\acute{y}}~\mathbf{L}_{\acute{z}c}~{\mathbf{H_u}}^a+\mathbf{L}_{\acute{w}a}~\mathbf{L}_{\acute{x}b}~\mathbf{Q}^{b\alpha}_{\acute{y}}~\mathbf{D}^{\mathsf c}_{\acute{z}\alpha}~{\mathbf{H_u}}^a\right]\nonumber\\
&&\times\left[\sqrt{2}\sum_{N=2}^{7}\frac{{{U}_{d_{3N}}}\left(\sum_{s=1}^{2}f^{(10_s+)}_{_{\acute{y}\acute{z}}}{{V}_{d_{sN}}}\right)}{\mathsf{m}_{d_N}}
-\frac{1}{\sqrt{3}}f^{(120-)}_{_{\acute{y}\acute{z}}}\sum_{N=2}^{7}\frac{{{U}_{d_{3N}}}{{V}_{d_{3N}}}}{\mathsf{m}_{d_N}}\right]\nonumber\\
&&+8 \left[\epsilon_{ab}~\mathbf{Q}^{a\alpha}_{\acute{w}}~\mathbf{D}^{\mathsf c}_{\acute{x}\alpha}~\mathbf{U}^{\mathsf c}_{\acute{y}\beta}~\mathbf{Q}^{b\beta}_{\acute{z}}+\mathbf{E}^{\mathsf c}_{\acute{w}}~\mathbf{L}_{\acute{x}a}~\mathbf{U}^{\mathsf c}_{\acute{y}\alpha}~\mathbf{Q}^{a\alpha}_{\acute{z}}\right]\nonumber\\
&&\times\left[2\sum_{N=2}^{7}\frac{\left\{\sum_{r=1}^{2}f^{(10_r+)}_{_{\acute{w}\acute{x}}}{U}_{d_{rN}}\right\}
\left\{\sum_{s=1}^{2}f^{(10_s+)}_{_{\acute{y}\acute{z}}}{V}_{d_{sN}}\right\}}{{\mathsf{m}_{d_N}}}+\sqrt{\frac{2}{3}}f^{(120-)}_{_{\acute{w}\acute{x}}}
\sum_{N=2}^{7}\frac{\left\{\sum_{r=1}^{2}f^{(10_r+)}_{_{\acute{y}\acute{z}}}{U}_{d_{rN}}\right\}
{{V}_{d_{3N}}}}{{\mathsf{m}_{d_N}}} \right].\nonumber\\
 \end{eqnarray}
 The coefficient ${\cal I}_{\acute{w}\acute{x}, \acute{y}\acute{z}}$ is defined in Appendix E.
\subsubsection{Evaluating \boldmath{$W_4^{TT}$}}
\begin{eqnarray}
\label{75}
W_4^{TT\prime}=&&
8{\cal J}_{\acute{w}\acute{x}, \acute{y}\acute{z}}\left[\epsilon^{\alpha\gamma\delta}\epsilon^{\beta\mu\nu}~\mathbf{D}^{\mathsf c}_{\acute{u}\alpha}~\mathbf{D}^{\mathsf c}_{\acute{v}\beta}~\mathbf{U}^{\mathsf c}_{\acute{w}\gamma}~\mathbf{D}^{\mathsf c}_{\acute{x}\delta}~\mathbf{U}^{\mathsf c}_{\acute{y}\mu}~\mathbf{D}^{\mathsf c}_{\acute{z}\nu}+\mathbf{D}^{\mathsf c}_{\acute{u}\alpha}~\mathbf{D}^{\mathsf c}_{\acute{v}\beta}~\mathbf{Q}^{a\alpha}_{\acute{w}}~\mathbf{L}_{\acute{x}a}~\mathbf{Q}^{b\beta}_{\acute{y}}~\mathbf{L}_{\acute{z}b}\right.\nonumber\\
&&\left.+\epsilon^{\alpha\gamma\delta}~\mathbf{D}^{\mathsf c}_{\acute{u}\alpha}~\mathbf{D}^{\mathsf c}_{\acute{v}\beta}~\mathbf{U}^{\mathsf c}_{\acute{w}\gamma}~\mathbf{D}^{\mathsf c}_{\acute{x}\delta}~\mathbf{Q}^{a\beta}_{\acute{y}}~\mathbf{L}_{\acute{z}a}+\epsilon^{\beta\gamma\delta}~\mathbf{D}^{\mathsf c}_{\acute{u}\alpha}~\mathbf{D}^{\mathsf c}_{\acute{v}\beta}~\mathbf{Q}^{a\alpha}_{\acute{w}}~\mathbf{L}_{\acute{x}a}~\mathbf{U}^{\mathsf c}_{\acute{y}\gamma}~\mathbf{D}^{\mathsf c}_{\acute{z}\delta}\right]\nonumber\\
&&+4 \left[\epsilon_{ab}~\mathbf{Q}^{a\alpha}_{\acute{u}}~\mathbf{Q}^{b\beta}_{\acute{v}}~\mathbf{U}^{\mathsf c}_{\acute{w}\alpha}~\mathbf{D}^{\mathsf c}_{\acute{x}\beta}+\mathbf{E}^{\mathsf c}_{\acute{u}}~\mathbf{U}^{\mathsf c}_{\acute{v}\alpha}~\mathbf{Q}^{a\alpha}_{\acute{w}}~\mathbf{L}_{\acute{x}a}
+\epsilon^{\alpha\beta\gamma}~\mathbf{E}^{\mathsf c}_{\acute{u}}~\mathbf{U}^{\mathsf c}_{\acute{v}\alpha}~\mathbf{U}^{\mathsf c}_{\acute{w}\beta}~\mathbf{D}^{\mathsf c}_{\acute{x}\gamma}
\right]\nonumber\\
&&\times\left[2\sum_{N=1}^{8}\frac{\left\{\sum_{r=1}^{2}\left({\cal G}_{\acute{u}\acute{v}}^{(r)}+{\cal G}_{\acute{v}\acute{u}}^{(r)}\right){U}_{t_{rN}}\right\}
\left\{\sum_{s=1}^{2}f^{(10_s+)}_{_{\acute{w}\acute{x}}}{V}_{t_{sN}}\right\}}{{\mathsf{m}_{t_N}}}\right.\nonumber\\
&&\left.-\sqrt{\frac{2}{3}}f^{(120-)}_{_{\acute{w}\acute{x}}}
\sum_{N=1}^{8}\frac{\left\{\sum_{r=1}^{2}\left({\cal G}_{\acute{u}\acute{v}}^{(r)}+{\cal G}_{\acute{v}\acute{u}}^{(r)}\right){U}_{t_{rN}}\right\}
{{V}_{t_{3N}}}}{{\mathsf{m}_{t_N}}} \right]\nonumber\\
&&+4\epsilon_{ab}\epsilon_{\alpha\beta\gamma}~\mathbf{Q}^{a\alpha}_{\acute{u}}~\mathbf{Q}^{b\beta}_{\acute{v}}~\mathbf{Q}^{c\gamma}_{\acute{w}}~\mathbf{L}_{\acute{x}c}\nonumber\\
&&\times\left[-2\sum_{N=1}^{8}\frac{\left\{\sum_{r=1}^{2}{\cal G}_{\acute{u}\acute{v}}^{(r)}{U}_{t_{rN}}\right\}
\left\{\sum_{s=1}^{2}f^{(10_s+)}_{_{\acute{w}\acute{x}}}{V}_{t_{sN}}\right\}}{{\mathsf{m}_{t_N}}}+\sqrt{\frac{2}{3}}f^{(120-)}_{_{\acute{w}\acute{x}}}\sum_{N=1}^{8}\frac{\left\{\sum_{r=1}^{2}{\cal G}_{\acute{u}\acute{v}}^{(r)}{U}_{t_{rN}}\right\}
{{V}_{t_{3N}}}}{{\mathsf{m}_{t_N}}} \right].\nonumber\\
\end{eqnarray}
The coefficient ${\cal J}_{\acute{w}\acute{x}, \acute{y}\acute{z}}$ is explicitly given in Appendix E.

\subsubsection{Evaluating \boldmath{$W_4^{DT}$}}
\begin{eqnarray}
\label{76}
W_4^{DT\prime}=&&
16\sqrt{\frac{2}{3}} f^{(120-)}_{_{\acute{y}\acute{z}}}\left[\epsilon^{\alpha\beta\gamma}~\mathbf{L}_{\acute{u}a}~\mathbf{D}^{\mathsf c}_{\acute{v}\alpha}~\mathbf{Q}^{a\delta}_{\acute{w}}~\mathbf{D}^{\mathsf c}_{\acute{x}\delta}~\mathbf{U}^{\mathsf c}_{\acute{y}\beta}~\mathbf{D}^{\mathsf c}_{\acute{z}\gamma}+\epsilon^{ab}~\mathbf{L}_{\acute{u}a}~\mathbf{D}^{\mathsf c}_{\acute{v}\alpha}~\mathbf{E}^{\mathsf c}_{\acute{w}}~\mathbf{L}_{\acute{x}b}~\mathbf{Q}^{c\alpha}_{\acute{y}}~\mathbf{L}_{\acute{z}c}\right.\nonumber\\
&&\left.+\mathbf{L}_{\acute{u}a}~\mathbf{D}^{\mathsf c}_{\acute{v}\alpha}~\mathbf{Q}^{a\beta}_{\acute{w}}~\mathbf{D}^{\mathsf c}_{\acute{x}\beta}~\mathbf{Q}^{b\alpha}_{\acute{y}}~\mathbf{L}_{\acute{z}b}+\epsilon^{ab}\epsilon^{\alpha\beta\gamma}
~\mathbf{L}_{\acute{u}a}~\mathbf{D}^{\mathsf c}_{\acute{v}\alpha}~\mathbf{E}^{\mathsf c}_{\acute{w}}~\mathbf{L}_{\acute{x}b}~\mathbf{U}^{\mathsf c}_{\acute{y}\beta}~\mathbf{D}^{\mathsf c}_{\acute{z}\gamma}\right]\nonumber\\
&&\times\left(\sum_{M=2}^{7}\frac{\left\{\sum_{r=1}^{2}{\cal G}_{\acute{u}\acute{v}}^{(r)}{U}_{d_{rM}}\right\}
\left\{\sum_{s=1}^{2}f^{(10_s+)}_{_{\acute{w}\acute{x}}}{V}_{d_{sM}}\right\}}{{\mathsf{m}_{d_M}}}\right)\left(\sum_{N=1}^{8}\frac{{{U}_{t_{3N}}}
{V}_{t_{3N}}}{{\mathsf{m}_{t_N}}}\right)\nonumber\\
&&+\imath \frac{4}{\sqrt{3}}\sum_{r=1}^{2}{U}_{d_{r1}}{\cal G}_{\acute{x}\acute{y}}^{(r)} f^{(120-)}_{_{\acute{y}\acute{z}}}\left[\epsilon^{\alpha\beta\gamma}~\mathbf{U}^{\mathsf c}_{\acute{w}\gamma}~\mathbf{D}^{\mathsf c}_{\acute{x}\alpha}~\mathbf{L}_{\acute{y}a}
~\mathbf{D}^{\mathsf c}_{\acute{z}\beta}~{\mathbf{H_u}}^a+\mathbf{Q}^{a\alpha}_{\acute{w}}~\mathbf{L}_{\acute{x}a}~\mathbf{L}_{\acute{y}b}~\mathbf{D}^{\mathsf c}_{\acute{z}\alpha}~{\mathbf{H_u}}^b\right]\times\sum_{N=1}^{8}\frac{{{V}_{t_{3N}}}
{{U}_{t_{3N}}}}{{\mathsf{m}_{t_N}}}\nonumber\\
 &&-16\left[\epsilon_{ab}~\mathbf{Q}^{a\alpha}_{\acute{w}}~\mathbf{D}^{\mathsf c}_{\acute{x}\alpha}~\mathbf{U}^{\mathsf c}_{\acute{y}\beta}~\mathbf{Q}^{b\beta}_{\acute{z}}+\mathbf{E}^{\mathsf c}_{\acute{w}}~\mathbf{L}_{\acute{x}a}~\mathbf{U}^{\mathsf c}_{\acute{y}\alpha}~\mathbf{Q}^{a\alpha}_{\acute{z}}\right]\sum_{N=2}^{7}\frac{\left\{\sum_{r=1}^{2}f^{(10_r+)}_{_{\acute{w}\acute{x}}}{U}_{d_{rN}}\right\}
\left\{\sum_{s=1}^{2}f^{(10_s+)}_{_{\acute{y}\acute{z}}}{V}_{d_{sN}}\right\}}{{\mathsf{m}_{d_N}}}.\nonumber\\
\end{eqnarray}

\subsubsection{Evaluating \boldmath{$W_4^{TD}$}}
\begin{eqnarray}
\label{77}
W_4^{TD\prime}=&&-16\sqrt{\frac{2}{3}}f^{(120-)}_{_{\acute{y}\acute{z}}}\left[\epsilon^{\alpha\beta\gamma}~\mathbf{D}^{\mathsf c}_{\acute{u}\alpha}~\mathbf{L}_{\acute{v}a}~\mathbf{U}^{\mathsf c}_{\acute{w}\beta}~\mathbf{D}^{\mathsf c}_{\acute{x}\gamma}~\mathbf{Q}^{a\rho}_{\acute{y}}~\mathbf{D}^{\mathsf c}_{\acute{z}\rho}
+\epsilon^{ab}\epsilon^{\alpha\beta\gamma}
~\mathbf{D}^{\mathsf c}_{\acute{u}\alpha}~\mathbf{L}_{\acute{v}a}~\mathbf{U}^{\mathsf c}_{\acute{w}\beta}~\mathbf{D}^{\mathsf c}_{\acute{x}\gamma}~\mathbf{E}^{\mathsf c}_{\acute{y}}~\mathbf{L}_{\acute{z}b}\right.\nonumber\\
&&\left.+\epsilon^{ac}~\mathbf{D}^{\mathsf c}_{\acute{u}\alpha}~\mathbf{L}_{\acute{v}a}~\mathbf{Q}^{b\alpha}_{\acute{w}}~\mathbf{L}_{\acute{x}b}~\mathbf{E}^{\mathsf c}_{\acute{y}}~\mathbf{L}_{\acute{z}c}+\mathbf{D}^{\mathsf c}_{\acute{u}\alpha}~\mathbf{L}_{\acute{v}a}~\mathbf{Q}^{b\alpha}_{\acute{w}}~\mathbf{L}_{\acute{x}b}
~\mathbf{Q}^{a\beta}_{\acute{y}}~\mathbf{D}^{\mathsf c}_{\acute{z}\beta}\right]\nonumber\\
&&\times \left(\sum_{M=1}^{8}\frac{\left\{\sum_{r=1}^{2}{\cal G}_{\acute{u}\acute{v}}^{(r)}{U}_{t_{rM}}\right\}
\left\{\sum_{s=1}^{2}f^{(10_s+)}_{_{\acute{w}\acute{x}}}{V}_{t_{sM}}\right\}}{{\mathsf{m}_{t_M}}}\right)\left(\sum_{N=2}^{7}\frac{{{U}_{d_{3N}}}
{V}_{d_{3N}}}{{\mathsf{m}_{d_N}}}\right)\nonumber\\
&&-\imath 4\sqrt{2}{U}_{d_{31}}\left[\epsilon^{\alpha\beta\gamma}~\mathbf{D}^{\mathsf c}_{\acute{w}\alpha}~\mathbf{L}_{\acute{x}a}~\mathbf{U}^{\mathsf c}_{\acute{y}\beta}~\mathbf{D}^{\mathsf c}_{\acute{z}\gamma}~{\mathbf{H_u}}^a+\mathbf{D}^{\mathsf c}_{\acute{w}\alpha}~\mathbf{L}_{\acute{x}a}~\mathbf{Q}^{b\alpha}_{\acute{y}}~\mathbf{L}_{\acute{z}b}~{\mathbf{H_u}}^a\right]\nonumber\\
&&\times\sum_{N=1}^{8}\frac{\left\{\sum_{r=1}^{2}{\cal G}_{\acute{w}\acute{x}}^{(r)}{U}_{t_{rN}}\right\}
\left\{\sum_{s=1}^{2}f^{(10_s+)}_{_{\acute{y}\acute{z}}}{V}_{t_{sN}}\right\}}{{\mathsf{m}_{t_N}}}\nonumber\\
&&+8\left[2\epsilon^{\alpha\beta\gamma}~\mathbf{U}^{\mathsf c}_{\acute{w}\alpha}~\mathbf{D}^{\mathsf c}_{\acute{x}\beta}~\mathbf{E}^{\mathsf c}_{\acute{y}}~\mathbf{U}^{\mathsf c}_{\acute{z}\gamma}+2\epsilon_{ab}~\mathbf{U}^{\mathsf c}_{\acute{w}\alpha}~\mathbf{D}^{\mathsf c}_{\acute{x}\beta}~\mathbf{Q}^{a\beta}_{\acute{y}}~\mathbf{Q}^{b\alpha}_{\acute{z}}+2~\mathbf{Q}^{a\alpha}_{\acute{w}}~\mathbf{L}_{\acute{x}a}~\mathbf{E}^{\mathsf c}_{\acute{y}}~\mathbf{U}^{\mathsf c}_{\acute{z}\alpha}\right.\nonumber\\
&&\left.
-\epsilon_{bc}\epsilon_{\alpha\beta\gamma}
~\mathbf{Q}^{a\alpha}_{\acute{w}}~\mathbf{L}_{\acute{x}a}~\mathbf{Q}^{b\beta}_{\acute{y}}~\mathbf{Q}^{c\gamma}_{\acute{z}}\right]\sum_{N=1}^{8}\frac{\left\{\sum_{r=1}^{2}f^{(10_r+)}_{_{\acute{w}\acute{x}}}{U}_{t_{rN}}\right\}
\left\{\sum_{s=1}^{2}f^{(10_s+)}_{_{\acute{y}\acute{z}}}{V}_{t_{sN}}\right\}}{{\mathsf{m}_{t_N}}}.
\end{eqnarray}
\subsection{Operators arising from \boldmath{$\mathsf{M}_{\acute{x}i}~\mathsf{H}^{(120)i}~\mathsf{M}_{\acute{y}j}~\mathsf{H}^{(\overline{126})j}$}}
\begin{eqnarray}
W_5=\frac{\displaystyle 1}{\displaystyle\mathbf S_{1_{_{\overline{126}}}}     }
\imath\sqrt{2}~\mathsf{M}_{\acute{x}i}~\mathsf{H}^{(120)i}~f^{(120-)}_{_{\acute{x}\acute{y}}}~\mathsf{M}_{\acute{y}j}~\mathsf{H}^{(\overline{126})j}.
\end{eqnarray}
This term does not generate any ${\mathsf{B-L}}=-2$ operators involving only the SM fields  because, firstly  ${U}_{d_{41}} =0$ and secondly because there is no
 $\mathsf{\overline{5}}$ of $\mathsf{SU(5)}$ in $\mathsf{\overline{126}}$ and thus a mass term involving $\mathsf{5}$ and  $\mathsf{\overline{5}}$ cannot be written.

 \subsection{Operators arising from \boldmath{$\mathsf{M}_{\acute{x}i}~\mathsf{H}^{(10_r)i}~\mathsf{M}_{\acute{y}j}~
\mathsf{H}^{(\overline{126})j}$}}
\begin{eqnarray}
W_6=\frac{\displaystyle1}{\displaystyle\mathbf S_{1_{_{\overline{126}}}}     }
\imath\sqrt{3}~\mathsf{M}_{\acute{x}i}~\mathsf{H}^{(10_r)i}~f^{(10_r+)}_{_{\acute{x}\acute{y}}}~\mathsf{M}_{\acute{y}j}~
\mathsf{H}^{(\overline{126})j}.
\end{eqnarray}
Just as in the preceding case this term also does not generate any ${\mathsf{B-L}}=-2$ operators involving only the Standard Model fields.

 \begin{table}[t]
\begin{center}
\begin{tabular}{|c|c|c|c|c|}
\hline
&Effective Operator&$B$&$L$ &       dim$L_{\rm SM}$         \\
\hline\hline
\multirow{1}{*}{4 Fields}&  $\mathbf{L}_{a}~\mathbf{L}_{b}~{\mathbf{H_u}}^a~{\mathbf{H_u}}^b$& $0$&$+2$&{$5$}\\
\hline
  \multirow{3}{*}{5 Fields} &  $\epsilon^{\alpha\beta\gamma}~\mathbf{D}^{\mathsf c}_{\alpha}
~\mathbf{D}^{\mathsf c}_{\beta}~\mathbf{U}^{\mathsf c}_{\gamma}~\mathbf{L}_{a}~{\mathbf{H_u}}^a$& $-1$&$+1$&$7$\\
  &$\epsilon^{ab}~\mathbf{L}_{a}
~\mathbf{L}_{b}~\mathbf{L}_{c}~\mathbf{E}^{\mathsf c}~{\mathbf{H_u}}^c$&$0$ &$+2$&$7$\\
 &$
\mathbf{D}^{\mathsf c}_{\alpha}~\mathbf{Q}^{a\alpha}~\mathbf{L}_{a}~\mathbf{L}_{b}~\mathbf{H_u}^b$&$0$ &$+2$&$7$\\
\hline
\multirow{6}{*}{6 Fields} &  $\epsilon^{\alpha\beta\gamma}\epsilon^{\rho\sigma\lambda}~\mathbf{D}^{\mathsf c}_{\alpha}
~\mathbf{D}^{\mathsf c}_{\beta}~\mathbf{U}^{\mathsf c}_{\gamma}~\mathbf{D}^{\mathsf c}_{\rho}~\mathbf{D}^{\mathsf c}_{\sigma}~\mathbf{U}^{\mathsf c}_{\lambda}$&$-2$ &$0$&$9$\\
&$\epsilon^{ab}\epsilon^{cd}~\mathbf{L}_{a}
~\mathbf{L}_{b}~\mathbf{L}_{c}~\mathbf{L}_{d}~\mathbf{E}^{\mathsf c}~\mathbf{E}^{\mathsf c}$&$0$ &$+2$&$9$\\
&$\epsilon^{\alpha\beta\gamma}\epsilon^{ab}~\mathbf{D}^{\mathsf c}_{\alpha}
~\mathbf{D}^{\mathsf c}_{\beta}~\mathbf{U}^{\mathsf c}_{\gamma}~\mathbf{L}_{a}~\mathbf{L}_{b}~\mathbf{E}^{\mathsf c}$& $-1$&$+1$&$9$\\
&$\epsilon^{\alpha\beta\gamma}~\mathbf{D}^{\mathsf c}_{\alpha}
~\mathbf{D}^{\mathsf c}_{\beta}~\mathbf{U}^{\mathsf c}_{\gamma}~\mathbf{D}^{\mathsf c}_{\rho}~\mathbf{Q}^{a\rho}~\mathbf{L}_{a}$&$-1$ &$+1$&$9$\\
&$\epsilon^{ab}~\mathbf{D}^{\mathsf c}_{\alpha}~\mathbf{Q}^{c\alpha}~\mathbf{L}_{a}
~\mathbf{L}_{b}~\mathbf{L}_{c}~\mathbf{E}^{\mathsf c}$&$0$ &$+2$&$9$\\
&$
\mathbf{D}^{\mathsf c}_{\alpha}~\mathbf{D}^{\mathsf c}_{\beta}~\mathbf{Q}^{a\alpha}~\mathbf{Q}^{b\beta}~\mathbf{L}_{a}~\mathbf{L}_{b}$&$0$ &$+2$&$9$\\
\hline
\multirow{4}{*}{4 Fields} &  $\epsilon_{\alpha\beta\gamma}\epsilon_{ab}~\mathbf{Q}^{a\alpha}~\mathbf{Q}^{b\beta}~\mathbf{Q}^{c\gamma}~\mathbf{L}_{c}$&$+1$ &$+1$&{$6$}\\
&  $\epsilon^{\alpha\beta\gamma}~\mathbf{D}^{\mathsf c}_{\alpha}~\mathbf{U}^{\mathsf c}_{\beta}~\mathbf{U}^{\mathsf c}_{\gamma}~\mathbf{E}^{\mathsf c}$& $-1$&$-1$ &{$6$}\\
&  $\epsilon_{ab}~\mathbf{D}^{\mathsf c}_{\alpha}~\mathbf{U}^{\mathsf c}_{\beta}~\mathbf{Q}^{a\alpha}~\mathbf{Q}^{b\beta}$&$0$ &$0$ &{$6$}\\
&  $\mathbf{U}^{\mathsf c}_{\alpha}~\mathbf{Q}^{a\alpha}~\mathbf{L}_{a}~\mathbf{E}^{\mathsf c}$& $0$&$0$ &{$6$}\\
\hline
\hline\hline
\end{tabular}
\caption{\small{Summary of the ${\mathsf{B-L}}=0$ {(bottom four entries)}
and
 ${\mathsf{B-L}}=-2$ {(all the remaining)}
  operators that arise from matter-Higgs interactions involving $\mathsf{10}$, $\mathsf{120}$ and $\mathsf{\overline{126}}$ of
Higgs fields.
The object
dim$L_{\rm SM}$  is the dimensionality of operators in the Lagrangian which have only the {Standard Model} particles
{with even R parity}. }}\label{qnos}
\end{center}
\end{table}

\section{Discussion of results\label{sec6}}

The analysis of Sec.(4)shows that there are  a large number of sources of {$\mathsf{B-L=-2}$}
 operators arising
from matter-Higgs interactions.  As discussed in Sec.(4), the cubic matter-Higgs interactions
consist of couplings of type $\mathsf{16\cdot16\cdot10}$, $\mathsf{16\cdot16\cdot120}$ and $\mathsf{16\cdot16\cdot\overline{126}}$.  After
 the singlet in $\mathsf{\overline{126}}$ develops a VEV of size the GUT scale,
the  singlet in the $\mathsf{16}$-plet of matter gains a GUT size mass via the $\mathsf{16\cdot16\cdot\overline{126}}$ coupling
which violates {$\mathsf{B-L}$} by two units.  The elimination of the singlet of the $\mathsf{16}$-plet results in a large
number of  {$\mathsf{B-L=-2}$} violating interactions, and  thus it is useful to indicate all the sources of such terms.
For convenience of the reader we summarize the sources of the {$\mathsf{B-L}$} violating  dimension 5,
dimension 7 and dimension 9 operators below:\\

{$\mathsf{B-L=-2}$}
 dimension 5 operator:
$\mathbf{L}_{a}~\mathbf{L}_{b}~{\mathbf{H_u}}^a~{\mathbf{H_u}}^b$}: This operator arises from a number of sources.
The total contribution to it can be gotten from Eqs.(\ref{55}, \ref{63},  \ref{74}).
Next we consider the {$\mathsf{B-L=-2}$} dimension 7 operators. Here we have the following {set of} operators:
(i) $\epsilon^{\alpha\beta\gamma}~\mathbf{D}^{\mathsf c}_{\alpha}
~\mathbf{D}^{\mathsf c}_{\beta}~\mathbf{U}^{\mathsf c}_{\gamma}~\mathbf{L}_{a}~{\mathbf{H_u}}^a$:
{Contribution to it arises from}
Eqs.(\ref{57},  \ref{65},  \ref{76}, \ref{77});  (ii)
 $\epsilon^{ab}~\mathbf{L}_{a}
~\mathbf{L}_{b}~\mathbf{L}_{c}~\mathbf{E}^{\mathsf c}~{\mathbf{H_u}}^c$:
{Contribution to it arises from}
Eqs.(\ref{55},  \ref{63},  \ref{74}); (iii)
$\mathbf{D}^{\mathsf c}_{\alpha}~\mathbf{Q}^{a\alpha}~\mathbf{L}_{a}~\mathbf{L}_{b}~\mathbf{H_u}^b$:
{Contribution to it arises from}
Eqs.(\ref{55},  \ref{57},  \ref{63}, \ref{65}, \ref{74},
\ref{76}, \ref{77}).
From Table 4 we see that there are six {$\mathsf{B-L=-2}$} dimension nine operators:
(a)   $\epsilon^{\alpha\beta\gamma}\epsilon^{\rho\sigma\lambda}~\mathbf{D}^{\mathsf c}_{\alpha}
~\mathbf{D}^{\mathsf c}_{\beta}~\mathbf{U}^{\mathsf c}_{\gamma}~\mathbf{D}^{\mathsf c}_{\rho}~\mathbf{D}^{\mathsf c}_{\sigma}~\mathbf{U}^{\mathsf c}_{\lambda}$: {Contribution to it arises from}
Eqs.(\ref{56},  \ref{64}, \ref{75}); (b)
 $\epsilon^{ab}\epsilon^{cd}~\mathbf{L}_{a}
~\mathbf{L}_{b}~\mathbf{L}_{c}~\mathbf{L}_{d}~\mathbf{E}^{\mathsf c}~\mathbf{E}^{\mathsf c}$:
{Contribution to it arises from}
Eqs.(\ref{55},  \ref{63}, \ref{74}); (c)
$\epsilon^{\alpha\beta\gamma}\epsilon^{ab}~\mathbf{D}^{\mathsf c}_{\alpha}
~\mathbf{D}^{\mathsf c}_{\beta}~\mathbf{U}^{\mathsf c}_{\gamma}~\mathbf{L}_{a}~\mathbf{L}_{b}~\mathbf{E}^{\mathsf c}$:
{Contribution to it arises from}
Eqs.(\ref{57}, \ref{65},
\ref{76}, \ref{77}); (d)
$\epsilon^{\alpha\beta\gamma}~\mathbf{D}^{\mathsf c}_{\alpha}
~\mathbf{D}^{\mathsf c}_{\beta}~\mathbf{U}^{\mathsf c}_{\gamma}~\mathbf{D}^{\mathsf c}_{\rho}~\mathbf{Q}^{a\rho}~\mathbf{L}_{a}$:
{Contribution to it arises from}
Eqs.(\ref{56}, \ref{57},
\ref{64}, \ref{65},  \ref{75},  \ref{76},  \ref{77}); (e)
 $\epsilon^{ab}~\mathbf{D}^{\mathsf c}_{\alpha}~\mathbf{Q}^{c\alpha}~\mathbf{L}_{a}
~\mathbf{L}_{b}~\mathbf{L}_{c}~\mathbf{E}^{\mathsf c}$:
{Contribution to it arises from}
Eqs.(\ref{55}, \ref{57},
\ref{63}, \ref{65}, \ref{74},  \ref{76},  \ref{77}); (f)
 $
\mathbf{D}^{\mathsf c}_{\alpha}~\mathbf{D}^{\mathsf c}_{\beta}~\mathbf{Q}^{a\alpha}~\mathbf{Q}^{b\beta}~\mathbf{L}_{a}~\mathbf{L}_{b}$:
{Contribution to it arises from}
Eqs.(\ref{55},   \ref{56}, \ref{57},
\ref{63}, \ref{64}, \ref{65}, \ref{74}, \ref{75}, \ref{76},  \ref{77}). \\

{We also} summarize the sources of  the {$\mathsf{B-L=0}$} operators. There  are four of them as shown in Table 4. We list their sources as follows:
(i) $\epsilon_{\alpha\beta\gamma}\epsilon_{ab}~\mathbf{Q}^{a\alpha}~\mathbf{Q}^{b\beta}~\mathbf{Q}^{c\gamma}~\mathbf{L}_{c}$:
{Contribution to it arises from}
Eqs.(\ref{57}, \ref{75},  \ref{77});
(ii) $\epsilon^{\alpha\beta\gamma}~\mathbf{D}^{\mathsf c}_{\alpha}~\mathbf{U}^{\mathsf c}_{\beta}~\mathbf{U}^{\mathsf c}_{\gamma}~\mathbf{E}^{\mathsf c}$: Contribution to it arise from Eqs.(\ref{57}, \ref{75},  \ref{77}); (iii)
 $\epsilon_{ab}~\mathbf{D}^{\mathsf c}_{\alpha}~\mathbf{U}^{\mathsf c}_{\beta}~\mathbf{Q}^{a\alpha}~\mathbf{Q}^{b\beta}$:
 Contribution to it arises from
Eqs.(\ref{55},  \ref{57},  \ref{74}, \ref{75}, \ref{76},
\ref{77}); (iv)
 $\mathbf{U}^{\mathsf c}_{\alpha}~\mathbf{Q}^{a\alpha}~\mathbf{L}_{a}~\mathbf{E}^{\mathsf c}$:
 Contribution to it arises from
Eqs.(\ref{55},  \ref{57},  \ref{74}, \ref{75}, \ref{76},
\ref{77}).\\

{It is instructive to trace back to the primitive $\mathsf{SU(5)}$ invariant effective structures  in the superpotential from which the set of operators listed in Table 4 arise.  Thus the $\mathsf{B-L=-2}$ four field interaction arise from a primitive $\mathsf{SU(5)}$ structure $\mathsf{M}_i\mathsf{M}_j \mathsf{H_u}^i \mathsf{H_u}^j$ while the  $\mathsf{B-L=-2}$ five field interactions arise from the $\mathsf{SU(5)}$ structure $\mathsf{M}_i \mathsf{M}_j \mathsf{M}_k \mathsf{M}^{ij}\mathsf{H_u}^k$ and the $\mathsf{B-L=-2}$
six field interactions arise from the primitive $\mathsf{SU(5)}$ structure $\mathsf{M}_i\mathsf{M}_j\mathsf{M}^{ij}\mathsf{M}_k\mathsf{M}_{\ell} \mathsf{M}^{k\ell}$. Here
$\mathsf{M}_i$ are the $\mathsf{\bar 5}$-plet of matter and $\mathsf{M}^{ij}$ are the $\mathsf{10}$-plet of matter fields and $\mathsf{H_u}^i$ is the $\mathsf{5}$-plet of
light Higgs fields. The $\mathsf{B-L}=0$ four field operators in Table 4 can be traced back to the primitive $\mathsf{SU(5)}$ structure
$ \epsilon_{ijklm} \mathsf{M}^{ij} \mathsf{M}^{kl} \mathsf{M}^{mn} \mathsf{M}_n$ in the superpotential. It is to be noted that the $\mathsf{B-L=-2}$ four field and five
field primitive $\mathsf{SU(5)}$ structures only contain  up-Higgs, i.e., $\mathsf{H_u}$ and the down-Higgs $\mathsf{H_d}$ does not appear.
This helps explain why only $\mathsf{H_u}$ appears in effective operators in Table 4.
We note that $\mathsf{SU(5)}$ is, however, broken at the
GUT scale, and thus these $\mathsf{SU(5)}$ structures are to be used only as mnemonics for book keeping to identify the origins of
the effective operators at low energies. \\
}

{There are different scales associated with the  effective operators listed in Table 4. Thus suppose $M$ stands for
 the GUT scale and denote all GUT scale masses by $M$. In this case the $\mathsf{B-L=-2}$ four field effective operator
 is suppressed by the factor $1/M$ which leads to neutrino masses $O(<H>^2/M)$ which can lie in the desirable sub eV
 region. The $\mathsf{B-L=-2}$ five field operator $\mathbf{D}^{\mathsf c}\mathbf{D}^{\mathsf c}\mathbf{U}^{\mathsf c} \mathbf{L}_a {\mathbf{H_u}}^a$ is suppressed by two powers of $M$ in the
 superpotential. After dressing with loops involving a sparticle it will generate a dimension 7 operator in the effective
 Lagrangian suppressed by the factor $1/(M^2 M_s)$ where $M_s$ is the effective weak SUSY scale. This operator
 can produce  $\mathsf{B-L=-2}$ nucleon decay modes such $p\to \nu \pi^+$. In the usual GUT models, they would be suppressed
 relative to the proton decay arising from $\mathsf{B-L=0}$   dimension six proton decay operators. The six field operator
$\mathbf{D}^{\mathsf c}\mathbf{D}^{\mathsf c}\mathbf{U}^{\mathsf c}\mathbf{D}^{\mathsf c}\mathbf{D}^{\mathsf c}\mathbf{U}^{\mathsf c}$ is suppressed by $1/M^3$ in the superpotential and by a factor $1/(M^3 M_s^2)$ in
the effective Lagrangian. It can induce $n-\bar n$ oscillations and such effects can be enhanced and become
observable in models with low scales $M$. }

\section{Conclusion}

The $\mathsf{SO(10)}$ models are among the prime candidates for a unified framework, which include
the strong, the weak, and the electromagnetic interactions.  However, like most grand unified
models, the $\mathsf{SO(10)}$ models also suffer from the so called doublet-triplet problem, which
means that after spontaneous breaking of the GUT symmetry, the Higgs doublets (as well
as the Higgs triplets) will all be superheavy, requiring a huge fine-tuning to make the Higgs doublets
light.
The missing partners mechanism is one of the ways in which a grand unified model can resolve
this severe doublet-triplet problem. Recently, a variety of $\mathsf{SO(10)}$ models were proposed
in a supersymmetric framework, which {include} a missing partner mechanism. Here, we discussed
the simplest of such models which contains a heavy sector consisting of  $\mathsf{126+\overline{126} + 210}$ Higgs multiplets,
which {breaks} the GUT symmetry down to $\mathsf{SU(3)_C\times SU(2)L\times U(1)_Y}$. Combined with
a light sector consisting of $2\times \mathsf{10 +120}$ and a mixing between the light and the heavy sectors,
one finds that the model contains just a pair of light Higgs doublets while the remaining exotic fields
in $2\times \mathsf{10 +120}$ become superheavy. We have carried out a detailed analysis of the
Higgs sector and identified the exact  linear combination of the fields that enter in the light Higgs {doublet}
fields for {this} $\mathsf{SO(10)}$ missing partner model.\\

Further, in this work we  have given a full classification of  {$\mathsf{B-L=-2}$} {operators}
 that arise
from the {matter}-Higgs interactions. A list  of these operators is given in Table 4 which include dimension five, seven and nine operators
all of which are $\mathsf{B-L}$ violating.  The dimension 5 operator is the well -known Weinberg operator which gives mass
to the neutrinos, while the other operators can enter in GUT scale baryogenesis. Further, these operators can generate
new kinds of proton decay modes~\cite{Babu:2012vb}. Thus the conventional
{$\mathsf{B-L=0}$}
 baryon and lepton number violating operators
give rise to the modes $p\to e^+\pi^0, p\to  \bar \nu K^+$
while ${\mathsf{B-L}}=-2$ operators can generate proton decay modes
such as $p\to \nu \pi^+$, $n\to e^- \pi^+, e^- K^+$.
{Further the $\mathsf{B-L}=-2$ dimension nine operator $\mathbf{D}^{\mathsf c}
\mathbf{D}^{\mathsf c}\mathbf{U}^{\mathsf c}\mathbf{D}^{\mathsf c}\mathbf{D}^{\mathsf c}\mathbf{U}^{\mathsf c}$ can induce $n-\bar n$ oscillations.}
{Further, as discussed in Sec.(4), the decays of the heavy Higgs given by  Eq.(\ref{4.11})- Eq.(\ref{4.18}) which are $\mathsf{B-L}$ violating
and which contain new sources of CP violation can be used to generate GUT scale baryogenesis which cannot be washed out by
sphaleron interactions at the electroweak scale.}
In addition to the
{$\mathsf{B-L=-2}$}
operators discussed here there are
a large number of
{$\mathsf{B-L=-2}$}
operators that arise from four field Higgs interactions such as from $(\mathsf{126}\times \mathsf{126})_r\cdot(X\times Y)_r$ and $(\mathsf{126}\times \mathsf{\overline{126}})_r\cdot(X\times Y)_r$
where $X,Y=\mathsf{10},~\mathsf{45},~\mathsf{54},~\mathsf{120},~\mathsf{126}, ~\mathsf{\overline{126}}, ~\mathsf{210}$  of Higgs fields(many of these operators are discussed in~\cite{Babu:2012iv}).
{ A complete analysis of   {$\mathsf{B-L}$}
 violating interactions from this set is outside the scope of this
work and requires a separate analysis.} \\

\noindent
{\bf  Acknowledgements:}\\

\noindent
We thank K. S.  Babu and
Rabindra N Mohapatra for comments. {A communication from Goran Senjanovic is acknowledged.}
This research was supported in part by the NSF Grant PHY-1314774.\\

\section*{Appendix A:  Notation}
{
In this Appendix we display the decomposition of
$\mathsf{16}-$plet of matter and $\mathsf{10}-$,
$\mathsf{120}-$, $\overline{\mathsf{126}}-$ and $\mathsf{210}-$plets of Higgs of $\mathsf{SO(10)}$ in terms of  $\mathsf{SU(5)}$ representations. Thus we have }
\begin{eqnarray}
\mathsf{16}&=&\mathsf{1}({-5}) \left[\mathsf{M}_{\acute{x}}\right] + \overline {\mathsf{5}}(3) \left[\mathsf{M}_{\acute{x}i}\right] + \mathsf{10} (-1) \left[\mathsf{M}^{ij}_{\acute{x}}\right],\nonumber\\
\mathsf{10_r}&=&\mathsf{5}(2)\left[\mathsf{H}^{(10_r)i}\right] + \overline{\mathsf{5}}(-2)\left[\mathsf{H}^{(10_r)}_i\right],\nonumber\\
\mathsf{120}&=&\mathsf{5}(2)\left[\mathsf{H}^{(120)i}\right] + \overline{\mathsf{5}}(-2)\left[\mathsf{H}^{(120)}_i\right]+\mathsf{10}(-6)\left[\mathsf{H}^{(120)ij}\right] + \overline{\mathsf{10}}(6)\left[\mathsf{H}^{(120)}_{ij}\right] +\mathsf{45}(2)\left[\mathsf{H}^{(120)ij}_k\right]\nonumber\\
&&+\overline{\mathsf{45}}(-2)\left[\mathsf{H}^{(120)k}_{ij}\right],\nonumber\\
\overline{\mathsf{126}}&=&\mathsf{1}(10)\left[\mathsf{H}^{(\overline{126})}\right] + \mathsf{5}(2)\left[\mathsf{H}^{(\overline{126})i}\right]+ \overline{\mathsf{10}}(6)\left[\mathsf{H}^{(\overline{126})}_{ij}\right] +\mathsf{15}(-6)\left[\mathsf{H}^{(\overline{126})ij}_{(S)}\right]+\overline{\mathsf{45}}(-2)\left[\mathsf{H}^{(\overline{126})k}_{ij}\right]\nonumber\\
&&+ \mathsf{50}(2)\left[\mathsf{H}^{(\overline{126})ijk}_{lm}\right],\nonumber\\
\mathsf{210}&=&\mathsf{1}(0)\left[\mathsf{H}^{(210)}\right]+\mathsf{5}(-8)\left[\mathsf{H}^{(210)i}\right] + \overline{\mathsf{5}}(8)\left[\mathsf{H}^{(210)}_i\right]+\mathsf{10}(4)\left[\mathsf{H}^{(210)ij}\right] + \overline{\mathsf{10}}(-4)\left[\mathsf{H}^{(210)}_{ij}\right]\nonumber\\ &&+\mathsf{24}(0)\left[\mathsf{H}^{(120)i}_j\right]+\mathsf{40}(-4)\left[\mathsf{H}^{(210)ijk}_l\right]+\overline{\mathsf{40}}(4)\left[\mathsf{H}^{(210)l}_{ijk}\right]+\mathsf{75}(0)\left[\mathsf{H}^{(210)ij}_{kl}\right],
\end{eqnarray}
where $i,~j=1,...,5$ are $\mathsf{SU(5)}$ indices, $\acute{u},~\acute{v},~\acute{w},~\acute{x},~\acute{y},~\acute{z}=1,~2,~3$ represent generation indices and $r,~s=1,~2$ {count}
 the number of $\mathsf{10}$ plet of $\mathsf{SO(10)}$ used in our model of {the}
missing partner mechanism.
The following are the Standard Model particle assignments
\begin{eqnarray}
{M}_{\acute{x}}={\bm\nu}_{\acute{x}}^{\mathtt c};~~~~~~
\mathsf{M}_{\acute{x}\alpha}={\bf D}_{\acute{x}\alpha}^{\mathtt c};~~~~~~\mathsf{M}_{\acute{x}a}={{\mathbf E}_{\acute{x}}\choose -{\bm\nu}_{\acute{x}}}=\mathbf{L}_{\acute{x}a};\nonumber\\
\mathsf{M}_{\acute{x}}^{a\alpha}={{\mathbf U}^{\alpha}_{\acute{x}}\choose {\mathsf D}^{\alpha}_{\acute{x}}}=\mathbf{Q}^{a\alpha}_{\acute{x}};~~~~~~\mathsf{M}_{\acute{x}}^{\alpha\beta}=\epsilon^{\alpha\beta\gamma}{\mathbf
U}_{\acute{x}\gamma}^{\mathtt c};~~~~~~{\mathsf M}_{\acute{x}}^{ab}=\epsilon^{ab}{\mathbf E}_{\acute{x}}^{\mathtt c},
\end{eqnarray}
where $\alpha,~ \beta,~ \gamma=1,~2,~3$ are $\mathsf{SU(3)}$ color indices, while $a,~b=1,~2$ are $\mathsf{SU(2)}$ weak indices and the
superscript $^{\mathtt c}$ denotes charge conjugation.
\section* {Appendix B:  Decomposition of {\boldmath$\mathsf{SU(5)}$} fields  in terms of Standard Model states\label{AppE}}
In this appendix we give a decomposition of  the $\mathsf{24}$, $\mathsf{45}$, $\mathsf{50}$ and $\mathsf{75}$-plets of $\mathsf{SU(5)}$ in terms of $\mathsf{SU(3)_C\times SU(2)_L \times U(1)_Y}$ fields. These fields are needed in the spontaneous breaking of GUT  and electroweak symmetry.
\subsection* {B1:  Decomposition of  {\boldmath$\mathsf{24}$}-plet of {\boldmath$\mathsf{SU(5)}$} \label{AppE}}
  The $\mathsf{24}$-plet of $\mathsf{SU(5)}$, residing in $\mathsf{210}$-plet of $\mathsf{SO(10)}$, has the following  $\mathsf{SU(3)_C\times SU(2)_L \times U(1)_Y}$ decomposition
    \begin{eqnarray}\label{A0}
\mathsf{H}^{(210)i}_j(\mathsf{24})=(\mathsf{1,1},0)\mathbf S_{24_{_{210}}}+(\mathsf{1,3},0){\mathbf U}_{b}^{a}+(\mathsf{8,1},0){\mathbf U}_{\beta}^{\alpha}+[(\mathsf{3,2},-5){\mathbf U}^{\alpha}_a + c.c.],
\end{eqnarray}
where we have defined
\begin{equation}\label{A1}
\mathsf{H}^{(210)\alpha}_{\alpha}=-\mathsf{H}^{(210)a}_{a}
\equiv \mathbf S_{24_{_{210}}}.
\end{equation}
The relationship above follows from the tracelessness condition on the tensor $\mathsf{H}^{(210)i}_{j}$.
The reducible tensors of the  $\mathsf{24}$-plet can be expressed in terms of the irreducible ones as follows:
\begin{eqnarray}\label{A2}
\mathsf{H}^{(210)a}_{b}={\mathbf U}_{b}^{a}-\frac{1}{2}\delta^a_b\mathbf S_{24_{_{210}}};~~~&&~~~\mathsf{H}^{(210)\alpha}_{\beta}=
{\mathbf U}_{\beta}^{\alpha}+\frac{1}{3}\delta^{\alpha}_{\beta}\mathbf S_{24_{_{210}}}.
\end{eqnarray}
The kinetic energy of the  $\mathsf{24}$-plet is given by
\begin{eqnarray}\label{A3}
-\partial_A\mathsf{H}^{(210)i}_{j}\partial^A\mathsf{H}^{(210)i\dagger}_{j}&=&-\left[\partial_A\mathbf S_{24_{_{210}}}\partial^A\mathbf S_{24_{_{210}}}^{\dagger}+\partial_A{\mathbf U}_{\beta}^{\alpha}
\partial^A{\mathbf U}_{\beta}^{\alpha\dagger}+\partial_A{\mathbf U}^{a}_{b}
\partial^A{\mathbf U}^{a\dagger}_{b}+\partial_A{\mathbf U}^{\alpha}_{a}
\partial^A{\mathbf U}^{\alpha\dagger}_{a}\right.\nonumber\\
&&\left.~~~~+\partial_A{\mathbf U}^{a}_{\alpha}\partial^A{\mathbf U}_{\alpha}^{a\dagger}\right],
\end{eqnarray}
 so that the SM fields are normalized according to
 \begin{eqnarray}\label{A5}
 \mathbf S_{24_{_{210}}}\rightarrow\sqrt{\frac{6}{5}}\mathbf S_{24_{_{210}}};~~~~~~{\mathbf U}_{\beta}^{\alpha}\rightarrow{\mathbf U}_{\beta}^{\alpha};~~~~~~
{\mathbf U}_{b}^{a}\rightarrow{\mathbf U}^{a}_{b};~~~~~~
 {\mathbf U}^{\alpha}_{a}\rightarrow {\mathbf U}^{\alpha}_{a};~~~~~~{\mathbf U}^{a}_{\alpha}\rightarrow{\mathbf U}^{a}_{\alpha}.
 \end{eqnarray}

\subsection* {B2:  Decomposition of {\boldmath$\mathsf{45}$}-plet of {\boldmath$\mathsf{SU(5)}$}\label{AppF}}
  The $\mathsf{45}$-plet of $\mathsf{SU(5)}$, residing in  $\mathsf{120}$ and  $\mathsf{126}$-plets of $\mathsf{SO(10)}$, has the following $\mathsf{SU(3)_C\times SU(2)_L \times U(1)_Y}$ decomposition
  \begin{eqnarray}\label{B0}
  {\mathsf{H}^{(120)ij}_{k}\choose\mathsf{H}^{(126)ij}_{k}}(\mathsf{45})&=&(\mathsf{1,2},3){{}^{({45}_{120})}\!{\mathsf D}^{a}\choose {}^{({45}_{126})}\!{\mathsf D}^{a}}+(\mathsf{3,1},-2){{}^{({45}_{120})}\!{\mathsf T}^{\alpha}\choose {}^{({45}_{126})}\!{\mathsf T}^{\alpha}}+(\mathsf{3,3},-2){\mathbf V}_{b}^{a\alpha}+(\mathsf{\overline{3},1},8){\mathbf V}_{\alpha} +(\mathsf{\overline{3},2},-7){\mathbf V}_{a\alpha}\nonumber\\
&&+~(\mathsf{\overline{6},1},-2){\mathbf V}^{\alpha\beta}_{\gamma}+(\mathsf{{8},2},3){\mathbf V}^{\alpha a}_{\beta},
\end{eqnarray}
where we have defined
\begin{equation}\label{B1}
{\mathsf{H}^{(120)\beta a}_{\beta}\choose \mathsf{H}^{(126)\beta a}_{\beta}}=-{\mathsf{H}^{(120)b a}_{b}\choose \mathsf{H}^{(120)b a}_{b}}
\equiv {{}^{({45}_{120})}\!{\mathsf D}^{a}\choose {}^{({45}_{126})}\!{\mathsf D}^{a}};~~~~~~~~~~~~~~~~~~~~{\mathsf{H}^{(120)\beta \alpha}_{\beta}\choose\mathsf{H}^{(126)\beta \alpha}_{\beta}}=-{\mathsf{H}^{(120)b \alpha}_{b}\choose\mathsf{H}^{(126)b \alpha}_{b}}
\equiv {{}^{({45}_{120})}\!{\mathsf T}^{\alpha}\choose {}^{({45}_{126})}\!{\mathsf T}^{\alpha}}.
\end{equation}
The relationship above follows from the tracelessness condition on the tensor $\mathsf{H}^{(120)ij}_{k}$ and $\mathsf{H}^{(126)ij}_{k}$.
The reducible tensors of the $\mathsf{45}$-plet can be expressed in terms of the irreducible ones as follows:
\begin{eqnarray}\label{B2}
{\mathsf{H}^{(120)a\alpha}_{b}\choose\mathsf{H}^{(126)a\alpha}_{b}}={\mathbf V}_{b}^{a\alpha}-\frac{1}{2}\delta^a_b{{}^{({45}_{120})}\!{\mathsf T}^{\alpha}\choose {}^{({45}_{126})}\!{\mathsf T}^{\alpha}};~~~~~~~~~~~~~~~~~~~~~~~~~~~~~~~~~~~~~~~~~~{\mathsf{H}^{(120)\alpha a}_{\beta}\choose\mathsf{H}^{(126)\alpha a}_{\beta}}=
{\mathbf V}_{\beta}^{\alpha a}+\frac{1}{3}\delta^{\alpha}_{\beta}{{}^{({45}_{120})}\!{\mathsf D}^{a}\choose {}^{({45}_{126})}\!{\mathsf D}^{a}};\nonumber\\
{\mathsf{H}^{(120)ab}_{\alpha}\choose\mathsf{H}^{(126)ab}_{\alpha}}=\epsilon^{ab}{\mathbf V}_{\alpha};~~~~~~~{\mathsf{H}^{(120)\alpha\beta}_{a}\choose\mathsf{H}^{(126)\alpha\beta}_{a}}=
\epsilon^{\alpha\beta\gamma}{\mathbf V}_{a\gamma};~~~~~~~{\mathsf{H}^{(120)ab}_{c}\choose\mathsf{H}^{(126)ab}_{c}}=\delta^{b}_{c}{{}^{({45}_{120})}\!{\mathsf D}^{a}\choose {}^{({45}_{126})}\!{\mathsf D}^{a}}-    \delta^{a}_{c}{{}^{({45}_{120})}\!{\mathsf D}^{b}\choose {}^{({45}_{126})}\!{\mathsf D}^{b}}\nonumber\\
{\mathsf{H}^{(120)\alpha\beta}_{\gamma}\choose\mathsf{H}^{(126)\alpha\beta}_{\gamma}}={\mathbf V}_{\gamma}^{\alpha\beta}+\frac{1}{2}\left[\delta^{\alpha}_{\gamma}{{}^{({45}_{120})}\!{\mathsf T}^{\beta}\choose {}^{({45}_{126})}\!{\mathsf T}^{\beta}}-    \delta^{\beta}_{\gamma}{{}^{({45}_{120})}\!{\mathsf T}^{\alpha}\choose {}^{({45}_{126})}\!{\mathsf T}^{\alpha}}\right].~~~~~~~~~~~~~~~~~~~~~~~~~~~~~~~
\end{eqnarray}
The kinetic energy of the $\mathsf{45}$-plet is given by
\begin{eqnarray}\label{B3}
-\partial_A{\mathsf{H}^{(120)ij}_{k}\choose\mathsf{H}^{(126)ij}_{k}}\partial^A{\mathsf{H}^{(120)ij\dagger}_{k}\choose\mathsf{H}^{(126)ij\dagger}_{k}}&=&
-\left[\partial_A{{}^{({45}_{120})}\!{\mathsf D}^{a}\choose {}^{({45}_{126})}\!{\mathsf D}^{a}}\partial^A{{}^{({45}_{120})}\!{\mathsf D}^{a\dagger}\choose {}^{({45}_{126})}\!{\mathsf D}^{a\dagger}}+\partial_A{{}^{({45}_{120})}\!{\mathsf T}^{\alpha}\choose {}^{({45}_{126})}\!{\mathsf T}^{\alpha}}\partial^A{{}^{({45}_{120})}\!{\mathsf T}^{\alpha\dagger}\choose {}^{({45}_{126})}\!{\mathsf T}^{\alpha\dagger}}\right.\nonumber\\
&&\left.~~~~+\partial_A{\mathbf V}_{b}^{a\alpha}\partial^A{\mathbf V}_{b}^{a\alpha\dagger}+\partial_A{\mathbf V}_{\alpha}\partial^A{\mathbf V}_{\alpha}^{\dagger} +\partial_A{\mathbf V}_{a\alpha}\partial^A{\mathbf V}_{a\alpha}^{\dagger}\right.\nonumber\\
&&\left.~~~~+\partial_A{\mathbf V}^{\alpha\beta}_{\gamma}\partial^A{\mathbf V}^{\alpha\beta\dagger}_{\gamma}+\partial_A{\mathbf V}^{\alpha a}_{\beta}\partial^A{\mathbf V}^{\alpha a\dagger}_{\beta}\right],
\end{eqnarray}
 so that the SM fields are normalized according to
 \begin{eqnarray}\label{B5}
 {{}^{({45}_{120})}\!{\mathsf D}^{a}\choose {}^{({45}_{126})}\!{\mathsf D}^{a}}\rightarrow\frac{1}{2}\sqrt{\frac{3}{2}}{{}^{({45}_{120})}\!{\mathsf D}^{a}\choose {}^{({45}_{126})}\!{\mathsf D}^{a}};~~~~~~~~~~~~~~~~~~~~~~~~~~~~~{{}^{({45}_{120})}\!{\mathsf T}^{\alpha}\choose {}^{({45}_{126})}\!{\mathsf T}^{\alpha}}\rightarrow\frac{1}{\sqrt{2}}{{}^{({45}_{120})}\!{\mathsf T}^{\alpha}\choose {}^{({45}_{126})}\!{\mathsf T}^{\alpha}};\nonumber\\
 {\mathbf V}_{\alpha}\rightarrow\frac{1}{\sqrt{2}} {\mathbf V}_{\alpha};~~~~{\mathbf V}_{a\alpha}\rightarrow\frac{1}{\sqrt{6}}{\mathbf V}_{a\alpha};~~~
 {\mathbf V}_{\beta}^{\alpha a}\rightarrow\frac{1}{\sqrt{2}}{\mathbf V}_{\beta}^{\alpha a};~~~
 {\mathbf V}_{b}^{a\alpha}\rightarrow\frac{1}{\sqrt{2}}{\mathbf V}_{b}^{a\alpha};~~~
 {\mathbf V}_{\gamma}^{\alpha\beta}\rightarrow\frac{1}{\sqrt{2}}{\mathbf V}_{\gamma}^{\alpha\beta}.
 \end{eqnarray}
 One can now extend the above results to $\mathsf{\overline{45}}$ of $\mathsf{SU(5)}$ contained in $\mathsf{120}$ and $\mathsf{\overline{126}}$ plets.
\subsection* {B3: Decomposition of {\boldmath$\mathsf{50}$}-plet of {\boldmath$\mathsf{SU(5)}$} \label{AppG}}
The $\mathsf{50}$-plet of $\mathsf{SU(5)}$, residing in $\overline{\mathsf{126}}$-plet of $\mathsf{SO(10)}$, has the following  $\mathsf{SU(3)_C\times SU(2)_L \times U(1)_Y}$ decomposition
 \begin{eqnarray}\label{C0}
 \mathsf{H}^{(\overline{126})ijk}_{lm}(\mathsf{50})&=&(\mathsf{1,1},-12){\mathbf W}+(\mathsf{3,1},-2){}^{({50}_{\overline{126}})}\!{\mathsf T}^{\alpha}+(\mathsf{\bar{3},2},-7){\mathbf W}^{\alpha\beta}_a
+(\mathsf{\bar{6},3},-2){\mathbf W}^{\alpha\beta a}_{\gamma b}+
(\mathsf{6,1},8){\mathbf W}^{\alpha}_{\beta\gamma}\nonumber\\
&&+~(\mathsf{8,2},3){\mathbf W}^{\alpha a}_{\beta},
 \end{eqnarray}
where we have defined
\begin{equation}\label{C1}
 \mathsf{H}^{(\overline{126})ab \gamma}_{ab}=\mathsf{H}^{(\overline{126})\alpha\beta \gamma}_{\alpha\beta}
=-\mathsf{H}^{(\overline{126})a\alpha \gamma}_{a\alpha}\equiv{}^{({50}_{\overline{126}})}\!{\mathsf T}^{\alpha};~~~~~~~~~\mathsf{H}^{(\overline{126})\gamma \alpha\beta}_{\gamma a}\equiv{\mathbf W}^{\alpha\beta}_{a};~~~~~~~~~\overline{\Delta}^{\gamma\alpha \beta}_{\gamma a}\equiv{\mathbf W}^{\alpha a}_{\beta}.
\end{equation}
Again the first relationship follows from the tracelessness condition on the $\mathsf{SU(5)}$ irreducible tensor $\mathsf{H}^{(\overline{126})ijk}_{lm}$:
\begin{equation}\label{C2}
\mathsf{H}^{(\overline{126})a\alpha i}_{a\alpha}=-\frac{1}{2}\left[\mathsf{H}^{(\overline{126})\alpha\beta i}_{\alpha\beta}+\mathsf{H}^{(\overline{126})abi}_{ab}\right].
\end{equation}
The reducible tensors of the $\mathsf{50}$-plet can be expressed in terms of the irreducible ones as follows:
\begin{eqnarray}\label{C3}
\mathsf{H}^{(\overline{126})\alpha\beta\gamma}_{\rho\sigma}=\frac{1}{2}\left[\delta^{\alpha}_{\rho}\delta^{\beta}_{\sigma}{}^{({50}_{\overline{126}})}\!{\mathsf T}^{\gamma}-
\delta^{\beta}_{\rho}\delta^{\alpha}_{\sigma}{}^{({50}_{\overline{126}})}\!{\mathsf T}^{\gamma}-
\delta^{\alpha}_{\rho}\delta^{\gamma}_{\sigma}{}^{({50}_{\overline{126}})}\!{\mathsf T}^{\beta}+
\delta^{\gamma}_{\rho}\delta^{\alpha}_{\sigma}{}^{({50}_{\overline{126}})}\!{\mathsf T}^{\beta}+
\delta^{\beta}_{\rho}\delta^{\gamma}_{\sigma}{}^{({50}_{\overline{126}})}\!{\mathsf T}^{\alpha}-
\delta^{\gamma}_{\rho}\delta^{\beta}_{\sigma}{}^{({50}_{\overline{126}})}\!{\mathsf T}^{\alpha}
\right];\nonumber\\
\mathsf{H}^{(\overline{126})\alpha\beta a}_{\gamma\sigma}=\delta^{\alpha}_{\gamma}{\mathbf W}^{\beta a}_{\sigma}-
\delta^{\alpha}_{\sigma}{\mathbf W}^{\beta a}_{\gamma}+\delta^{\beta}_{\sigma}{\mathbf W}^{\alpha a}_{\gamma}
-\delta^{\beta}_{\gamma}{\mathbf W}^{\alpha a}_{\sigma};~~~~~~~~~~\mathsf{H}^{(\overline{126})\alpha\beta\gamma}_{ab}=\epsilon^{\alpha\beta\gamma}\epsilon_{ab}{\mathbf W};~~~~~~~~~~~~~~~~~~~~\nonumber\\
\mathsf{H}^{(\overline{126})\alpha\beta\gamma}_{\sigma a}=\delta^{\gamma}_{\sigma}{\mathbf W}^{\alpha\beta}_{a}-
\delta^{\beta}_{\sigma}{\mathbf W}^{\alpha\gamma}_{a}+\delta^{\alpha}_{\sigma}{\mathbf W}^{\beta \gamma}_{a}
;~~~~~~~~~~\mathsf{H}^{(\overline{126})\alpha ab}_{\beta\gamma}=\epsilon^{ab}{\mathbf W}^{\alpha}_{\beta\gamma};~~~~~~~~~~~~~~~~~~~~\nonumber\\
\mathsf{H}^{(\overline{126})\alpha\beta a}_{\gamma b}={\mathbf W}^{\alpha\beta a}_{\gamma b}+\frac{1}{4}\delta_b^a\left[
\delta^{\alpha}_{\gamma}{}^{({50}_{\overline{126}})}\!{\mathsf T}^{\beta}-\delta^{\beta}_{\gamma}{}^{({50}_{\overline{126}})}\!{\mathsf T}^{\alpha}\right];~~~~~~~~~~
\mathsf{H}^{(\overline{126})ab \alpha}_{c\beta}=\delta^{a}_{c}{\mathbf W}^{\alpha b}_{\beta}-\delta^{b}_{c}{\mathbf W}^{\alpha a}_{\beta};~~~~~~~~~~~~~~~~~~~~\nonumber\\
\mathsf{H}^{(\overline{126})ab \alpha}_{cd}=\frac{1}{2}\left[\delta_c^a\delta_d^b-\delta_d^a\delta_c^b\right]{}^{({50}_{\overline{126}})}\!{\mathsf T}^{\alpha}
;~~~~~~~~~~\mathsf{H}^{(\overline{126})\alpha\beta a}_{bc}=\delta^{a}_{c}{\mathbf W}^{\alpha \beta}_b-\delta^{a}_{b}{\mathbf W}^{\alpha\beta}_{c};~~~~~~~~~~~~~
\end{eqnarray}
The kinetic energy of the $\mathsf{50}$-plet is given by
\begin{eqnarray}\label{C4}
-\partial_A\mathsf{H}^{(\overline{126})ijk}_{lm}\partial^A\mathsf{H}^{(\overline{126})ijk\dagger}_{lm}&=&
-\left[\partial_A{\mathbf W}\partial^A{\mathbf W}^{\dagger}+\partial_A{}^{({50}_{\overline{126}})}\!{\mathsf T}^{\alpha}
{}^{({50}_{\overline{126}})}\!{\mathsf T}^{\alpha}+\frac{1}{2!}\partial_A{\mathbf W}^{\alpha\beta}_{a}
\partial^A{\mathbf W}^{\alpha\beta\dagger}_{a}+\frac{1}{2!}\partial_A{\mathbf W}^{\alpha a}_{\beta}
\partial^A{\mathbf W}^{\alpha a\dagger}_{\beta}\right.\nonumber\\
&&\left.~~~~+\frac{1}{3!}\frac{1}{2!}\partial_A{\mathbf W}^{\alpha\beta a}_{\gamma b}
\partial^A{\mathbf W}^{\alpha\beta a\dagger}_{\gamma b}+\frac{1}{2!}\partial_A{\mathbf W}^{\alpha}_{\beta\gamma}
\partial^A{\mathbf W}^{\alpha\dagger}_{\beta\gamma}\right],
\end{eqnarray}
 so that the SM fields are normalized according to
 \begin{eqnarray}\label{C5}
 {\mathbf W}\rightarrow\frac{1}{2\sqrt{3}}{\mathbf W};~~~~~~~~~&&~~~{}^{({50}_{\overline{126}})}\!{\mathsf T}^{\alpha}
 \rightarrow\frac{1}{{3}}{}^{({50}_{\overline{126}})}\!{\mathsf T}^{\alpha};~~~~~~~~~~~~~~~~~~
{\mathbf W}^{\alpha\beta}_{a}\rightarrow\frac{1}{2\sqrt{6}}{\mathbf W}^{\alpha\beta}_{a};\nonumber\\
 {\mathbf W}^{\alpha a}_{\beta}\rightarrow\frac{1}{4\sqrt{3}}{\mathbf W}^{\alpha a}_{\beta};~~~&&~~~ {\mathbf W}^{\alpha\beta a}_{\gamma b}\rightarrow\frac{1}{6\sqrt{2}}{\mathbf W}^{\alpha\beta a}_{\gamma b};~~~~~~~~~ {\mathbf W}^{\alpha}_{\beta\gamma}\rightarrow\frac{1}{2
 \sqrt{3}}{\mathbf W}^{\alpha}_{\beta\gamma}.
 \end{eqnarray}
 One can now extend the above results to $\mathsf{\overline{50}}$ of $\mathsf{SU(5)}$ contained in ${\mathsf{126}}$ plet.\\
\subsection* {B4:  Decomposition of {\boldmath$\mathsf{75}$}-plet of {\boldmath$\mathsf{SU(5)}$}\label{AppH}}
  The $\mathsf{75}$-plet of $\mathsf{SU(5)}$, residing in $\mathsf{210}$-plet of $\mathsf{SO(10)}$, has the following  $\mathsf{SU(3)_C\times SU(2)_L \times U(1)_Y}$ decomposition
  \begin{eqnarray}\label{D0}
\mathsf{H}^{(210)ij}_{kl}(\mathsf{75})&=&(\mathsf{1,1},0){\mathbf S_{75_{_{210}}}}+(\mathsf{8,1},0){\mathbf X}_{\beta}^{\alpha}+(\mathsf{8,3},0){\mathbf X}_{\beta b}^{\alpha a}+[(\mathsf{3,2},-5){\mathbf X}^{\alpha}_a+(\mathsf{\bar{6},2},-5){\mathbf X}_{\gamma a}^{\alpha\beta}\nonumber\\
&&+~(\mathsf{\bar{3},1},-10){\mathbf X}_{\alpha}+c.c.],
  \end{eqnarray}
where we have defined
\begin{equation}\label{D1}
\mathsf{H}^{(210)ab}_{ab}=\mathsf{H}^{(210)\alpha\beta}_{\alpha\beta}
=-\mathsf{H}^{(210)\alpha a}_{\alpha a}\equiv \mathbf S_{75_{_{210}}};~~~~~~~~~\mathsf{H}^{(210)ab}_{\alpha b}\equiv {\mathbf X}_{a}^{\alpha} ;~~~~~~~~~{\mathbf X}_{\beta}^{\alpha}\equiv \mathsf{H}^{(210)\alpha a}_{\beta a}+\frac{1}{3}\delta^{\alpha}_{\beta}\mathbf S_{75_{_{210}}}.
\end{equation}
The first relationship above follows from the double tracelessness condition on the tensor $\mathsf{H}^{(210)ij}_{kl}$:
\begin{equation}\label{D2}
\mathsf{H}^{(210)\alpha a}_{\alpha a}=-\frac{1}{2}\left(\mathsf{H}^{(210)\alpha\beta}_{\alpha\beta}+\mathsf{H}^{(210)ab}_{ab}\right).
\end{equation}
The reducible tensors of the $\mathsf{75}$-plet can be expressed in terms of the irreducible ones as follows:
\begin{eqnarray}\label{D3}
\mathsf{H}^{(210)ab}_{cd}=\frac{1}{2}\left(\delta^a_c\delta^b_d-\delta^a_d\delta^b_c\right)\mathbf S_{75_{_{210}}};~~~&&~~~\mathsf{H}^{(210)\alpha \beta}_{\gamma \sigma}=
\frac{1}{2}\left(\delta^{\alpha}_{\sigma}{\mathbf X}_{\gamma}^{\beta}-\delta^{\alpha}_{\gamma}{\mathbf X}_{\sigma}^{\beta}\right)
+\frac{1}{6}\left(\delta^{\alpha}_{\gamma}\delta^{\beta}_{\sigma}-\delta^{\alpha}_{\sigma}\delta^{\beta}_{\gamma}\right)\mathbf S_{75_{_{210}}};\nonumber\\
\mathsf{H}^{(210)\alpha a}_{\beta b}={\mathbf X}_{\beta b}^{\alpha a}+\frac{1}{2}\delta^a_b{\mathbf X}_{\beta }^{\alpha}-\frac{1}{6}\delta^a_b\delta^{\alpha}_{\beta}\mathbf S_{75_{_{210}}};~~~&&~~~\mathsf{H}^{(210)\alpha\beta}_{\gamma a}={\mathbf X}_{\gamma a}^{\alpha \beta}-\frac{1}{2}\left(\delta^{\alpha}_{\gamma}{\mathbf X}^{\beta}_a-\delta^{\beta}_{\gamma}{\mathbf X}^{\alpha}_a\right);\nonumber\\
\mathsf{H}^{(210)\alpha \beta}_{ab}=\epsilon_{ab}\epsilon^{\alpha\beta\gamma}{\mathbf X}_{\gamma};~~~&&~~~\mathsf{H}^{(210)a\alpha}_{bc}=\delta^a_b{\mathbf X}^{\alpha}_c-
\delta^a_c{\mathbf X}^{\alpha}_b.
\end{eqnarray}
The kinetic energy of the $\mathsf{75}$-plet is given by
\begin{eqnarray}\label{D4}
-\partial_A\mathsf{H}^{(210)ij}_{kl}\partial^A\mathsf{H}^{(210)ij\dagger}_{kl}&=&-\left[\partial_A\mathbf S_{75_{_{210}}}\partial^A\mathbf S_{75_{_{210}}}^{\dagger}+\partial_A{\mathbf X}_{\alpha}
\partial^A{\mathbf X}_{\alpha}^{\dagger}+\partial_A{\mathbf X}^{\alpha}
\partial^A{\mathbf X}^{\alpha\dagger}+\partial_A{\mathbf X}^{\alpha}_{\beta}
\partial^A{\mathbf X}^{\alpha\dagger}_{\beta}\right.\nonumber\\
&&\left.~~~~+\partial_A{\mathbf X}^{\alpha}_{a}
\partial^A{\mathbf X}^{\alpha\dagger}_{a}+\partial_A{\mathbf X}_{\alpha}^{a}
\partial^A{\mathbf X}^{a\dagger}_{\alpha}+
+\frac{1}{2!}\frac{1}{2!}\partial_A{\mathbf X}^{\alpha\beta}_{\gamma a}
\partial^A{\mathbf X}^{\alpha\beta\dagger}_{\gamma a}\right.\nonumber\\
&&\left.~~~~+\frac{1}{2!}\frac{1}{2!}\partial_A{\mathbf X}_{\alpha\beta}^{\gamma a}
\partial^A{\mathbf X}_{\alpha\beta}^{\gamma a\dagger}+\frac{1}{2!}\frac{1}{2!}\partial_A{\mathbf X}^{\alpha a}_{\beta b}
\partial^A{\mathbf X}^{\alpha a\dagger}_{\beta b}\right],
\end{eqnarray}
 so that the SM fields are normalized according to
 \begin{eqnarray}\label{D5}
 \mathbf S_{75_{_{210}}}\rightarrow\frac{1}{\sqrt{2}}\mathbf S_{75_{_{210}}};~~~~~~&&~~~{\mathbf X}_{\alpha}\rightarrow\frac{1}{{2}}{\mathbf X}_{\alpha};~~~~~~~~~~~~~~~~~{\mathbf X}^{\alpha}\rightarrow\frac{1}{{2}}{\mathbf X}^{\alpha};~~~~~~~~~~~~~~~~~{\mathbf X}^{\alpha}_{\beta}\rightarrow\frac{1}{\sqrt{3}}{\mathbf X}^{\alpha}_{\beta};\nonumber\\
 {\mathbf X}^{\alpha}_{a}\rightarrow\frac{1}{\sqrt{6}}{\mathbf X}^{\alpha}_{a};~~~~~~&&~~~{\mathbf X}_{\alpha}^{a}\rightarrow\frac{1}{\sqrt{6}}{\mathbf X}_{\alpha}^{a};~~~~~~~~~~~~~~~{\mathbf X}^{\alpha\beta}_{\gamma a}\rightarrow\frac{1}{2\sqrt{2}}{\mathbf X}^{\alpha\beta}_{\gamma a};~~~~~~~~~~{\mathbf X}_{\alpha\beta}^{\gamma a}\rightarrow\frac{1}{2\sqrt{2}}{\mathbf X}_{\alpha\beta}^{\gamma a};\nonumber\\
 ~~~~~~~&&~~~~~~~~~~~~~~~~~~~{\mathbf X}^{\alpha a}_{\beta b}\rightarrow\frac{1}{{4}}{\mathbf X}^{\alpha a}_{\beta b}.
 \end{eqnarray}

\section* {Appendix C:  Details of GUT symmetry breaking}\label{AppA}
Here, we give further details of the GUT symmetry breaking discussed in Sec.(\ref{sec2}).  As mentioned in
Sec.(\ref{sec2}) the fields that enter in the GUT symmetry breaking are $\mathsf{126+\ov{126}+210}$. A decomposition of
the $\mathsf{SO(10)}$ invariant interaction for these fields: $\mathsf{210\cdot210}$, $\mathsf{126\cdot\overline{126}}$, $\mathsf{210\cdot210\cdot 210}$ and $\mathsf{210\cdot126\cdot\overline{126}}$ in $\mathsf{SU(5)}$ fragments is exhibited below.
\begin{eqnarray}
  W_{{GUT}}&=& {m_{\Phi}}\Phi_{\mu\nu\sigma\xi}\Phi_{\mu\nu\sigma\xi}+{m_{\Delta}}\Delta_{\mu\nu\sigma\xi\zeta}\overline{\Delta}_{\mu\nu\sigma\xi\zeta}
  +{\lambda}
  \Phi_{\mu\nu\sigma\xi}\Phi_{\sigma\xi\rho\tau}\Phi_{\rho\tau\mu\nu}+ {\eta}\Phi_{\mu\nu\sigma\xi}\Delta_{\mu\nu\rho\tau\zeta}\overline{\Delta}_{\sigma\xi\rho\tau\zeta}\nonumber\\
  &=&{m_{\Phi}}\frac{1}{2^4}\left[6\Phi_{c_ic_j\bar{c}_k\bar{c}_l}\Phi_{c_kc_l\bar{c}_i\bar{c}_j}+\cdots\right]+
  {m_{\Delta}}\frac{1}{2^5}\left[\Delta_{c_ic_j{c}_k{c}_lc_m}\overline{\Delta}_{\bar{c}_i\bar{c}_j\bar{c}_k\bar{c}_l\bar{c}_m}+\cdots\right]\nonumber\\
 &&+ \lambda\frac{1}{2^6}\left[2\Phi_{c_ic_j\bar{c}_k\bar{c}_l}\Phi_{c_kc_l\bar{c}_m\bar{c}_n}\Phi_{c_mc_n\bar{c}_i\bar{c}_j}
 +8\Phi_{c_ic_j\bar{c}_k\bar{c}_l}\Phi_{c_mc_k\bar{c}_n\bar{c}_i}\Phi_{c_lc_n\bar{c}_j\bar{c}_m}+\cdots\right]\nonumber\\
 &&+\eta\frac{1}{2^7}\left[\Phi_{c_ic_j\bar{c}_k\bar{c}_l}\Delta_{c_kc_l{c}_m{c}_nc_p}\overline{\Delta}_{\bar{c}_i\bar{c}_j\bar{c}_m\bar{c}_n\bar{c}_p}
 +\cdots\right]\nonumber\\
 &=&{m_{\Phi}}\left[\frac{3}{8}\mathsf{H}^{(210)ij}_{kl}~{\mathsf{H}^{(210)kl}_{ij}}+\frac{1}{2}\mathsf{H}^{(210)i}_{j}~\mathsf{H}^{(210)j}_{i}+\frac{3}{80}\mathbf S_{1_{_{{210}}}}                   \mathbf S_{1_{_{{210}}}}                  +\cdots\right]+
 {m_{\Delta}}\left[\frac{15}{2}\mathbf S_{1_{_{{126}}}}                   \mathbf S_{1_{_{\overline{126}}}}     +\cdots\right]\nonumber\\
 &&+\lambda\left[\frac{1}{32}\mathsf{H}^{(210)ij}_{kl}~\mathsf{H}^{(210)kl}_{mn}~\mathsf{H}^{(210)mn}_{ij}+\frac{1}{8}{\mathsf{H}^{(210)ij}_{kl}}~\mathsf{H}^{(210)ln}_{jm}~
 \mathsf{H}^{(210)mk}_{ni}
 -\frac{1}{8}\mathsf{H}^{(210)ij}_{kl}~\mathsf{H}^{(210)kl}_{jm}~\mathsf{H}^{(210)m}_{i}\right.\nonumber\\
 &&\left.~~~~~~+\frac{1}{24}\mathsf{H}^{(210)ij}_{kl}~\mathsf{H}^{(210)k}_{i}~\mathsf{H}^{(210)l}_{j}
 +\frac{1}{80}\mathsf{H}^{(210)ij}_{kl}~\mathsf{H}^{(210)kl}_{ij}~\mathbf S_{1_{_{{210}}}}                  +\frac{7}{108}\mathsf{H}^{(210)i}_{j}~\mathsf{H}^{(210)j}_{k}~\mathsf{H}^{(210)k}_{i}\right.\nonumber\\
 &&\left.~~~~~~
 -\frac{1}{160}\mathsf{H}^{(210)i}_{j}~\mathsf{H}^{(210)j}_{i}~\mathbf S_{1_{_{{210}}}}                  -\frac{3}{3200}\mathbf S_{1_{_{{210}}}}                  ^3+\cdots\right]\nonumber\\
&&+\eta\left[-\frac{3}{16}\mathbf S_{1_{_{{126}}}}                   \mathbf S_{1_{_{\overline{126}}}}     \mathbf S_{1_{_{{210}}}}                  +\cdots\right].\label{superpotential gut su(5)tensor intermediate}
\end{eqnarray}
We show only those fields whose VEVs enter in the spontaneous breaking of the GUT symmetry.
These include the  $\mathsf{75}$ ($\mathsf{H}^{(210)ij}_{kl}$) and
$\mathsf{24}$ ($\mathsf{H}^{(210)i}_{j}$) in the $\mathsf{210}$-plet, and the singlets  $\mathbf S_{1_{210}}$, $\mathbf S_{1_{126}}$,  $\mathbf S_{1_{\ov{126}}}$ in the  $\mathsf{210}, $ $\mathsf{126}$-plet,
 $\mathsf{\ov{126}}$-plet, respectively. A further reduction of Eq.(\ref{superpotential gut su(5)tensor intermediate}) gives Eq.(\ref{superpotential gut su(5)tensor}).

{Finally, we redisplay Eqs.(\ref{Determination S_24_210}-\ref{Determination S_1_126}) showing their explicit dependence on the parameters $\eta$ and $\lambda$:
\begin{eqnarray}
\mathbf{S}_{24_{_{{210}}}}\left[9\eta^3\lambda^3\mathbf{S}_{24_{_{{210}}}}^3
-24\eta^2\lambda^2\mathbf{S}_{24_{_{{210}}}}^2\left(-28\lambda m_{\Delta}+45\eta m_{\Phi}\right)+64\eta\lambda \mathbf{S}_{24_{_{{210}}}}\left(320\lambda^2 m_{\Delta}^2-279\eta\lambda m_{\Delta}m_{\Phi}+972\eta^2m_{\Phi}^2\right)\right.\nonumber\\
\left.+3824\left(\lambda m_{\Delta}-2\eta m_{\Phi}\right)\left(4\lambda m_{\Delta}+3\eta m_{\Phi}\right)^2\right]=0,~~~~~~~~~~~~~~~~
\end{eqnarray}
\begin{eqnarray}
\mathbf S_{75_{_{{210}}}}&=&\frac{5\left[\eta\lambda\mathbf S_{24_{_{{210}}}}^2+24\mathbf S_{24_{_{{210}}}}\left(\lambda{{m}}_{\Delta}-2\eta{{m}}_{\Phi}\right)\right]}
{6\left[\eta\lambda\mathbf S_{24_{_{{210}}}}+8\left(4\lambda{{m}}_{\Delta}+3\eta{{m}}_{\Phi}\right)\right]},
\end{eqnarray}

\begin{eqnarray}
\mathbf S_{1_{_{{126}}}}\cdot\mathbf S_{1_{_{\overline{126}}}}&=&
\frac{1}{{{216}\eta^3\left(\eta\lambda\mathbf S_{24_{_{{210}}}}+{32}\lambda{{m}}_{\Delta}+{24}\eta{{m}}_{\Phi}\right)}}
\left[{5}\eta^3\lambda^2\mathbf S_{24_{_{{210}}}}^3-32\eta^2\lambda\mathbf S_{24_{_{{210}}}}^2\left(8\lambda{{m}}_{\Delta}+{39}\eta{{m}}_{\Phi}\right)\right.\nonumber\\
&&\left.~-{1728}\eta\mathbf S_{24_{_{{210}}}}                  \left({7}\lambda^2{{m}}_{\Delta}^2-{7}\eta\lambda{{m}}_{\Delta}{{m}}_{\Phi}{-6}\eta^2{{m}}_{\Phi}^2\right)\right.\nonumber\\
&&\left.~-{13824}{{m}}_{\Delta}\left(3\lambda{{m}}_{\Delta}-2\eta{{m}}_{\Phi}\right)\left(4\lambda{{m}}_{\Delta}+3\eta{{m}}_{\Phi}\right)\right].
\end{eqnarray}

}

\section* {Appendix D:  Couplings of light and heavy Higgs sectors}\label{AppB}
As discussed in Sec.(\ref{sec4}), the doublet-triplet splitting involves
couplings of light and heavy fields. {In this appendix, we give further details of the couplings that enter in Eqs.(\ref{doublet mass matrix}) and (\ref{triplet mass matrix}).}
There are eight such couplings: $\mathsf{10\cdot126\cdot120}$,   $\mathsf{10\cdot\overline{126}\cdot210}$, $\mathsf{120\cdot126\cdot210}$, $\mathsf{120\cdot\overline{126}\cdot210}$, {$\mathsf{126\cdot\overline{126}}$, $\mathsf{210\cdot{210}}$, $\mathsf{210\cdot{210}\cdot210}$ and $\mathsf{210\cdot 126\cdot\overline{126}}$}.
In the analysis below we first give the $\mathsf{SU(5)}$ decomposition of the relevant parts
of {these} $\mathsf{SO(10)}$ invariant couplings and then further reduce them in the $\mathsf{SU(3)_C\times SU(2)_L\times U(1)_Y}$ invariant form
to exhibit the Higgs doublets and Higgs triplets and their mixings.
We here note that an analysis of these couplings  in the Pati-Salam decomposition has been carried out
 in the first two papers of \cite{4}. We give now the details of our analysis in $SU(5)\times U(1)$
 decomposition.\\

\subsection*{Mixing of light and heavy sector}
\subsubsection* {D1: {\boldmath$\mathsf{10\cdot126\cdot210}$  Coupling}}
\begin{eqnarray}
\mathrm{A}~{}^{1}\Gamma_{\mu}\Delta_{\mu\nu\sigma\xi\zeta}\Phi_{\mu\sigma\xi\zeta}&=&\frac{\imath}{5!}\mathrm{A}\left[\frac{1}{2\sqrt{5}}
\mathsf{H}^{(10_1)i}~\mathsf{H}^{(126)}~\mathsf{H}^{(210)}_i+\frac{1}{4\sqrt{10}}
\mathsf{H}^{(10_1)i}~\mathsf{H}^{(126)lm}_{ijk}~\mathsf{H}^{(210)jk}_{lm}\right.\nonumber\\
&&\left.~~~~~+\frac{1}{2\sqrt{10}}
\mathsf{H}^{(10_1)i}~\mathsf{H}^{(126)}_{j}~\mathsf{H}^{(210)j}_i
-\frac{\sqrt{3}}{10}
\mathsf{H}^{(10_1)i}~\mathsf{H}^{(126)}_{i}~\mathsf{H}^{(210)}\right.\nonumber\\
&&\left.~~~~~+~\frac{1}{4\sqrt{5}}
\mathsf{H}^{(10_1)}_i~\mathsf{H}^{(126)jk}_{l}~\mathsf{H}^{(210)il}_{jk}
+\frac{1}{\sqrt{30}}\mathsf{H}^{(10_1)}_i~\mathsf{H}^{(126)ij}_{k}~\mathsf{H}^{(210)k}_j+\cdots\right]\nonumber\\
&=&\frac{\imath}{5!}\mathrm{A}\left[\left(\frac{1}{2\sqrt{5}}\mathbf S_{1_{_{126}}}\right){}^{(\overline{5}_{210})}\!{\mathsf D}_{a}{}^{({{5}}_{10_1})}\!{\mathsf D}^{a}+\left(-\frac{\sqrt{3}}{10}\mathbf S_{1_{_{210}}}-\frac{\sqrt{3}}{{20}}\mathbf S_{24_{_{210}}}\right){}^{(\overline{5}_{126})}\!{\mathsf D}_{a}{}^{({{5}}_{10_1})}\!{\mathsf D}^{a}\right.\nonumber\\
&&\left.~~~~~+~\left(-\frac{1}{4\sqrt{6}}\mathbf S_{24_{_{210}}}+\frac{1}{4\sqrt{15}}\mathbf S_{75_{_{210}}}\right){}^{({\overline{5}}_{10_1})}\!{\mathsf D}_{a}{}^{({45}_{126})}\!{\mathsf D}^{a}\right.\nonumber\\
&&\left.~~~~~+~\left(\frac{1}{2\sqrt{5}}\mathbf S_{1_{_{126}}}\right){}^{(\overline{5}_{210})}\!{\mathsf T}_{\alpha}{}^{({{5}}_{10_1})}\!{\mathsf T}^{\alpha}+\left(-\frac{\sqrt{3}}{10}\mathbf S_{1_{_{210}}}+\frac{1}{10\sqrt{3}}\mathbf S_{24_{_{210}}}\right){}^{(\overline{5}_{126})}\!{\mathsf T}_{\alpha}{}^{({{5}}_{10_1})}\!{\mathsf T}^{\alpha}\right.\nonumber\\
&&\left.~~~~~+~\left(-\frac{1}{6\sqrt{2}}\mathbf S_{24_{_{210}}}{-\frac{1}{12\sqrt{5}}}\mathbf S_{75_{_{210}}}\right){}^{({\overline{5}}_{10_1})}\!{\mathsf T}_{\alpha}{}^{({45}_{126})}\!{\mathsf T}^{\alpha}\right.\nonumber\\
&&\left.~~~~~+\left({\frac{1}{12\sqrt{5}}}\mathbf S_{75_{_{210}}}\right){}^{(\overline{50}_{126})}\!{\mathsf T}_{\alpha}{}^{({{5}}_{10_1})}\!{\mathsf T}^{\alpha}+\cdots\right].
\end{eqnarray}

\subsubsection* {D2: \boldmath$\mathsf{10\cdot\overline{126}\cdot210}$  Coupling}
\begin{eqnarray}
\mathrm{B_r}~{}^{r}\Gamma_{\mu}\overline{\Delta}_{\mu\nu\sigma\xi\zeta}\Phi_{\mu\sigma\xi\zeta}&=&-\frac{\imath}{5!}\mathrm{B_r}\left[\frac{1}{2\sqrt{5}}
\mathsf{H}^{(10_r)}_i~\mathsf{H}^{(\overline{126})}~\mathsf{H}^{(210)i}+\frac{1}{4\sqrt{10}}
\mathsf{H}^{(10_r)}_i~\mathsf{H}^{(\overline{126})ijk}_{lm}~\mathsf{H}^{(210)lm}_{jk}\right.\nonumber\\
&&\left.~~~~~+\frac{1}{2\sqrt{10}}
\mathsf{H}^{(10_r)}_i~\mathsf{H}^{(\overline{126})j}~\mathsf{H}^{(210)i}_j-\frac{\sqrt{3}}{10}
\mathsf{H}^{(10_r)}_i~\mathsf{H}^{(\overline{126})i}~\mathsf{H}^{(210)}\right.\nonumber\\
&&\left.~~~~~
+~\frac{1}{4\sqrt{5}}
\mathsf{H}^{(10_r)i}~\mathsf{H}^{(\overline{126}){l}}_{jk}~\mathsf{H}^{(210)jk}_{il}
+\frac{1}{\sqrt{30}}
\mathsf{H}^{(10_r)i}~\mathsf{H}^{(\overline{126})k}_{ij}~\mathsf{H}^{(210)j}_k+\cdots\right]\nonumber\\
&=&-\frac{\imath}{5!}\mathrm{B_r}\left[\left(\frac{1}{2\sqrt{5}}\mathbf S_{1_{_{\overline{126}}}}     \right){}^{({\overline{5}}_{10_r})}\!{\mathsf D}_{a}{}^{({5}_{210})}\!{\mathsf D}^{a}+\left(-\frac{\sqrt{3}}{10}\mathbf S_{1_{_{210}}}-\frac{\sqrt{3}}{{20}}\mathbf S_{24_{_{210}}}\right){}^{({\overline{5}}_{10_r})}\!{\mathsf D}_{a}{}^{({5}_{\overline{126}})}\!{\mathsf D}^{a}\right.\nonumber\\
&&\left.~~~~~+~\left(-\frac{1}{4\sqrt{6}}\mathbf S_{24_{_{210}}}+\frac{1}{4\sqrt{15}}\mathbf S_{75_{_{210}}}\right){}^{(\overline{45}_{\overline{126}})}\!{\mathsf D}_{a}{}^{({{5}}_{10_r})}\!{\mathsf D}^{a}\right.\nonumber\\
&&\left.~~~~~+~\left(\frac{1}{2\sqrt{5}}\mathbf S_{1_{_{\overline{126}}}}     \right){}^{({\overline{5}}_{10_r})}\!{\mathsf T}_{\alpha}{}^{({5}_{210})}\!{\mathsf T}^{\alpha}+\left(-\frac{\sqrt{3}}{10}\mathbf S_{1_{_{210}}}+\frac{1}{10\sqrt{3}}\mathbf S_{24_{_{210}}}\right){}^{({\overline{5}}_{10_r})}\!{\mathsf T}_{\alpha}{}^{({5}_{\overline{126}})}\!{\mathsf T}^{\alpha}\right.\nonumber\\
&&\left.~~~~~+~\left(-\frac{1}{6\sqrt{2}}\mathbf S_{24_{_{210}}}{-\frac{1}{12\sqrt{5}}}\mathbf S_{75_{_{210}}}\right){}^{(\overline{45}_{\overline{126}})}\!{\mathsf T}_{\alpha}{}^{({{5}}_{10_r})}\!{\mathsf T}^{\alpha}\right.\nonumber\\
&&\left.~~~~~+\left({\frac{1}{12\sqrt{5}}}\mathbf S_{75_{_{210}}}\right){}^{({\overline{5}}_{10_r})}\!{\mathsf T}_{\alpha}{}^{({50}_{\overline{126}})\!{\mathsf T}^{\alpha}}+\cdots\right].
\end{eqnarray}

\subsubsection* {D3: {\boldmath$\mathsf{120\cdot126\cdot210}$ Coupling}}
\begin{eqnarray}
\mathrm{C}~\Sigma_{\mu\nu\sigma}\Delta_{\nu\sigma\xi\zeta\rho}
\Phi_{\mu\xi\zeta\rho}&=&\frac{\imath}{5!}\mathrm{C}\left[\frac{1}{4\sqrt{15}}
\mathsf{H}^{(120)ij}_k~\mathsf{H}^{(126)kl}_{jmn}~\mathsf{H}^{(210)mn}_{il}+\frac{1}{8\sqrt{15}}\mathsf{H}^{(120)jk}_i~\mathsf{H}^{(126)mn}_{jkl}~\mathsf{H}^{(210)il}_{mn}
\right.\nonumber\\
&&\left.~~~~~+\frac{1}{12\sqrt{10}}
\mathsf{H}^{(120)ij}_k~\mathsf{H}^{(126)kl}_{ijm}~\mathsf{H}^{(210)m}_{l}-\frac{1}{12\sqrt{10}}
\mathsf{H}^{(120)ij}_k~\mathsf{H}^{(126)}_l~\mathsf{H}^{(210)kl}_{ij}\right.\nonumber\\
&&\left.~~~~~
-\frac{1}{12\sqrt{15}}\mathsf{H}^{(120)ij}_k~\mathsf{H}^{(126)}_i~\mathsf{H}^{(210)k}_{j}-\frac{1}{4\sqrt{30}}
\mathsf{H}^{(120)k}_{ij}~\mathsf{H}^{(126)lm}_k~\mathsf{H}^{(210)ij}_{lm}\right.\nonumber\\
&&\left.~~~~~+~\frac{1}{8\sqrt{5}}
\mathsf{H}^{(120)k}_{ij}~\mathsf{H}^{(126)ij}_l~\mathsf{H}^{(210)l}_{k}+\frac{1}{12\sqrt{5}}
\mathsf{H}^{(120)k}_{ij}~\mathsf{H}^{(126)jl}_k~\mathsf{H}^{(210)i}_{l}\right.\nonumber\\
&&\left.~~~~~-\frac{1}{20\sqrt{6}}
\mathsf{H}^{(120)k}_{ij}~\mathsf{H}^{(126)ij}_k~\mathsf{H}^{(210)}-\frac{1}{8}\sqrt{\frac{3}{5}}
\mathsf{H}^{(120)i}~\mathsf{H}^{(126)}_j~\mathsf{H}^{(210)j}_{i}\right.\nonumber\\
&&\left.~~~~~-\frac{1}{10\sqrt{2}}
\mathsf{H}^{(120)i}~\mathsf{H}^{(126)}_i~\mathsf{H}^{(210)}-\frac{1}{4\sqrt{30}}
\mathsf{H}^{(120)}_i~\mathsf{H}^{(126)jk}_l~\mathsf{H}^{(210)il}_{jk}\right.\nonumber\\
&&\left.~~~~~{+~\frac{1}{12\sqrt{5}}}
\mathsf{H}^{(120)}_i~\mathsf{H}^{(126)ij}_k~\mathsf{H}^{(210)k}_{j}{-\frac{1}{\sqrt{30}}}
\mathsf{H}^{(120)i}~\mathsf{H}^{(126)}~\mathsf{H}^{(210)}_{i}+\cdots\right]\nonumber\\
&=&\frac{\imath}{5!}\mathrm{C}\left[\left(\frac{1}{48\sqrt{3}}\mathbf S_{24_{_{210}}}+\frac{1}{12\sqrt{30}}\mathbf S_{75_{_{210}}}\right){}^{(\overline{5}_{126})}\!{\mathsf D}_{a}{}^{({{45}}_{120})}\!{\mathsf D}^{a}\right.\nonumber\\
&&\left.~~~~~+~\left(-\frac{1}{20\sqrt{6}}\mathbf S_{1_{_{210}}}-\frac{1}{240\sqrt{6}}\mathbf S_{24_{_{210}}}-\frac{1}{12\sqrt{15}}\mathbf S_{75_{_{210}}}\right){}^{(\overline{45}_{120})}\!{\mathsf D}_{a}{}^{({{45}}_{126})}\!{\mathsf D}^{a}\right.\nonumber\\
&&\left.~~~~~+~\left(-\frac{1}{10\sqrt{2}}\mathbf S_{1_{_{210}}}+\frac{3}{40\sqrt{2}}\mathbf S_{24_{_{210}}}\right){}^{(\overline{5}_{126})}\!{\mathsf D}_{a}{}^{({{5}}_{120})}\!{\mathsf D}^{a}\right.\nonumber\\
&&\left.~~~~~+~\left({-\frac{1}{48}}\mathbf S_{24_{_{210}}}-\frac{1}{12\sqrt{10}}\mathbf S_{75_{_{210}}}\right){}^{(\overline{5}_{120})}\!{\mathsf D}_{a}{}^{({{45}}_{126})}\!{\mathsf D}^{a}\right.\nonumber\\
&&\left.~~~~~{-}~\left({\frac{1}{\sqrt{30}}}\mathbf S_{1_{_{126}}}\right){}^{(\overline{5}_{210})}\!{\mathsf D}_{a}{}^{({{5}}_{120})}\!{\mathsf D}^{a}\right.\nonumber\\
&&\left.~~~~~+~\left(\frac{1}{72}\mathbf S_{24_{_{210}}}-\frac{1}{36\sqrt{10}}\mathbf S_{75_{_{210}}}\right){}^{(\overline{5}_{126})}\!{\mathsf T}_{\alpha}{}^{({{45}}_{120})}\!{\mathsf T}^{\alpha}\right.\nonumber\\
&&\left.~~~~~+~\left(-\frac{1}{20\sqrt{6}}\mathbf S_{1_{_{210}}}-\frac{1}{40\sqrt{6}}\mathbf S_{24_{_{210}}}\right){}^{(\overline{45}_{120})}\!{\mathsf T}_{\alpha}{}^{({{45}}_{126})}\!{\mathsf T}^{\alpha}\right.\nonumber\\
&&\left.~~~~~+~\left(-\frac{1}{10\sqrt{2}}\mathbf S_{1_{_{210}}}-\frac{1}{20\sqrt{2}}\mathbf S_{24_{_{210}}}\right){}^{(\overline{5}_{126})}\!{\mathsf T}_{\alpha}{}^{({{5}}_{120})}\!{\mathsf T}^{\alpha}\right.\nonumber\\
&&\left.~~~~~+~\left({-\frac{1}{24\sqrt{3}}\mathbf S_{24_{_{210}}}-\frac{1}{12\sqrt{30}}\mathbf S_{75_{_{210}}}}\right){}^{(\overline{5}_{120})}\!{\mathsf T}_{\alpha}{}^{({{45}}_{126})}\!{\mathsf T}^{\alpha}\right.\nonumber\\
&&\left.~~~~~+~\left(-\frac{1}{60\sqrt{6}}\mathbf S_{24_{_{210}}}{+\frac{1}{36\sqrt{15}}\mathbf S_{75_{_{210}}}}\right){}^{(\overline{50}_{126})}\!{\mathsf T}_{\alpha}{}^{({{45}}_{120})}\!{\mathsf T}^{\alpha}\right.\nonumber\\
&&\left.~~~~~~{-}\left({\frac{1}{\sqrt{30}}}\mathbf S_{1_{_{126}}}\right){}^{(\overline{5}_{210})}\!{\mathsf T}_{\alpha}{}^{({{5}}_{120})}\!{\mathsf T}^{\alpha}
+~\cdots\right].
\end{eqnarray}

\subsubsection* {D4: {\boldmath$\mathsf{120\cdot\overline{126}\cdot210}$ Coupling}}
\begin{eqnarray}
\overline{\mathrm{C}}~\Sigma_{\mu\nu\sigma}\overline{\Delta}_{\nu\sigma\xi\zeta\rho}
\Phi_{\mu\xi\zeta\rho}&=&-\frac{\imath}{5!}\overline{\mathrm{C}}\left[\frac{1}{4\sqrt{15}}
\mathsf{H}^{(120)k}_{ij}~\mathsf{H}^{(\overline{126})jmn}_{kl}~\mathsf{H}^{(210)il}_{mn}
+\frac{1}{8\sqrt{15}}\mathsf{H}^{(120)i}_{jk}~\mathsf{H}^{(\overline{126})jkl}_{mn}~\mathsf{H}^{(210)mn}_{il}\right.\nonumber\\
&&\left.~~~~~
+\frac{1}{12\sqrt{10}}
\mathsf{H}^{(120)k}_{ij}~\mathsf{H}^{(\overline{126})ijm}_{kl}~\mathsf{H}^{(210)l}_{m}-~\frac{1}{12\sqrt{10}}
\mathsf{H}^{(120)k}_{ij}~\mathsf{H}^{(\overline{126})l}~\mathsf{H}^{(210)ij}_{kl}\right.\nonumber\\
&&\left.~~~~~
-\frac{1}{12\sqrt{15}}\mathsf{H}^{(120)k}_{ij}~\mathsf{H}^{(\overline{126})i}~\mathsf{H}^{(210)j}_{k}-\frac{1}{4\sqrt{30}}
\mathsf{H}^{(120)ij}_{k}~\mathsf{H}^{(\overline{126})k}_{lm}~\mathsf{H}^{(210)lm}_{ij}\right.\nonumber\\
&&\left.~~~~~+~\frac{1}{8\sqrt{5}}
\mathsf{H}^{(120)ij}_{k}~\mathsf{H}^{(\overline{126})l}_{ij}~\mathsf{H}^{(210)k}_{l}+\frac{1}{12\sqrt{5}}
\mathsf{H}^{(120)ij}_{k}~\mathsf{H}^{(\overline{126})k}_{jl}~\mathsf{H}^{(210)l}_{i}\right.\nonumber\\
&&\left.~~~~~-\frac{1}{20\sqrt{6}}
\mathsf{H}^{(120)ij}_{k}~\mathsf{H}^{(\overline{126})k}_{ij}~\mathsf{H}^{(210)}-~\frac{1}{8}\sqrt{\frac{3}{5}}
\mathsf{H}^{(120)}_{i}~\mathsf{H}^{(\overline{126})j}~\mathsf{H}^{(210)i}_{j}\right.\nonumber\\
&&\left.~~~~~-\frac{1}{10\sqrt{2}}
\mathsf{H}^{(120)}_{i}~\mathsf{H}^{(\overline{126})i}~\mathsf{H}^{(210)}-\frac{1}{4\sqrt{30}}
\mathsf{H}^{(120)i}~\mathsf{H}^{(\overline{126})l}_{jk}~\mathsf{H}^{(210)jk}_{il}\right.\nonumber\\
&&\left.~~~~~{+~\frac{1}{12\sqrt{5}}}
\mathsf{H}^{(120)i}~\mathsf{H}^{(\overline{126})k}_{ij}~\mathsf{H}^{(210)j}_{k}{-\frac{1}{\sqrt{30}}}
\mathsf{H}^{(120)}_i~\mathsf{H}^{(\overline{126})}~\mathsf{H}^{(210)i}+\cdots\right]\nonumber\\
&=&-\frac{\imath}{5!}\overline{\mathrm{C}}\left[\left(\frac{1}{48\sqrt{3}}\mathbf S_{24_{_{210}}}+\frac{1}{12\sqrt{30}}\mathbf S_{75_{_{210}}}\right){}^{({\overline{45}}_{120})}\!{\mathsf D}_{a}{}^{({5}_{\overline{126}})}\!{\mathsf D}^{a}\right.\nonumber\\
&&\left.~~~~~+~\left(-\frac{1}{20\sqrt{6}}\mathbf S_{1_{_{210}}}-\frac{1}{240\sqrt{6}}\mathbf S_{24_{_{210}}}-\frac{1}{12\sqrt{15}}\mathbf S_{75_{_{210}}}\right){}^{({\overline{45}}_{\overline{126}})}\!{\mathsf D}_{a}{}^{({45}_{120})}\!{\mathsf D}^{a}\right.\nonumber\\
&&\left.~~~~~+~\left(-\frac{1}{10\sqrt{2}}\mathbf S_{1_{_{210}}}+\frac{3}{40\sqrt{2}}\mathbf S_{24_{_{210}}}\right){}^{({\overline{5}}_{120})}\!{\mathsf D}_{a}{}^{({5}_{\overline{126}})}\!{\mathsf D}^{a}\right.\nonumber\\
&&\left.~~~~~+~\left({-\frac{1}{48}}\mathbf S_{24_{_{210}}}-\frac{1}{12\sqrt{10}}\mathbf S_{75_{_{210}}}\right){}^{({\overline{45}}_{\overline{126}})}\!{\mathsf D}_{a}{}^{({5}_{120})}\!{\mathsf D}^{a}\right.\nonumber\\
&&\left.~~~~~{-}~\left({\frac{1}{\sqrt{30}}}\mathbf S_{1_{_{\overline{126}}}}     \right){}^{({\overline{5}}_{120})}\!{\mathsf D}_{a}{}^{({5}_{210})}\!{\mathsf D}^{a}\right.\nonumber\\
&&\left.~~~~~+~\left(\frac{1}{72}\mathbf S_{24_{_{210}}}-\frac{1}{36\sqrt{10}}\mathbf S_{75_{_{210}}}\right){}^{({\overline{45}}_{120})}\!{\mathsf T}_{\alpha}{}^{({5}_{\overline{126}})}\!{\mathsf T}^{\alpha}\right.\nonumber\\
&&\left.~~~~~+~\left(-\frac{1}{20\sqrt{6}}\mathbf S_{1_{_{210}}}-\frac{1}{40\sqrt{6}}\mathbf S_{24_{_{210}}}\right){}^{({\overline{45}}_{\overline{126}})}\!{\mathsf T}_{\alpha}{}^{({45}_{120})}\!{\mathsf T}^{\alpha}\right.\nonumber\\
&&\left.~~~~~+~\left(-\frac{1}{10\sqrt{2}}\mathbf S_{1_{_{210}}}-\frac{1}{20\sqrt{2}}\mathbf S_{24_{_{210}}}\right){}^{({\overline{5}}_{120})}\!{\mathsf T}_{\alpha}{}^{({5}_{\overline{126}})}\!{\mathsf T}^{\alpha}\right.\nonumber\\
&&\left.~~~~~+~\left({-\frac{1}{24\sqrt{3}}\mathbf S_{24_{_{210}}}-\frac{1}{12\sqrt{30}}\mathbf S_{75_{_{210}}}}\right){}^{({\overline{45}}_{\overline{126}})}\!{\mathsf T}_{\alpha}{}^{({5}_{120})}\!{\mathsf T}^{\alpha}\right.\nonumber\\
&&\left.~~~~~+~\left(-\frac{1}{60\sqrt{6}}\mathbf S_{24_{_{210}}}{+\frac{1}{36\sqrt{15}}\mathbf S_{75_{_{210}}}}\right){}^{({\overline{45}}_{120})}\!{\mathsf T}_{\alpha}{}^{({50}_{\overline{126}})}\!{\mathsf T}^{\alpha} \right.\nonumber\\
&&\left.~~~~~~{-}\left({\frac{1}{\sqrt{30}}}\mathbf S_{1_{_{126}}}\right){}^{(\overline{5}_{210})}\!{\mathsf T}_{\alpha}{}^{({{5}}_{120})}\!{\mathsf T}^{\alpha}
+~\cdots\right].
\end{eqnarray}
{
\subsection*{Heavy sector}
\subsubsection* {D5: {\boldmath$\mathsf{126\cdot\overline{126}}$ Coupling}}
\begin{eqnarray}
{m_{\Delta}}\Delta_{\mu\nu\sigma\xi\zeta}\overline{\Delta}_{\mu\nu\sigma\xi\zeta}&=&{m_{\Delta}}\left[{2}\mathsf{H}^{(126)}_i~\mathsf{H}^{(\overline{126})i}
+\mathsf{H}^{(126)ij}_k~\mathsf{H}^{(\overline{126})k}_{ij}+\frac{1}{6}\mathsf{H}^{(126)lm}_{ijk}~\mathsf{H}^{(\overline{126})ijk}_{lm}+~\cdots\right]\nonumber\\
&=&{m_{\Delta}}\left[{2}{}^{({\overline{5}}_{126})}\!{\mathsf D}_{a}{}^{({{5}}_{\overline{126}})}\!{\mathsf D}^{a}+
{}^{({{45}}_{126})}\!{\mathsf D}_{a}{}^{(\overline{{45}}_{\overline{126}})}\!{\mathsf D}^{a}+{2}{}^{({\overline{5}}_{126})}\!{\mathsf T}_{\alpha}{}^{({{5}}_{\overline{126}})}\!{\mathsf T}^{\alpha} \right.\nonumber\\
&&\left.~~~~~+
{}^{({{45}}_{126})}\!{\mathsf T}_{\alpha}{}^{(\overline{{45}}_{\overline{126}})}\!{\mathsf T}^{\alpha}+
\frac{1}{6}{}^{({\overline{50}}_{126})}\!{\mathsf T}_{\alpha}{}^{({{50}}_{\overline{126}})}\!{\mathsf T}^{\alpha} +~\cdots\right].
\end{eqnarray}

\subsubsection* {D6: {\boldmath$\mathsf{210\cdot{210}}$ Coupling}}

\begin{eqnarray}
{m_{\Phi}}\Phi_{\mu\nu\sigma\xi}\Phi_{\mu\nu\sigma\xi}&=&{m_{\Phi}}\left[2\mathsf{H}^{(210)}_i~\mathsf{H}^{({210})i}+~\cdots\right]\nonumber\\
&=&{m_{\Phi}}\left[2~{}^{({\overline{5}}_{210})}\!{\mathsf D}_{a}{}^{({{5}}_{{210}})}\!{\mathsf D}^{a} +2~{}^{({\overline{5}}_{210})}\!{\mathsf T}_{\alpha}{}^{({{5}}_{{210}})}\!{\mathsf T}^{\alpha}+~\cdots\right].
\end{eqnarray}

\subsubsection* {D7: {\boldmath$\mathsf{210\cdot{210}\cdot210}$ Coupling}}
\begin{eqnarray}
{\lambda}
  \Phi_{\mu\nu\sigma\xi}\Phi_{\sigma\xi\rho\tau}\Phi_{\rho\tau\mu\nu}&=&\lambda\left[\frac{1}{\sqrt{2}}\mathsf{H}^{(210)}_i~\mathsf{H}^{({210})j}~\mathsf{H}^{({210})i}_j
{-\sqrt{\frac{3}{5}}}\mathsf{H}^{(210)}_i~\mathsf{H}^{({210})i}~\mathsf{H}^{({210})}+~\cdots\right]\nonumber\\
&=&-\lambda\left[\left({\frac{3}{10\sqrt{2}}}\mathbf S_{1_{_{210}}}+\frac{1}{2}\sqrt{\frac{3}{5}}\mathbf S_{24_{_{210}}}\right)
{}^{({\overline{5}}_{210})}\!{\mathsf D}_{a}{}^{({5}_{{210}})}\!{\mathsf D}^{a}\right.\nonumber\\
&&\left.~~~~~+\left({-\frac{3}{10\sqrt{2}}}\mathbf S_{1_{_{210}}}+\frac{1}{\sqrt{15}}\mathbf S_{24_{_{210}}}\right)
{}^{({\overline{5}}_{210})}\!{\mathsf T}_{\alpha}{}^{({5}_{{210}})}\!{\mathsf T}^{\alpha}+~\cdots\right].
\end{eqnarray}

\subsubsection* {D8: {\boldmath$\mathsf{210\cdot{126}\cdot\overline{126}}$ Coupling}}
\begin{eqnarray}
  {\eta}\Phi_{\mu\nu\sigma\xi}\Delta_{\mu\nu\rho\tau\zeta}\overline{\Delta}_{\sigma\xi\rho\tau\zeta}&=&\eta\left[\frac{1}{5}
  \mathsf{H}^{(210)}_i~\mathsf{H}^{(\overline{126})i}~\mathsf{H}^{({126})}+\frac{1}{5}
  \mathsf{H}^{({126})}_i~\mathsf{H}^{({210})i}~\mathsf{H}^{(\overline{126})}\right.\nonumber\\
&&\left.~~
+\frac{1}{10\sqrt{3}}
  \mathsf{H}^{({126})kl}_m~\mathsf{H}^{(\overline{126})m}_{ij}~\mathsf{H}^{({210})ij}_{kl}+{\frac{1}{10\sqrt{2}}}
  \mathsf{H}^{({126})kl}_i~\mathsf{H}^{(\overline{126})j}_{kl}~\mathsf{H}^{({210})i}_{j}\right.\nonumber\\
&&\left.~~
+{\frac{1}{15\sqrt{2}}}
  \mathsf{H}^{({126})jk}_l~\mathsf{H}^{(\overline{126})l}_{ik}~\mathsf{H}^{({210})i}_{j}+{\frac{3}{10\sqrt{2}}}
  \mathsf{H}^{(\overline{126})j}~\mathsf{H}^{({126})}_{i}~\mathsf{H}^{({210})i}_{j}\right.\nonumber\\
&&\left.~~-\frac{2}{5\sqrt{15}}
  \mathsf{H}^{(\overline{126})i}~\mathsf{H}^{({126})}_{i}~\mathsf{H}^{({210})}
-{\frac{1}{30\sqrt{2}}}
  \mathsf{H}^{(\overline{126})klm}_{ij}~\mathsf{H}^{({126})}_{m}~\mathsf{H}^{({210})ij}_{kl}\right.\nonumber\\
&&\left.~~-{\frac{1}{30\sqrt{2}}}
  \mathsf{H}^{(\overline{126})m}~\mathsf{H}^{({126})kl}_{ijm}~\mathsf{H}^{({210})ij}_{kl}
  +{\frac{1}{120\sqrt{3}}}
  \mathsf{H}^{(\overline{126})mnp}_{ij}~\mathsf{H}^{({126})kl}_{mnp}~\mathsf{H}^{({210})ij}_{kl}\right.\nonumber\\
&&\left.~~+{\frac{1}{40\sqrt{3}}}
  \mathsf{H}^{(\overline{126})klm}_{np}~\mathsf{H}^{({126})np}_{ijm}~\mathsf{H}^{({210})ij}_{kl}
  +\frac{1}{10\sqrt{3}}
  \mathsf{H}^{(\overline{126})kmn}_{ip}~\mathsf{H}^{({126})lp}_{jmn}~\mathsf{H}^{({210})ij}_{kl}\right.\nonumber\\
&&\left.~~{+\frac{1}{45\sqrt{2}}
  \mathsf{H}^{(\overline{126})klm}_{in}~\mathsf{H}^{({126})jn}_{klm}~\mathsf{H}^{({210})i}_j}+\frac{1}{60\sqrt{15}}
  \mathsf{H}^{(\overline{126})ijk}_{lm}~\mathsf{H}^{({126})lm}_{ijk}~\mathsf{H}^{({210})} +~\cdots\right]\nonumber\\
  &=&\eta\left[\left(\frac{1}{5}\mathbf S_{1_{_{126}}}\right)
{}^{({\overline{5}}_{{210}})}\!{\mathsf D}_{a}{}^{({5}_{\overline{126}})}\!{\mathsf D}^{a}+\left(\frac{1}{5}\mathbf S_{1_{_{\overline{126}}}}\right)
{}^{({\overline{5}}_{{126}})}\!{\mathsf D}_{a}{}^{({5}_{{210}})}\!{\mathsf D}^{a}\right.\nonumber\\
&&\left.~~+\left(-{\frac{1}{6\sqrt{15}}}\mathbf S_{24_{_{210}}}+{\frac{1}{15\sqrt{6}}}\mathbf S_{75_{_{210}}}\right)
{}^{(\overline{45}_{\overline{126}})}\!{\mathsf D}_{a}{}^{({{45}}_{126})}\!{\mathsf D}^{a}\right.\nonumber\\
&&\left.~~-\left({\frac{2}{5\sqrt{15}}}\mathbf S_{1_{_{210}}}+{\frac{3}{{20}}\sqrt{\frac{3}{5}}}\mathbf S_{24_{_{210}}}\right)
{}^{(\overline{5}_{{126}})}\!{\mathsf D}_{a}{}^{({{5}}_{\overline{126}})}\!{\mathsf D}^{a}\right.\nonumber\\
&&\left.~~+\left(\frac{1}{5}\mathbf S_{1_{_{126}}}\right)
{}^{({\overline{5}}_{{210}})}\!{\mathsf T}_{\alpha}{}^{({5}_{\overline{126}})}\!{\mathsf T}^{\alpha}+\left(\frac{1}{5}\mathbf S_{1_{_{\overline{126}}}}\right)
{}^{({\overline{5}}_{{126}})}\!{\mathsf T}_{\alpha}{}^{({5}_{{210}})}\!{\mathsf T}^{\alpha}\right.\nonumber\\
&&\left.~~+\left(-{\frac{2}{5\sqrt{15}}}\mathbf S_{1_{_{210}}}+{\frac{1}{10}\sqrt{\frac{3}{5}}}\mathbf S_{24_{_{210}}}\right)
{}^{(\overline{5}_{{126}})}\!{\mathsf T}_{\alpha}{}^{({{5}}_{\overline{126}})}\!{\mathsf T}^{\alpha}\right.\nonumber\\
&&\left.~~-\left({\frac{2}{45}}\mathbf S_{75_{_{210}}}\right)
{}^{({\overline{5}}_{{126}})}\!{\mathsf T}_{\alpha}{}^{({50}_{\overline{126}})}\!{\mathsf T}^{\alpha}-\left({\frac{2}{45}}\mathbf S_{75_{_{{210}}}}\right)
{}^{({\overline{50}}_{{126}})}\!{\mathsf T}_{\alpha}{}^{({5}_{\overline{126}})}\!{\mathsf T}^{\alpha}\right.\nonumber\\
&&\left.~~+\left(\frac{1}{60\sqrt{15}}\mathbf S_{1_{_{210}}}{+\frac{1}{180\sqrt{15}}\mathbf S_{24_{_{210}}}}+{\frac{1}{60\sqrt{6}}}\mathbf S_{75_{_{210}}}\right)
{}^{({\overline{50}}_{{126}})}\!{\mathsf T}_{\alpha}{}^{({50}_{\overline{126}})}\!{\mathsf T}^{\alpha}
+~\cdots\right].
\end{eqnarray}

}

{

}
\section* {Appendix E:  Coefficients of {\boldmath$\mathsf{B-L=-2}$} operators}\label{AppE}
In this appendix we exhibit some of the coefficients of the $\mathsf{B-L=-2}$ operators appearing in Sec.(\ref{sec:5}). The coefficients ${\cal I}_{\acute{w}\acute{x}, \acute{y}\acute{z}}$ and ${\cal J}_{\acute{w}\acute{x}, \acute{y}\acute{z}}$ appearing in Eqs.(\ref{74}) and (\ref{75})} are given by
\begin{eqnarray}
{\cal I}_{\acute{w}\acute{x}, \acute{y}\acute{z}}=
&&2\left(\sum_{M=2}^{7}\frac{\left\{\sum_{q=1}^{2}{\cal G}_{\acute{u}\acute{v}}^{(q)}{U}_{d_{qM}}\right\}
\left\{\sum_{r=1}^{2}f^{(10_r+)}_{_{\acute{w}\acute{x}}}{V}_{d_{rM}}\right\}}{{\mathsf{m}_{d_M}}}\right)\left(\sum_{N=2}^{7}\frac{{{U}_{d_{3N}}}
\left\{\sum_{s=1}^{2}f^{(10_s+)}_{_{\acute{y}\acute{z}}}{V}_{d_{sN}}\right\}}{{\mathsf{m}_{d_N}}}\right)\nonumber\\
&&+\frac{1}{3}f^{(120-)}_{_{\acute{w}\acute{x}}}f^{(120-)}_{_{\acute{y}\acute{z}}}\left(\sum_{M=2}^{7}\frac{\left\{\sum_{r=1}^{2}{\cal G}_{\acute{u}\acute{v}}^{(r)}{U}_{d_{rM}}\right\}
{{V}_{d_{3M}}}}{{\mathsf{m}_{d_M}}}\right)\left(\sum_{N=2}^{7}\frac{{{U}_{d_{3N}}}
{{V}_{d_{3N}}}}{{\mathsf{m}_{d_N}}}\right)\nonumber\\
&&-\sqrt{\frac{2}{3}}f^{(120-)}_{_{\acute{y}\acute{z}}}\left(\sum_{M=2}^{7}\frac{\left\{\sum_{r=1}^{2}{\cal G}_{\acute{u}\acute{v}}^{(r)}{U}_{d_{rM}}\right\}
\left\{\sum_{s=1}^{2}f^{(10_s+)}_{_{\acute{w}\acute{x}}}{V}_{d_{sM}}\right\}}{{\mathsf{m}_{d_M}}}\right)\left(\sum_{N=2}^{7}\frac{{{U}_{d_{3N}}}
{{V}_{d_{3N}}}}{{\mathsf{m}_{d_N}}}\right)\nonumber\\
&&-\sqrt{\frac{2}{3}}f^{(120-)}_{_{\acute{w}\acute{x}}}\left(\sum_{M=2}^{7}\frac{\left\{\sum_{r=1}^{2}{\cal G}_{\acute{u}\acute{v}}^{(r)}{U}_{d_{rM}}\right\}
{{V}_{d_{3M}}}}{{\mathsf{m}_{d_M}}}\right)\left(\sum_{N=2}^{7}\frac{{{U}_{d_{3N}}}
\left\{\sum_{s=1}^{2}f^{(10_s+)}_{_{\acute{y}\acute{z}}}{V}_{d_{sN}}\right\}}{{\mathsf{m}_{d_N}}}\right).
\end{eqnarray}

\begin{eqnarray}
{\cal J}_{\acute{w}\acute{x}, \acute{y}\acute{z}}=
&&2\left(\sum_{M=1}^{8}\frac{\left\{\sum_{q=1}^{2}{\cal G}_{\acute{u}\acute{v}}^{(q)}{U}_{t_{qM}}\right\}
\left\{\sum_{r=1}^{2}f^{(10_r+)}_{_{\acute{w}\acute{x}}}{V}_{t_{rM}}\right\}}{{\mathsf{m}_{t_M}}}\right)\left(\sum_{N=1}^{8}\frac{{{U}_{t_{3N}}}
\left\{\sum_{s=1}^{2}f^{(10_s+)}_{_{\acute{y}\acute{z}}}{V}_{t_{sN}}\right\}}{{\mathsf{m}_{t_N}}}\right)\nonumber\\
&&+\frac{1}{3}f^{(120-)}_{_{\acute{w}\acute{x}}}f^{(120-)}_{_{\acute{y}\acute{z}}}\left(\sum_{M=1}^{8}\frac{\left\{\sum_{r=1}^{2}{\cal G}_{\acute{u}\acute{v}}^{(r)}{U}_{t_{rM}}\right\}
{{V}_{t_{3M}}}}{{\mathsf{m}_{t_M}}}\right)\left(\sum_{N=1}^{8}\frac{{{U}_{t_{3N}}}
{{V}_{t_{3N}}}}{{\mathsf{m}_{t_N}}}\right)\nonumber\\
&&-\sqrt{\frac{2}{3}}f^{(120-)}_{_{\acute{y}\acute{z}}}\left(\sum_{M=1}^{8}\frac{\left\{\sum_{r=1}^{2}{\cal G}_{\acute{u}\acute{v}}^{(r)}{U}_{t_{rM}}\right\}
\left\{\sum_{s=1}^{2}f^{(10_s+)}_{_{\acute{w}\acute{x}}}{V}_{t_{sM}}\right\}}{{\mathsf{m}_{t_M}}}\right)\left(\sum_{N=1}^{8}\frac{{{U}_{t_{3N}}}
{{V}_{t_{3N}}}}{{\mathsf{m}_{t_N}}}\right)\nonumber\\
&&-\sqrt{\frac{2}{3}}f^{(120-)}_{_{\acute{w}\acute{x}}}\left(\sum_{M=1}^{8}\frac{\left\{\sum_{r=1}^{2}{\cal G}_{\acute{u}\acute{v}}^{(r)}{U}_{t_{rM}}\right\}
{{V}_{t_{3M}}}}{{\mathsf{m}_{t_M}}}\right)\left(\sum_{N=1}^{8}\frac{{{U}_{t_{3N}}}
\left\{\sum_{r=1}^{2}f^{(10_r+)}_{_{\acute{y}\acute{z}}}{V}_{t_{rN}}\right\}}{{\mathsf{m}_{t_N}}}\right).
\end{eqnarray}

\pagebreak

\end{document}